# Bit Patterned Magnetic Recording: Theory, Media Fabrication, and Recording Performance

Thomas R. ALBRECHT, Hitesh ARORA, Vipin AYANOOR-VITIKKATE, Jean-Marc BEAUJOUR, Daniel BEDAU, David BERMAN, Alexei L. BOGDANOV, Yves-Andre CHAPUIS, Julia CUSHEN, Elizabeth E. DOBISZ, Gregory DOERK, He GAO, Michael GROBIS, Bruce GURNEY, *Fellow, IEEE*, Weldon HANSON, Olav HELLWIG, Toshiki HIRANO, Pierre-Olivier JUBERT, Dan KERCHER, Jeffrey LILLE, Zuwei LIU, C. Mathew MATE, Yuri OBUKHOV, Kanaiyalal C. PATEL, Kurt RUBIN, Ricardo RUIZ, Manfred SCHABES, Lei WAN, Dieter WELLER, Tsai-Wei WU, and En YANG

HGST, a Western Digital Company, San Jose, California, USA

Bit Patterned Media (BPM) for magnetic recording provides a route to thermally stable data recording at >1 Tb/in$^2$ and circumvents many of the challenges associated with extending conventional granular media technology. Instead of recording a bit on an ensemble of random grains, BPM is comprised of a well ordered array of lithographically patterned isolated magnetic islands, each of which stores one bit. Fabrication of BPM is viewed as the greatest challenge for its commercialization. In this article we describe a BPM fabrication method which combines rotary-stage e-beam lithography, directed self-assembly of block copolymers, self-aligned double patterning, nanoimprint lithography, and ion milling to generate BPM based on CoCrPt alloy materials at densities up to 1.6 Td/in$^2$ (teradot/inch$^2$). This combination of novel fabrication technologies achieves feature sizes of <10 nm, which is significantly smaller than what conventional nanofabrication methods used in semiconductor manufacturing can achieve. In contrast to earlier work which used hexagonal close-packed arrays of round islands, our latest approach creates BPM with rectangular bitcells, which are advantageous for integration of BPM with existing hard disk drive technology. The advantages of rectangular bits are analyzed from a theoretical and modeling point of view, and system integration requirements such as provision of servo patterns, implementation of write synchronization, and providing for a stable head-disk interface are addressed in the context of experimental results. Optimization of magnetic alloy materials for thermal stability, writeability, and tight switching field distribution is discussed, and a new method for growing BPM islands from a specially patterned underlayer – referred to as "templated growth" – is presented. New recording results at 1.6 Td/in$^2$ (roughly equivalent to 1.3 Tb/in$^2$) demonstrate a raw error rate <10$^{-2}$, which is consistent with the recording system requirements of modern hard drives. Extendibility of BPM to higher densities, and its eventual combination with energy assisted recording are explored.

## I. INTRODUCTION

SINCE the hard disk drive was introduced in 1956, its storage capacity has increased tremendously – from 5 megabytes for the first IBM 350 disk storage unit to multiple terabytes for 2014 disk drives. This dramatic increase in storage capacity has been made possible by the disk drive industry sustaining for nearly 60 years average annual increases > 30% in the number of bits/in$^2$ stored in the layer of magnetic media on the rotating disks surfaces. In recent years, however, these annual increases in areal density have slowed to < 15%, primarily due to challenges in further extending the capability of conventional recording media, which consists of a continuous thin film of granular magnetic material.

To maintain adequate signal-to-noise ratio (SNR) to recover recorded data with an acceptably low error rate, the grain sizes within the magnetic film generally need to scale with size of the recorded bits. For a number of years, it was assumed that the maximum achievable density for granular media would be reached when scaling grains to smaller volume would result in the loss of thermal stability. This understanding has often been described as a "trilemma" in which thermal stability, media writeability, and media signal-to-noise ratio (SNR) drive mutually exclusive objectives in media design [1]. More recently, however, it has become apparent that even before this long-predicted trilemma is reached, scaling granular media properties to achieve sufficient SNR to support an areal density (AD) of >1 Tb/in$^2$ is already very challenging.

While a number of recent recording system innovations, such as shingled recording, two dimensional magnetic recording (TDMR) and the use of array heads [2] show promise for modest increases in AD, only two concepts have been proposed to fundamentally address the shortcomings of conventional media. Energy assisted recording – including heat-assisted magnetic recording (HAMR) [3] and microwave-assisted magnetic recording [4] – enables further scaling of granular media by increasing the effective write field and write field gradient available and allowing the use of higher coercivity small grain media with good thermal stability. The other option is bit patterned magnetic recording (BPMR), which achieves high SNR and thermal stability by replacing the random grains of conventional media with lithographically patterned magnetic islands which are significantly larger than, and thereby more thermally stable than conventional media grains [5], [6], [7]. Although both technologies offer the possibility of substantial AD gain, FePt granular media for HAMR still presents SNR difficulties typical of granular media, making it challenging to realize the potential of HAMR for large areal density gains.

In this article, we provide a comprehensive description of BPMR as developed by our company. This description includes the theory and modeling of BPMR, the innovative fabrication processes developed for realizing bit patterned media (BPM), and methods for integrating BPMR into a working recording system. Here we demonstrate a BPMR AD of 1.6 Td/in² (teradot/inch²), which is roughly equivalent to 1.3 Tb/in² when recording system overhead is taken into account. We also discuss how BPMR could be extended to areal densities several times this.

As shown in Fig. 1(a), a bit recorded on granular media is stored in an ensemble of grains. BPM, as shown in Fig. 1(b), consists of lithographically defined magnetic islands on the disk surface, and each island is magnetized to store an individual bit. For granular media, the SNR increases with the number of grains per bit; use of too few grains results in positional jitter due to rough boundaries arising from finite grain size and the magnetic indivisibility of individual grains. For BPM, SNR no longer depends on the number of grains per bit, but instead depends on fabrication tolerances that govern the positional jitter and the addressability of the islands during writing.

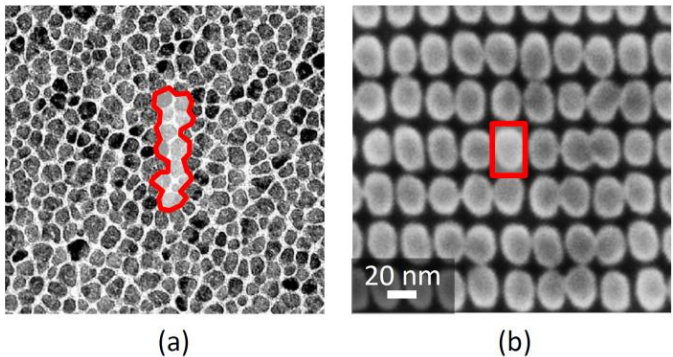

Fig. 1. Comparison of (a) granular and (b) bit patterned media. On granular media, an individual bit is recorded on an ensemble of grains (as represented by the red outline), while on bit patterned media, each island stores one bit.

Conceptually, BPMR is relatively simple to understand and model (particularly in comparison to granular media), and experiments correlate well with modeling results. Magnetic materials for BPM are less complex than those typically used for granular media; simple alloys or multilayers generally suffice with underlayers of modest complexity. On the other hand, integration of BPMR into the recording system of a hard disk drive (HDD) introduces significant new complexity by requiring a write synchronization system and servo technology suitable for following eccentric, prepatterned tracks at track pitch of ~20 nm.

The grand challenge for BPMR, however, is high volume low cost manufacturing of BPM with good magnetic properties and tight fabrication tolerances. What makes this challenge particularly daunting is the small feature size required – generally <20 nm full pitch (< 10 nm feature size) in the down-track direction, which is beyond the capabilities of conventional lithographic methods as practiced by the semiconductor industry. Consequently, we devote major portions of this article to presenting innovative fabrication strategies for BPM. These innovations include the implementation of rotary-stage e-beam lithography for initial pattern definition, the development of block copolymer self-assembly and self-aligned double patterning to increase the areal density of the master template, extension of nanoimprinting for the transfer the nanoscale patterns from templates to resist films, and the use of suitable masking and etching processes for transferring these nanoscale patterns into magnetic media.

The importance of nanofabrication methods developed for BPM may well extend beyond magnetic recording. Other applications where similar nanofabrication strategies may find applications include semiconductor devices (particularly memory and storage devices where highly periodic structures are used), solar energy conversion (where nanostructuring may increase efficiency), biomaterials and pharmaceuticals (in which specific nanoscale physical shapes may enhance specificity of target interactions), and catalysis (where nanostructuring can improve selectivity and efficiency and perhaps enable routes to synthesize entirely new molecular species).

## II. BPM Data Storage Requirements

The goal of using BPM as an HDD recording medium has driven many of the design choices taken in the development of BPM. It is thus useful to understand the vision and requirements for how BPM would be used in HDDs. On a very basic level an HDD consists of a spinning data storage medium (the disk), an element that reads and writes data to the disk (the head), and a mechanical arm that moves the head around the disk (the actuator). In addition to these three components is a series of electrical and mechanical subsystems that enable the drive to function within a greater computing or storage ecosystem. These systems will be described in more detail in a later section.

### A. Sector Failures and Bit Error Rates

In order for a recording system to be commercially viable it needs to store and retrieve customer data with exceptional consistency and reliability over the lifetime of the drive. While ideally the probability of data loss for drive would be zero, in practice a finite loss rate is tolerable as long as the rate is sufficiently low that the customer can effectively provision for it. The data loss rate is usually specified as an unrecoverable bit error rate, which for HDDs is typically stated as <1 error in $10^{14}$-$10^{16}$ bits read depending on the application. For the subsequent discussion we will assume that bits are grouped into 4kB sectors on the disk and that not recovering a single bit is equivalent to losing the whole sector. We will use the mid-range unrecoverable sector failure rate (SFR) of $10^{-11}$ or one bad sector for every 400 TB written.

A typical recording system has a raw bit error rate that far exceeds the unrecoverable bit error rate. The read channel employs error correcting codes (ECC) that can correct erroneous bits at the expense of additional storage overhead. At the detector stage of the read channel an error rate of $10^{-2}$ is well within the recovery threshold of the ECC. Read errors due to low SNR are easier to manage than hard errors due to an incorrectly written bit, as a hard error does not leave any residual information that can be used to determine the original data. As a result, the allowed write error rate for BPM will be slightly lower than an acceptable read error rate on conventional granular media. The exact value will depend on optimization and tolerated storage overhead [8]. To simplify this discussion we assume that a write error rate exceeding $10^{-2}$ will cause an unrecoverable sector failure error in the channel.

Unrecoverable sector failures stem from four main sources: mis-synchronization, adjacent track erasure, defects and thermally activated reversal. In our simplified sector failure rate model, if less than 1 in 100 bits are in the incorrect state or defective, the ECC will correctly decode the sector. Roughly, the bit error rate (BER) for written-in errors can be expressed as a sum of these four sources with a correlation term

$$BER_{total} = BER_{sync} + BER_{ATE} + \rho_{defect} + \rho_{thermal} - c_{BER} . \quad (1)$$

The bits affected by synchronization, track misregistration (TMR), and thermal stability come predominantly from smaller subpopulation in the tails of the parameter distributions. The correlation term $c_{BER}$ reflects the fact that a bit can only contribute one error despite the many ways it could have changed its state. While readback errors can occur, subsequent rereading may recover such errors. Defects can be missing or merged islands, as well as bits far outside the normal distribution of island properties. These defects are assumed to be randomly distributed around the disk as sectors with high concentrations of such defects can be detected and omitted. How these various sources arise will be discussed in this section.

### B. BPM Recording Theory and Concepts

Many of the theoretical aspects the recording physics of BPMR are well developed and have an extensive record in the literature [9], [10], [11], [12]. If the island has a single magnetic layer, each island switches quasi-uniformly and is generally well-described by Stoner-Wohlfarth theory [13]. In exchange spring media [14], [15], two or more materials are combined to render the islands writeable while maintaining thermal stability at sufficiently small island volume for the desired areal bit density. Yet even in the case of exchange spring media the underlying recording physics of BPMR media is relatively simple, since there are no complications from the traditional granular media noise, grain size distributions, etc., of unpatterned media. Challenges for BPMR designs become evident, however, when the media design questions are linked to BPM fabrication capabilities, where limitations in lithography, imprint technology, etch processes, etc. emerge as major constraints.

#### 1) Effective bit position jitter & bit error rates

Unlike conventional recording, bit patterned media requires precise registration between the passing of patterned bits under the write head and the switching of the write current. Mis-synchronization and TMR will cause bit errors that can lead to sector failures. There is a finite margin for how much the write phase can drift from ideal in order to maintain low BER rates. The margin can be thought of as arising from an effective bit position jitter of the magnetic islands. This effective position jitter that has three main sources: lithographic jitter $\sigma_{litho}$, magnetic switching jitter, and write clock jitter $\sigma_{sync}$. The magnetic switching jitter is the magnetic switching field distribution normalized by the head field gradient. There are two contributions to the magnetic switching field distribution. The first is the intrinsic switching field distribution $\sigma_{iSFD}$, and the second is the data dependent dipolar interaction field distribution $\sigma_{dipole}$. The interaction fields in BPM are typically only dipolar, though it is possible to create exchange bridging among the islands [16], [17]. If the various sources are uncorrelated Gaussian distributed variables, we find

$$\sigma_{eff}^2 = \left(\sigma_{iSFD}^2 + \sigma_{dipole}^2\right) / \left(dH/dx\right)^2 + \sigma_{litho}^2 + \sigma_{sync}^2 . \quad (2)$$

In the effective bit jitter model, a bit can be written incorrectly if its effective jitter places it in its neighbor's write window, as shown in Fig. 2.

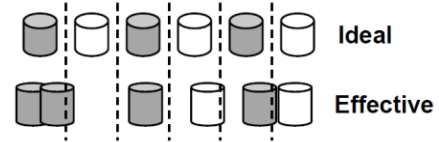

Fig. 2. Ideal placement vs. effective bit placement. The center of the second bit is effectively in the first bit's write window, indicated by the dotted lines and will be written incorrectly.

The main contributor to Eq. 2 is the iSFD. Dipolar interactions are usually less important, since they average to a small value in a randomly magnetized island array with roughly 50% up and 50% down magnetized islands (HDDs use coding and randomization to enforce roughly even distributions, regardless of actual data patterns recorded). In addition, shielding by the write head reduces the effect of dipolar fields during the write process. The synchronization jitter is assumed to vary slowly in a drive and is more appropriately treated as a phase offset for the sector write. The residual fast variation is an insignificant contributor to the effective bit position jitter.

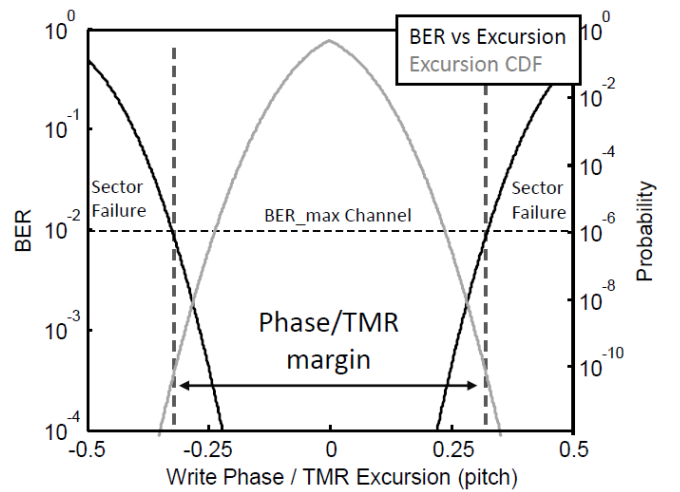

Fig. 3. Bit error rate (black) vs. write phase or TMR excursion. The gray curve shows the cumulative distribution function (CDF) representing the probability that a write phase or TMR excursion occurs beyond the amount indicated on the x-axis. The dotted lines show the margin assuming the channel fails when the error rate exceeds $10^{-2}$.

For a Gaussian distribution of errors, the BER is described by a sum of two error functions, corresponding to overwriting previously written bits and to having insufficient field to write the target bits:

$$BER = \alpha \left( 1 - \frac{1}{2} erf\left(\frac{\delta + 1/2}{\sigma_{eff}}\right) + \frac{1}{2} erf\left(\frac{\delta - 1/2}{\sigma_{eff}}\right) \right) \quad (3)$$

Here, the parameter $\delta$ represents the excursion of the write phase from optimal value. An analogous equation can be written for TMR excursions. The normalization factor $\alpha$ is equal to ½ for random data, but more generally $\alpha = \frac{1}{2}(1-c)$, where c is the correlation between successive values of bits in the sequence. The excursion $\delta$ and effective bit position jitter have been normalized by the lattice constant. The resulting BER is shown in Fig. 3 for the case of uncorrelated data ($\alpha = \frac{1}{2}$).

### 2) Thermal stability

The thermally activated reversal rate depends on the bit volume V, the anisotropy energy density $K_U$, Boltzman's constant $k_B$, and the drive temperature T. The typical way to estimate the rate is to use the Arrhenius switching model in which the reversal rate is given by $R = f_0 \exp(-K_U V / k_B T)$. Using the BER criteria of $10^{-2}$, in a sector failure rate of $<10^{-11}$ will be maintained after 10 years if the $K_U V / k_B T > 52$. However, a population of islands will have a distribution of volumes and anisotropy. It is reasonable to estimate the required average $K_U V / k_B T$ by requiring that only at most $10^{-2}$ of the islands have a $K_U V / k_B T < 52$. If we assume a Gaussian distribution for the product $K_U V$ with a 10% sigma, we find that $<K_U V / k_B T> > 67$ will allow thermally stable media. The thermally activated reversal rate is accelerated by adjacent track writing due to presence of the residual write fields on the adjacent track. Residual write fields are temporary but will accumulate near tracks that are written frequently. Dipolar fields can also accelerate thermal decay, especially for islands in areas of uniform magnetization. The impact on the thermal decay can be partially mitigated by imposing restrictions on allowed 2D patterns.

### C. Data track writing architecture

There are several manners in which data can be written to the disk as illustrated in Fig. 4(a-c). These are centered track recording, shingled magnetic recording [18], and hypertrack recording [19], [20]. In centered track recording the write head writes a single track in a single pass, without overwriting the adjacent tracks. Centered track recording requires a match between the physical dimensions of the writer and the track pitch. At small track pitches, conventional writer designs have a difficulty generating sufficient fields. The narrow writer problem can be overcome by writing in a shingled magnetic recording (SMR) or hypertrack magnetic recording (HTMR) fashion as illustrated in Fig. 4(b,c). In SMR the writer spans several tracks and each track is written sequentially like overlapping shingles. In HTMR two tracks are written simultaneously using a writer than spans two tracks.

At first glance, by writing and reading two tracks at a time, HTMR would appear to render the system equivalent to one with a 4X larger bit aspect ratio (BAR), which would generally be considered advantageous from a recording system integration point of view (see Section II-D). While the need for a very narrow writer is alleviated in HTMR, the requirement to carefully match the writer to the track pitch is not. Furthermore, although the effective track pitch is doubled, the write head side gradient and track-following servo requirements to prevent adjacent track writing are not correspondingly relaxed. An additional concern is that in order to compensate for write head skew, as present anywhere but the middle of the disks, the track needs to be skewed correspondingly to properly space the write field transitions.

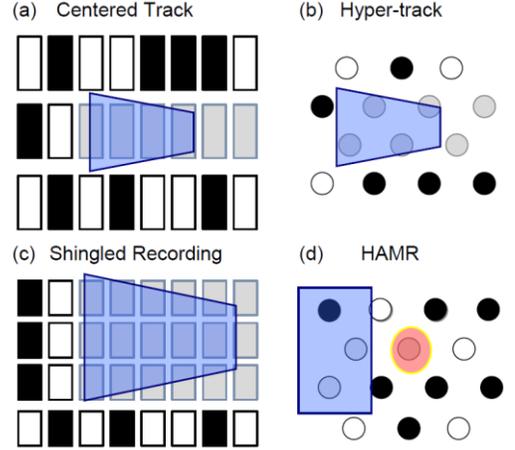

Fig. 4. Four implementations of BPR showing (a) centered track, (b) hypertrack, (c) shingled, and (d) heat assisted magnetic recording. The blue quadrangle represents the write pole and the red spot represents the heating location.

Combining heat assisted magnetic recording (HAMR) with BPM provides many synergistic benefits, some of which will be discussed later. From the standpoint of the recording system, HAMR can provide a very tight thermal spot, the extent of which can be controlled in order to provide an optimal match to the track pitch (Fig. 4(d)).

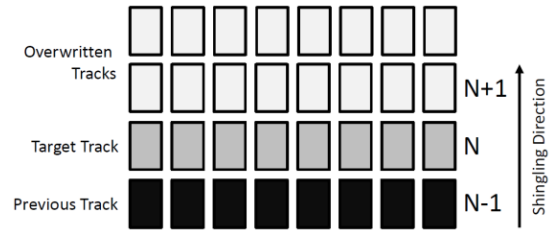

Fig. 5. Schematic of shingled recording. In this example data is written from bottom to top. The head write width spans the target track and overwritten tracks.

SMR has the advantage of allowing relatively large write heads with higher effective write fields compared to the narrower heads of centered writing and is the preferred architecture for implementing BPM with conventional magnetic recording heads. The shingled write process [2] writes islands with a head that is wider than the track pitch. Data are organized into large blocks such that multiple adjacent tracks are written sequentially (rather than a subset of blocks on a single track). Fig. 5 shows the basic geometry of down-track pitch, cross-track pitch, island size, etc.

### D. BAR considerations

When considering manufacturing of BPM, it becomes evident that the aspect ratio of the bit-cell (BAR) must be kept relatively small compared to traditional designs on granular media, where the smallest down-track bit spacing can be reduced without concomitant expense of a reduction of lithographic feature sizes. A relatively low BAR in BPMR helps avoid the need for fabricating excessively small feature sizes in the down-track direction. However, relatively little has been previously published regarding optimization of the BAR for BPMR, especially with shingled write schemes.

Even though the equations describing down-track and cross-track deviations in writing are symmetrical, there exists a subtle asymmetry between these two types of deviations. The first is that adjacent track erasures are in general more costly than mis-synchronization errors. If a mis-synchronization event is detected, the sector can be rewritten, but an adjacent track erasure event loses customer data unless the adjacent track data is already known. The second asymmetry arises from the fact that TMR is harder to maintain with tight tolerances than write phase, since electrical feedback can have much higher bandwidth than mechanical feedback. The asymmetry leads to a desire to have higher BAR in the recording system. As a result, a BAR such as 0.87 for a hexagonal island array is undesirable. In addition, write head designs typically have not had equal down-track and cross-track gradients. The down-track gradient in conventional heads is usually at least 50% larger than the cross-track gradient, which further increases the optimal BAR. We note that not all write head designs have such asymmetric gradients. Section V-E further explores BAR considerations for BPMR.

## III. MAGNETIC MATERIALS FOR BPM

### A. Prepatterned Substrates vs. Etching Continuous Films

The initial approach for BPM fabrication was to prepattern the substrate with pillar features, upon which magnetic films with perpendicular anisotropy were deposited [21], [22], [23]. Co/Pt and Co/Pd multilayer systems were the preferred materials for this fabrication approach, as those systems obtain their perpendicular anisotropy from multilayer interfaces and therefore do not rely on crystal anisotropy as much as other well-known high perpendicular anisotropy systems, such as CoCrPt alloys and $L1_0$ ordered FePt compounds. The prepatterning approach allows the pillars to be etched into substrate or underlayer materials favorable for reactive ion etching (RIE) and avoids the possibility of damage to the magnetic layer that might arise from patterning it directly by etching. Disadvantages of prepatterning, however, include the presence of magnetic material in the trenches between pillars [24] and difficulties adapting this approach for magnetic materials other than multilayers, which generally require thick seed and underlayers not compatible with deposition onto small pillars [25].

As lithographic and patterning techniques improved, the order of fabrication steps was reversed and a continuous film of magnetic material was deposited prior to the patterning steps.

This greatly expanded the variety of magnetic materials that could be used, including CoCrPt alloys, which have been extensively studied and currently appear in conventional perpendicular magnetic recording (PMR) media [26], [27], [28].

Regardless of which fabrication strategy is pursued, it is essential to include a continuous (not patterned) soft magnetic underlayer (SUL) below the magnetic seed layer structure in order to close the magnetic flux from the write head and to keep the magnetic flux lines at the recording point more perpendicular to the layer structure in a manner similar to conventional perpendicular recording.

### B. Magnetic Properties and Switching Field Distribution

To create etched BPM the initial films must be continuous, uniform and smooth in order to be patterned into highly uniform magnetic islands. In practice there can be significant differences in physical structure and materials properties from island to island, which cause variations in the switching field of the islands, usually referred to as the switching field distribution (SFD) [29], [30], [31], [32]. As discussed above in Section II-B, two distinct components contribute to the SFD: 1) the intrinsic switching field distribution (iSFD), which is caused by variations in the materials properties and microstructure (such as defects, grain boundaries, crystallite misorientation and also different degrees of island edge damage due to patterning [33], [34]) and 2) the dipolar contribution, which is an additional broadening that originates from different dipolar fields at each island arising from the specific magnetic state of nearby islands [35]. In order to ensure that an island being addressed by the writer properly reverses without disturbing the state of its neighbors it is important that the iSFD be kept as small as possible.

In order to evaluate and compare the reversal properties of different BPM systems, which can have different average island coercivity (i.e. average island switching field), it is useful to consider iSFD normalized to the coercivity, i.e. iSFD/$H_C$, as a figure of merit. We have summarized in Fig. 6 the evolution of SFD/$H_C$ (earlier in time) and iSFD/$H_C$ values (later in time, as the measurement technique became available) for various BPM magnetic material systems as they evolved with increasing areal density (AD). With prepatterned substrates and Pt or Co/Pd multilayers, SFD/$H_C$ increases from a ~3% for large micrometer sized islands to 12-18% for diameters of 30 nm or less. This trend arises from the fact that as islands become smaller, properties are averaged over less magnetic material.

In order to lower SFD/$H_C$ alternative magnetic structures were investigated, including exchange spring and exchange coupled composite magnetic multilayer systems [36], [15]. As shown in Fig. 6, for 45 nm pitch island arrays (320 Gd/in²), such heterogeneous magnetic structures decreased SFD/$H_C$ from >11% down to 7.5%.

As density increased and island diameters dropped below 30 nm, it became clear that lithographically introduced distributions in island placement and diameter were significant contributors to the SFD. Adoption of block copolymer self-assembly (see Section IV-B) resulted in tighter fabrication tolerances and an immediate improvement in SFD/$H_C$ (see Fig. 6) [17]. At an AD of 500 Gd/in² (hexagonal pitch = 39 nm) SFD/$H_C$ dropped from 12-18% down to 8-9%.

At that point in time, new methods were introduced, which allowed separate measurement of SFD and iSFD using minor loop techniques [29], [31]. For typical Co/Pd and Co/Pt multilayers the saturation magnetization $M_S$ is quite low (around 700 emu/cc) and thus the iSFD contribution to the SFD dominates. At 500 Gd/in$^2$ the best iSFD/$H_C$ values obtained were around 6%.

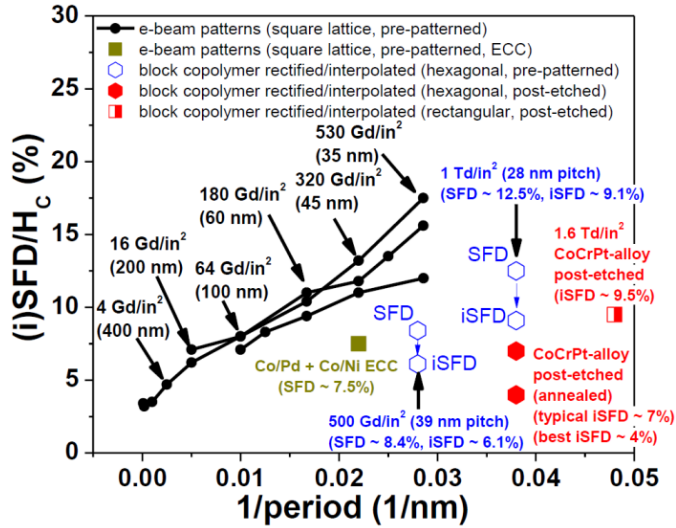

Fig. 6. Evolution of SFD/$H_C$ for various BPM generations of processing and materials. Initially BPM fabrication was based on e-beam-patterned substrates and subsequent deposition of Co/Pd multilayers (black circles). Exchange coupled composite structures with hard magnetic Co/Pt and soft magnetic Co/Ni multilayers (gold square) achieved lowest SFD/$H_C$. Later, block copolymer patterns were used (open blue hexagons) to achieve tighter patterning tolerances, which reduced (i)SFD/$H_C$ consistently. Finally, pattern transfer and hard mask removal techniques were improved, permitting direct etching of a continuous magnetic thin film (red solid hexagons and red half-filled square). With this approach, CoCrPt thin films on Ru underlayers similar to those used in PMR media could be explored for BPM applications and achieved lower iSFD/$H_C$ than previously used Co/Pd multilayer systems. Most recently iSFD/$H_C$ values of 4% [26] and 9.5% (see also Fig. 7) have been reached for post-etched CoCrPt based BPM at densities of 1 Td/in$^2$ and 1.6 Td/in$^2$ respectively.

When first moving to 1 Td/in$^2$ densities, prepatterned substrates (with block copolymer patterning) were still used and achieved SFD/$H_C$ and iSFD/$H_C$ around 12.5% and 9% respectively [23]. However, at the same time direct milling and etching capabilities of initially continuous magnetic thin films had made significant progress [27], [37]. Depositing the magnetic media layer onto a flat substrate with no limitations in underlayer materials and thickness before applying the actual patterning process was shown to be highly advantageous. However, it was found that the milling of multilayers caused sufficient edge damage, which at densities above 500 Gd/in$^2$, where a significant fraction of the island volume is close to the perimeter, degraded the media significantly. Therefore the move to etching full films into islands also drove a shift from multilayers towards conventional alloy based materials like CoCrPt with Ru based seed layers similar to granular PMR media. Over the past 5 years the trend from prepatterned substrates to etched films has required improvements in etching, film processing and materials choices.

Compositions of CoCrPt alloys considered for BPM applications are similar to those used in PMR media, but without any segregant material, i.e. Co$_{100-65}$Cr$_{0-20}$Pt$_{0-25}$ (subscripts indicate atomic %). Here we report on two different alloy compositions: Co$_{70}$Cr$_{18}$Pt$_{12}$, with a lower uniaxial perpendicular anisotropy $K_U$ and higher Cr content, and Co$_{80}$Cr$_{10}$Pt$_{10}$, with a higher anisotropy and lower Cr content. For both alloy compositions we found that the introduction of a very thin discontinuous Ta-oxide onset layer in between the low pressure Ru seed and the CoCrPt media layer decreased iSFD/$H_C$ significantly (by about 30-50%) [38]. The structures using alloys reported on in this section use Ta-oxide onset layers in the range 0.2-0.6 nm, which create a network of misfit dislocations on a length scale significantly smaller than the island diameter within the CoCrPt media layer, thus creating islands which are magnetically more similar with a narrower iSFD/$H_C$.

In addition to improving magnetic properties by modifying the layers in the deposited stack, we have found that vacuum annealing applied to the patterned CoCrPt islands after hard mask removal improves the anisotropy, thermal stability and iSFD/$H_C$ of the islands [39]. This improvement in island magnetic properties can be mainly attributed to a thermally activated Cr diffusion process that transforms the patterned CoCrPt islands into a high moment and high anisotropy Cr depleted CoPt-rich island core with a Cr/Cr-oxide rich protective island shell. Cr diffusion within CoCrPt media is well-known and was previously exploited in early Longitudinal Magnetic Recording (LMR) media generations to magnetically separate grains. However, in the discrete island BPM geometry, with distinct trench regions in between islands, the thermally activated Cr diffusion is much more extensive. In a previously reported annealing study we focused on Cr-rich Co$_{70}$Cr$_{18}$Pt$_{12}$ media and optimized annealing temperature and time for that specific case [39]. Here we note that with other CoCrPt media layer compositions fabricated at 1 Td/in$^2$ BPM we have obtained $H_C$ above 8 kOe and iSFD/$H_C$ values below 5% (see Fig. 6).

## C. Magnetic Materials for High Areal Density

For areal densities beyond 1.2 Td/in$^2$, additional media layer improvements were implemented to keep iSFD/$H_C$ below 10%. As island volume V shrank further, it also became increasingly important to maintain a high thermal stability ratio $K_U V/k_B T > 80$ (where $k_B$ is Boltzmann's constant and T is the temperature). While recording layer thicknesses of 8 nm and thicker have been successfully integrated in BPM systems below 1.2 Td/in$^2$, at higher density the ability to etch increasingly narrow trenches without affecting the target island size limits the magnetic layer thickness that can be used, and therefore the $K_U V/k_B T$ ratio. Furthermore, the perpendicular magnetic anisotropy density in CoCrPt alloy films decreases as a function of thickness, which lowers thermal stability even further as we move to thinner films. For example, the effective perpendicular anisotropy density of a Co$_{80}$Cr$_{10}$Pt$_{10}$ alloy film decreases by about 15% when the film thickness is reduced from 8 nm to 4 nm. The reduction of the perpendicular anisotropy density is associated with a strain-induced surface anisotropy of opposite sign to the volume anisotropy. The thickness dependence (in the 4-8 nm range) of the magnetic parameters of patterned Co$_{80}$Cr$_{10}$Pt$_{10}$ magnetic layers at 1.0 Td/in$^2$ reflects this deterioration of the perpendicular anisotropy density with decreasing magnetic layer thickness. $H_C$ decreases and iSFD broadens with decreasing film thickness, leading to unacceptable iSFD/$H_C$ values of ~20% for 4 nm thick $Co_{80}Cr_{10}Pt_{10}$ layers. The thermal stability ratio decreases from 150 for 8 nm media down to 50 at 4 nm. In order to compensate for the loss in magnetic volume and the decrease of the perpendicular anisotropy density with thinner recording layers, a natural approach is to vary the alloy composition and in particular employ alloys with higher Pt content and therefore higher $K_U$ [40]. However, BPM made of such alloys usually exhibit broader iSFD, which may be caused by an increase in stacking fault density and other defects in high Pt content cobalt alloys.

A magnetic stack combining a hard magnetic layer and a softer magnetic layer that are strongly exchange coupled was also evaluated, namely $Co_{70}Cr_7Pt_{25}$/ $Co_{80}Cr_{10}Pt_{10}$ bilayers. In this composite system, the $Co_{70}Cr_7Pt_{25}$ provides an increase of the overall perpendicular anisotropy density, while the softer magnetic layer keeps the effective iSFD value low [15], [36], [41]. A study of $H_C$ and iSFD as a function of the thickness ratio of the hard and soft magnetic layer, where the total magnetic layer thickness was kept constant (6 nm), showed a linear increase of $H_C$ with the thickness ratio. BPM fabricated at 1.0 Td/in$^2$ incorporating 3 nm $Co_{70}Cr_7Pt_{25}$ / 3 nm $Co_{80}Cr_{10}Pt_{10}$ bilayer showed $H_C$ = 6.5 kOe and $K_U V/k_B T$ ~ 190, compared to a 6 nm $CoCr_{10}Pt_{10}$ media with $H_C$ = 4.0 kOe and $K_U V/k_B T$ ~ 120. However, this improvement in the magnetic properties came at the expense of increasing iSFD, which broadened significantly with increasing the hard/soft layer thickness ratio. Media with a bilayer of 3 nm $Co_{70}Cr_7Pt_{25}$ / 3 nm $Co_{80}Cr_{10}Pt_{10}$ exhibited iSFD of 550 Oe, compared to 220 Oe for 6 nm $Co_{80}Cr_{10}Pt_{10}$ (i.e., iSFD/$H_C$: 8.5% vs. 5.5%).

An alternative approach has been undertaken to fabricate media at an AD around 1 Td/in$^2$ that contains only a single Co-alloy layer with relatively low Pt content (10-12%), and which uses the effect of interface induced surface anisotropy to increase the overall effective anisotropy beyond that available from bulk anisotropy alone, and without increasing iSFD. In particular, we have found that an ultrathin Pt layer (less than 1 nm thick) between the Ru underlayer and the magnetic alloy can increase anisotropy by up to 30%. Similarly to previously reported Co/Pt multilayers [42], we believe that the enhancement of the anisotropy is an electronic effect associated with the enhancement of the Co orbital moment at the Pt/CoCrPt interface and the magnetic polarization of the Pt underlayer. This ultrathin layer does not affect the growth of the magnetic alloy layer significantly, which maintains a c-axis distribution of about 2.6°, similar to the value obtained when the alloy is grown directly on Ru. Incorporating this composite structure into the BPM stack, we observe an increase of $H_C$ by about 2 kOe and $K_U V/k_B T$ increases by about 50. While the iSFD also increased, the iSFD/$H_C$ remained low. Additionally, because Pt is not magnetic unless it is next to the Co alloy, the etch depth does not need to extend into the Pt layer in order to magnetically decouple the islands, keeping the overall etch depth small.

In further improvement of this structure, we have found that the anisotropy can be additionally increased by replacing the capping layer (e.g. $SiN_x$, or similar protective overcoat) with another Pt layer, which adds a second CoCrPt/Pt interface. Such media was etched using a 1.6 Td/in$^2$ template (Fig. 7) to obtain the media described in Section VI. At this density, the media had magnetic properties measured by polar Kerr magnetometry of $H_C$ ~ 6 kOe, iSFD/$H_C$ < 10% and $K_U V/k_B T$ > 100. On its own, this type of media provides an alternative to the post-etch in-vacuum annealing approach described earlier [7], [39]. We expect, however, that combining the annealing with Pt interlayers can achieve even higher thermal stability and lower iSFD/$H_C$. We have found already that after annealing the $H_C$ of Pt/$Co_{70}Cr_{18}Pt_{12}$ can increase from 5 to 8 kOe. Overall, single CoCrPt magnetic films with Pt or other interlayers combined with additional post-etch annealing steps are a class of promising candidates for BPR media with areal densities well beyond 1.6 Td/in$^2$.

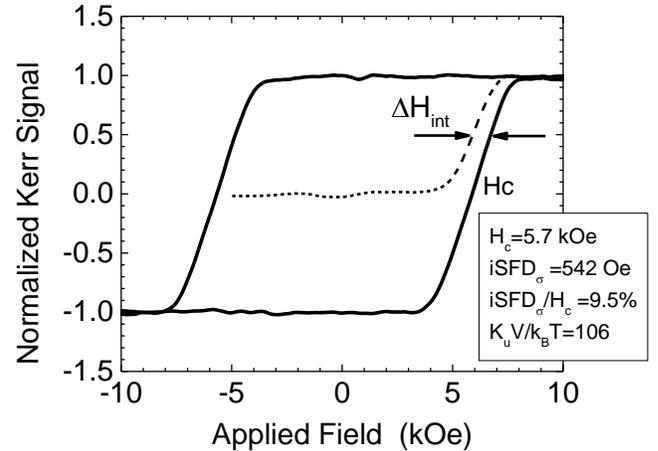

Fig. 7. Polar Kerr characterization of patterned media at 1.6 Td/in$^2$ areal density. Media is 0.6 nm Pt / 6 nm $Co_{80}Cr_{10}Pt_{10}$/ 2 nm Pt. iSFD$_\sigma$=$\Delta H_{int}$/1.35 was determined from the partial loop (dashed line) and the thermal stability parameter was extracted from the fit of $H_C$ versus the field sweep rate (not shown) using the Sharrock formula.

*D. Magnetic Materials for Templated Growth*

Templated growth (TG) provides an alternative means to generate ultra-high density BPM. In TG BPM, nucleation sites with the desired pattern are generated by nanolithographic patterning. A sequence of underlayers, magnetic layers and overcoat layers are then deposited on the nucleation features. The magnetic layers can consist of co-sputtered magnetic alloy and oxide segregants to obtain a periodic array of magnetic islands in a matrix of oxides [43]. The presence of the oxide matrix is a key factor distinguishing this approach from the prepatterned substrate method discussed previously.

In order for TG BPM to obtain the best recording media properties it is desirable to start with nucleation features with consistent size, shape, and surface conformation. They must encourage growth of magnetically decoupled islands with no decoupled sub-grains within one island; excellent perpendicular texture with a tight c axis orientation distribution (e.g. FWHM less than 2.5 degrees); and a tight switching field distribution across the disk. Using lithographic techniques to generate nucleation features in a manner similar to that used for forming islands in etched BPR media (see Section IV-H), we have demonstrated growth of islands with good magnetic properties.

Fig. 8(a) shows the TEM cross-section electron energy loss spectroscopy (EELS) map of a 1 Td/in$^2$ TG BPR media structure. In this sample, a hexagonal array of Cr nucleation features were generated by nanoimprint and etch back processes. Various thin underlayers, followed a thicker Ru layer, were then deposited onto the templated surface. The CoCrPt alloy magnetic layer was then cosputtered with an oxide segregant on top of the Ru layer. The Ru growth was dominated by the self-shadowing effect, [44] resulting in a columnar shaped topographic structure where each nucleation feature generated an island on which the magnetic alloy could be sputtered. The dome shape of the column is believed to arise from the difference in surface free energy and surface diffusion constants of the metallic and oxide phases, and is believed to aid in their segregation. It is important to note from the figure that after magnetic layer was deposited, the gaps between the magnetic islands are largely filled by non-magnetic oxide segregants. The final media has a continuous surface and is expected to have a more reliable head-disk interface compared to conventional etched BPR media. This trench free surface should not only enhance corrosion resistance (by eliminating exposed island sidewalls), but also reduce modulation of the slider flying height and therefore improve recording performance.

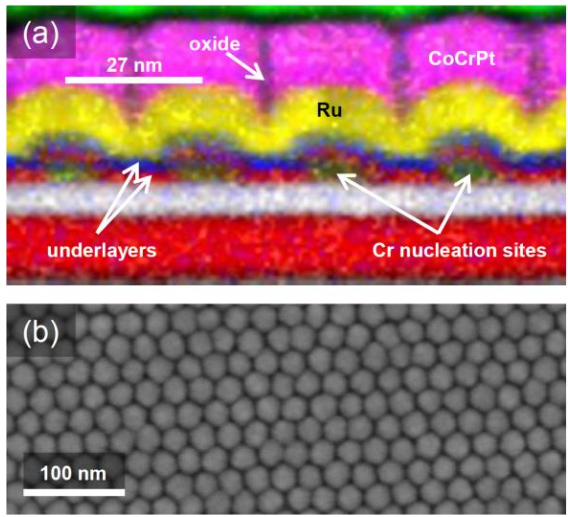

Fig. 8. (a) TEM Cross-section EELS map of a 1 Td/in$^2$ (27 nm pitch hexagonal array) TG BPR media, (b) plan view SEM image of the same media.

Fig. 8(b) shows the SEM plan view of the same sample in Fig. 8(a). It is clear that uniform hexagon shaped magnetic islands are formed with large filling factors (here about 75%). The location of the original hexagonal array of nucleation sites is directly mapped onto media islands. Since templated growth is an additive process, no etching damage is introduced. Also, without the limitations of etching the overall thickness can be larger than for etched media.

A major challenge for TG BPM is to control the microstructure and magnetic properties of each magnetic island during the growth process. When materials are deposited on amorphous topographic features multiple nucleation and growth fronts on each feature are introduced during the initial growth. Because these growth fronts have different orientations, when they merge they form a set of subgrains on each nucleation feature. For fcc materials each nucleation site and growth front tends to have its c axis normal to the local surface orientation. On a dome shaped nucleation feature this results in a set of subgrains whose c axis orientation forms a fan shape, which results in a large X-ray diffraction rocking angle (we have observed FWHM > 7 degrees), and therefore results in a large switching field distribution (we have observed iSFD/H$_C$ > 50%) [43].

Rather than choosing to grow on an amorphous nucleation feature, if a nucleation feature with a preferred crystallographic orientation [43], [45] is used, better orientation of grains (with correspondingly better structural and magnetic properties) can be obtained. In this epitaxial templated growth (eTG) process, the growth proceeds in a pseudo epitaxial manner so that subgrains initiated on different parts of the nucleation feature will have registry with the initial layer and will have a better texture. In particular, using a Pt layer with [111] texture perpendicular to the disk surface, we have formed nucleation features on which hexagonal close-packed Ru can be grown with c-axis also perpendicular to the surface.

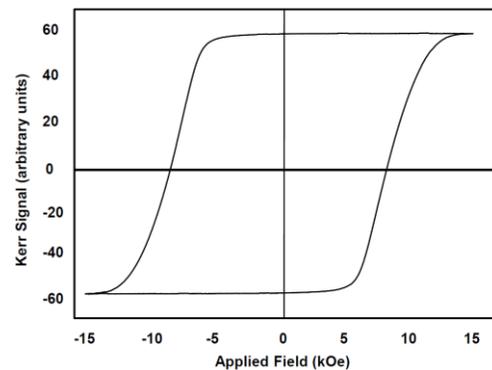

Fig. 9. Magnetic hysteresis loop of an epitaxial grown TG BPM with density of 1 Td/in$^2$. Vertical axis uses arbitrary units.

We have grown stacks which include Ru/CoCrPt+oxide and obtained a tight c axis orientation distribution, (rocking curve X-ray measurement of Ru with FWHM < 2.5 degrees). We can obtain magnetic islands that are well defined and magnetically decoupled, with no sub-grain growth observed within the islands. We have exercised this concept [43] with nucleation features generated from a 1 Td/in$^2$ imprint template and obtained media a high coercivity of ~8kOe, high thermal stability (K$_U$V/k$_B$T ~270), and a switching field distribution of iSFD/H$_C$ around 10%, as shown in Fig. 9. Because TG BPM generally has a higher fill factor than etched media we find eTG media can have a higher coercivity and thermal stability factor. So far TG BPM has a wider switching field distribution than conventional etched BPM, possibly due to variations in the nucleation features that are reflected in magnetic variations from island to island. As a consequence, recording performance of TG BPR media is not as good as etched media, and further optimization of the nucleation features, growth conditions, layer structure and materials composition are needed in order to achieve the full potential of eTG media.

It is worth noting that etched media has been demonstrated at areal densities as high as 5 Td/in$^2$, which indicates that methods to fabricate nucleation features at ultra-high densities are possible. TG BPM has the potential of reaching much higher AD without facing the scaling challenges associated with etching. Therefore, TG BPM is a promising candidate for ultra-high density bit patterned recording media.

## IV. MEDIA FABRICATION

Fabrication of BPM diverges significantly from nanofabrication processes practiced by the semiconductor industry for multiple reasons: 1) BPM features sizes (generally < 20 nm full pitch in the down-track direction) are beyond the capability of conventional lithographic methods. 2) BPM requires single-shot full-disk lithography to avoid stitching errors. 3) BPM patterns are circular in nature and highly periodic. 4) Cost targets are much lower and total area to be patterned is much larger than for semiconductors. 5) BPMR can tolerate a relatively high defect rate (~$10^{-3}$ defective islands).

Taking these factors into account has led to the fabrication process shown in Fig. 10. To provide full-disk stitch-free patterns with circular symmetry, a rotary-stage e-beam lithography system is used to produce guiding patterns for block copolymer (BCP) directed self-assembly (DSA). DSA typically provides a factor of 2 (or more) in pitch division, followed by self-aligned double patterning (SADP) for a further 2X reduction to sub-20 nm pitch (which is needed in the down-track direction, but optional for the cross-track direction).

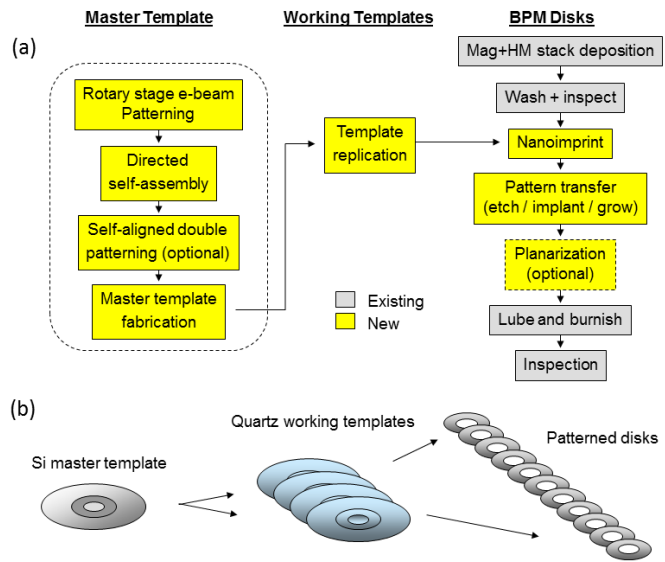

Fig. 10. (a) Process flow for fabrication of the master template, working templates, and BPM disks. (b) Two-generation nanoimprint strategy for duplicating the pattern from a single master template via replicated working templates to a large number of disks.

Early demonstrations of BPM fabricated by block copolymer lithography commonly employed hexagonal patterns of round islands because they can be made in a single BCP DSA process. However, as discussed in Sections II and V, hexagonal patterns are not ideal for BPMR and have therefore been replaced by patterns of rectangular islands in which the down-track pitch is less than the cross-track pitch, for a bit aspect ratio (BAR) > 1. As shown in Fig. 11, this type of pattern is created by combining a sub-master pattern of curved radial lines (divided up into multiple zones which reset the pitch near the target value as radius increases) and a sub-master pattern of circular tracks.

Upon completion of the master template, the master may be replicated by nanoimprint lithography to quartz working templates, which in turn are used to imprint BPM disks, as shown in Fig. 10(b). This two generation replication process allows millions of disks to be patterned from a single master template, provided that the master and working templates each have a lifetime of more than a thousand imprints.

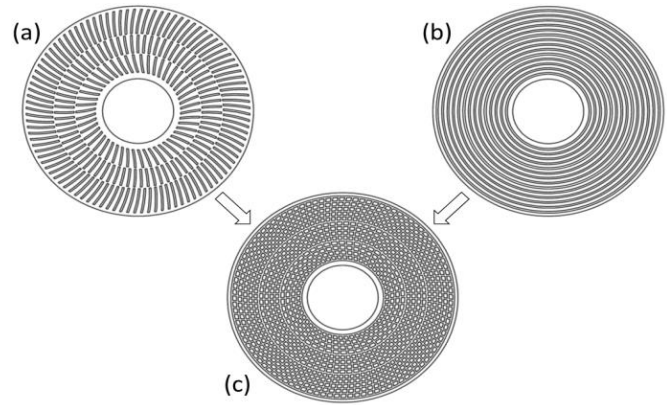

Fig. 11. The complete master pattern (c) is formed by intersecting a zoned radial line pattern (a) and a circumferential track pattern (b).

### A. Rotary-Stage e-Beam Lithography

BPM patterns are fundamentally circular. The biggest problem of conventional (X-Y) electron beam lithography (EBL) systems, when writing circular features, is the need to do Cartesian to polar coordinate system transformation both for the beam deflector and stage motion patterns. The amount of data produced for generation of circular patterns in X-Y EBL systems greatly exceeds the amount of data representing dots on a template for BPM and therefore the exposure data files for BPM template writing can become unmanageable due to their enormous size. Typically, one dot on the master template would require at least 40 bits of the exposure data file.

Except for a very few experimental setups, X-Y EBL systems move their stage in an orthogonal manner and therefore stitching of the exposed fields for circular patterns is required in both X and Y directions. Field stitching results in stitching errors of the order of 10 nm (3σ), which are unacceptably large for BPM, as well as significant writing time overhead for X-Y stage settling.

Naturally, rotary stage e-beam writers do not need field stitching to create circular patterns. In essence, one writing field of a rotary EB tool is a circular band/ring ~10 μm wide. The typical number of the bands needed to cover the whole disk is a few thousand. Radial-only field stitching (which is permissible for BPM) allows rotary EB writers to work in an open-loop mode; i.e. no periodic checks of e-beam drift are required, which further boosts the throughput.

The minimum spot size of an electron beam column is the main parameter defining the ultimate resolution. The minimum spot size depends on the objective lens optical aberrations, beam current, electron emitter current density, and beam acceleration voltage. Generally, minimum spot size increases with the beam current, but decreases for higher acceleration voltage. The rate of recording is directly proportional to the beam current and inversely proportional to the minimum resist exposure dose. Resist sensitivity is roughly inversely proportional to the acceleration voltage; *e.g.*, for positive resist ZEP-520 sensitivity at 50 kV is 120-150 μC/cm², whilst at 100 kV it is 250-300 μC /cm².

High resist sensitivity may have resolution tradeoffs. Resists which can be exposed with fewer electrons per unit volume produce a noisier image (due to shot noise), which results in increased line edge roughness (LER) at the nm scale. Due to the slower resist speed, use of higher accelerating voltage (> 50 kV) and slightly higher current has the advantage of exposing with the same resolution and rate, but with less shot noise than lower voltage systems.

Total exposure time for a rotary stage EBL tool can be calculated as:

$$t_{exp} = \pi \left( R_1{}^2 - R_0{}^2 \right) \frac{S f}{I} + \frac{R_1 - R_0}{W_b} \, t_{set} \,, \qquad (4)$$

where $R_0$ and $R_1$ are start and end recording radii respectfully, $I$ is beam current, $S$ is resist sensitivity expressed in charge per unit area, $f$ is pattern fill factor, *i.e.*, ratio of exposed area to the total area, $W_b$ is writing band width, and $t_{set}$ is the settling time between bands[1].

In order to be able to generate complicated patterns, like bit patterns or HDD servo areas, a rotary EBL tool incorporates a formatting unit. The formatter generates of deflection and blanking signals for the beam within each written band. The parameter data for each sector in a band are sent to the formatter *in real time,* and therefore there is no need to store massive amounts of pattern data as it would be the case for an X-Y EBL system.

Writing of a bit patterns or radial lines requires alternating between exposed and unexposed segments along a circular track. The simplest approach for this is to periodically blank the beam. However, this results in a loss of total beam time available for writing and therefore lower throughput. Our rotary EBL system (Elionix EBW7000C) uses a *blankerless* approach. The idea is to follow an exposure spot with circumferential beam deflection and after enough dose is deposited quickly jump to the next exposure spot (see Fig. 12). Beam blanking is virtually eliminated and throughput is maximized by this strategy.

Our formatter is capable of writing radial and circumferential lines, dot arrays, and various servo patterns. For high BAR patterned media the important criterion is high fidelity and placement accuracy of e-beam writing for radial and circumferential lines. To minimize stochastic errors a *multi-pass* writing strategy is applied in both cases. We typically use 4 to 16 passes to achieve required placement accuracy. For statistically independent events, the standard deviation of the normal distribution function is inversely proportional to the square root of the number of repetitions $N_r$, and thus the placement errors decrease as $\sqrt{N_r}$. Practical implementation of multi-pass writing is done by dividing tracks into a number of sub-tracks and then compressing them by beam deflection into overlapping passes forming tracks with the desired track pitch (see Fig. 13).

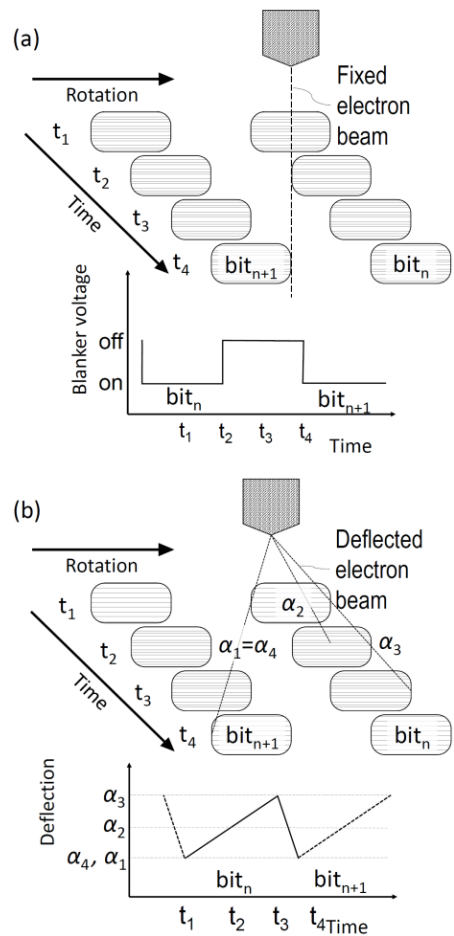

Fig. 12. a.) Blanked writing – in this case half of the beam time is wasted; b.) Blankerless dot following with circumferential deflection – the beam is never blanked and 100% of the beam time is utilized.

Radial line writing is the same as writing of bit arrays, except that exposed dots are assigned to sub-tracks with pitch smaller than the e-beam spot size. This causes adjacent dots to overlap in the radial direction, producing 'beaded' radial lines as shown in Fig. 13(b1). Such beading is mitigated by applying high frequency (>100 MHz) dither deflection in the radial direction, to effectively elongate the spot shape in the radial direction as shown in Fig. 13(b2).

Requirements for resists used in BPM master writing are very stringent. In order to produce high quality resist masks at ~ 20 nm half pitch, resists should satisfy several conditions:

- Resist resolution (effective grain size) is better than 1-2 nm.
- Resist speed is high enough to support reasonable throughput, but not so high as to result in any significant shot noise.

---

[1] An example for a master pattern for a full 65 mm disk: $R_0$ =14 mm, $R_1$ = 31 mm, $I$ = 5 nA, $S$ = 280 μC/cm², $f$ = 0.5, $W_b$ = 0.01 mm, $t_{set}$ = 40 s, the exposure time is approximately 206 hours (8 days 14 hours). The same writing conditions for a 95 mm master pattern with $R_0$ =17 mm, $R_1$ = 46 mm results in $t_{exp}$ = 479 hours (19 days 23 hours).

- When used for BCP DSA, resist must be chemically compatible with the BCP materials.
- Resist must have acceptable resistance to plasma etching and yet be trimmable.

Chemically amplified resists in use today are not desirable due to the very long writing time of BPM masters, during which the diffusion of the catalyzer can degrade the latent image contrast.

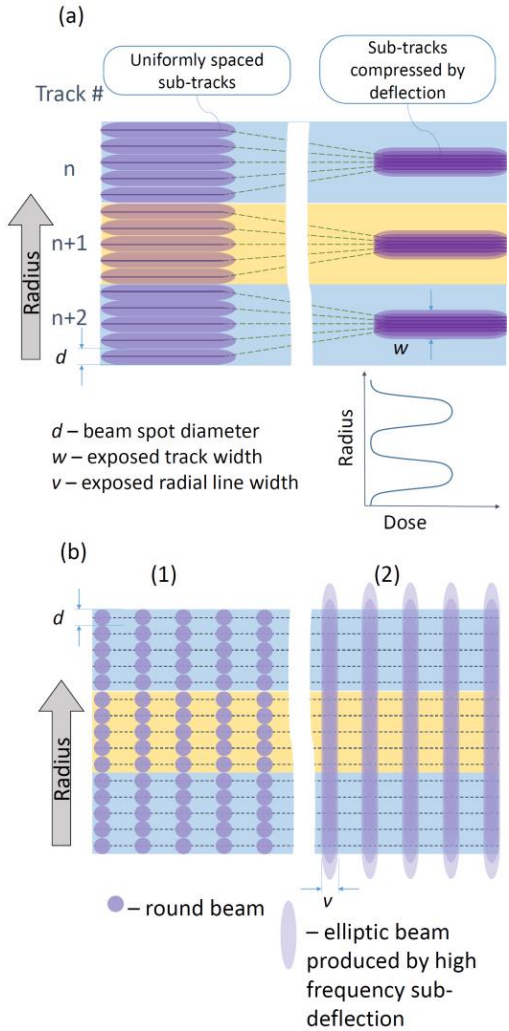

*d* – beam spot diameter
*w* – exposed track width
*v* – exposed radial line width

Fig. 13. a.) Circumferential line writing with multiple passes and sub-track compression. b.) Radial line writing with (1) standard and (2) overlapping elliptic beams produced by radial high frequency sub-deflection.

Among commercially available e-beam resists, the closest to optimal are positive resists ZEP-520A from Nippon Zeon Ltd. and SX AR-P by Allresist GmbH. If there were a negative resist with sufficient resolution and sensitivity, it would be preferable for DSA applications, as it would not require plasma trimming, which is a must for positive resist to achieve target line vs. space dimensions.

EBL has insufficient pattern resolution and fidelity to directly produce the small feature sizes needed for BPM. Therefore methods of pattern density multiplication, such as self-aligned line doubling (SADP) and/or directed self-assembly (DSA) of block copolymers (BCPs) are necessary.

## B. Directed Self-Assembly of Block Copolymers

### 1) Introduction to Self-Assembly for BPM

Relevant areal densities for potential BPM products begin above 1.5 Td/in² which translates to lithographic islands having a critical dimension (CD) <10 nm with a full pitch <20 nm if we restrict our design to rectangular islands with a BAR > 1 for the reasons discussed earlier. Because sub-10 nm lithography is not available by any conventional lithographic technique, block copolymer lithography has evolved as an organic part of the BPM patterning solution not only because of the ability of BCPs to form sub-lithographic features, but because of their flexibility to comply with other important BPM design requirements such as conforming to zoned periodic features on circular tracks at constant angular pitch.

Linear di-block copolymers consist of two immiscible polymeric blocks joined by a covalent bond. They spontaneously phase separate towards a minimum interaction volume driven by a segregation strength, $\chi N$, where $\chi$ represents the segment-segment Flory-Huggins interaction parameter and N, the degree of polymerization [46]. In block copolymer lithography, a BCP thin film is used in lieu of a resist layer with the self-assembling pattern forming the latent image. Selectively removing one of the two blocks leaves a sacrificial mask that can be used for pattern transfer in a similar way as developed resists are used in conventional lithography. Spheres, cylinders and lamellae represent the BCP morphologies most commonly used in lithographic applications to form arrays of either parallel stripes or hexagonally packed dots [47]. The most direct path to form rectangular islands from block copolymer patterns is by combining two separate line/space arrays and make them intersect orthogonally as shown in Fig. 11.

Thin films of lamellae-forming block copolymers where the lamellae orient perpendicular to the substrate are preferred over monolayer films of parallel oriented cylinders that can also form striped arrays. Perpendicular lamellae are more tolerant to thickness variations whereas monolayer films of cylinders may result in double layers or empty holes if the thickness is not commensurate with that of a single layer of cylinders. Because the domains of perpendicularly oriented lamellae span uniformly from top to bottom with both block materials contacting the bottom of the film, perpendicularly oriented domains are generally better suited for chemical contrast DSA. Additionally, it is also assumed that a uniform profile from top to bottom better resembles a lithographic mask than that of a floating cylinder whose cross sectional profile changes in width across the film thickness. However it is also known that successful demonstrations of line/space patterns have been reported using cylindrical domains [48], [49].

Poly(styrene-block-methyl methacrylate) (PS-b-PMMA) is the BCP most likely to be introduced first in a manufacturing environment. It has been extensively studied and a large body of creative inventions already exists for domain orientation and directed self-assembly [50]–[54]. However, the interaction parameter $\chi$ is relatively low. Given that segregation strength $\chi N$ for lamellar phase needs to be larger than about 10.5 to induce microphase separation and that the pitch scales with N, the achievable dimensions in PS-b-PMMA are limited to about 19-20 nm full pitch [55], although successful demonstrations of pattern transfer currently stop at 22 nm full pitch [56] indicating that PS-b-PMMA is likely to supply only a partial solution to BPM patterning. Fig. 14 shows the achievable areal densities for a given combination of track pitch (y-axis) and bit pitch (x-axis). It is apparent that the bulk of dimensions relevant to BPM require patterning solutions below 20 nm pitch where PS-b-PMMA cannot presently reach.

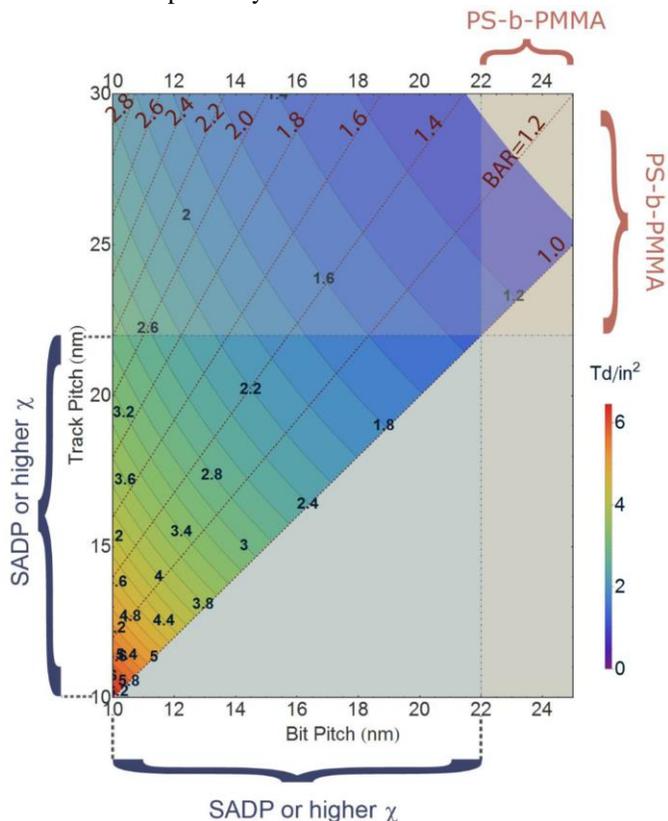

Fig. 14. Available areal densities for a given combination of track pitch (y-axis) and bit pitch (x-axis). The coloring indicates the AD and the diagonal lines designate the BAR. The plot is limited to the region where BAR ≥ 1.

There are currently two alternatives to stretch to the sub-20 nm space. 1) By using a "higher- χ" block copolymer material it is possible to achieve a full pitch even below 10 nm [57], [58]. However, potential solutions with "higher- χ" BCPs are still developing and will require additional innovations to overcome current challenges in controlling interfacial energies, DSA and pattern transfer. These materials will be briefly discussed at the end of this section. 2) By contrast, an extension of the well-known self-aligned double patterning (SADP) can be readily applied to block copolymer lithography to further subdivide the pitch obtained by PS-b-PMMA block copolymers. In principle, this provides a path to reach 11 nm full pitch with currently available PS-b-PMMA BCPs. However, SADP for sub-20nm pitch is a complex solution with multiple challenges in dimension control, spacer deposition and materials design. It is worth noting that with SADP, currently available materials and methods can reach to >3 Td/in$^2$ by further optimizing currently available processes.

The rest of this section will be devoted to current state of the art block copolymer patterning suitable for BPM master templates up to 2 Td/in$^2$ with BAR > 1. We will first cover patterning of line/space arrays using PS-b-PMMA from 41 to

22 nm pitch. Next we discuss solutions below 22 nm pitch including current research in "higher- χ" BCPs and in PS-b-PMMA patterning for SADP. We will then review a list of the minimum specifications that block copolymers need to satisfy for BPM patterning followed by a review of the status on pattern transfer. Lastly, we finish with a discussion on extendibility of BCP lithography to higher areal densities.

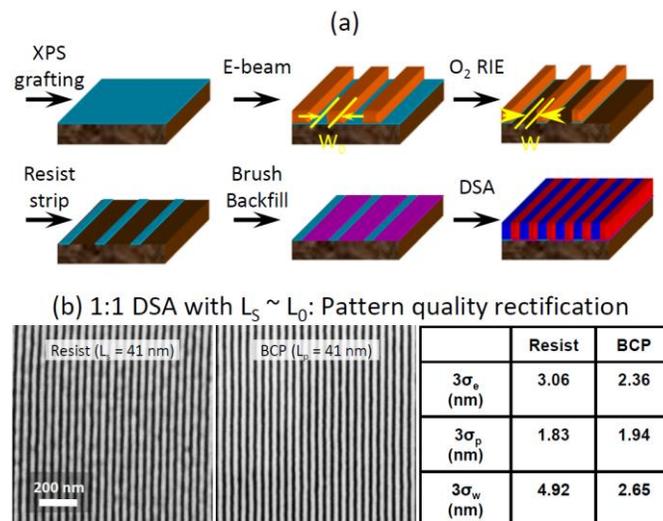

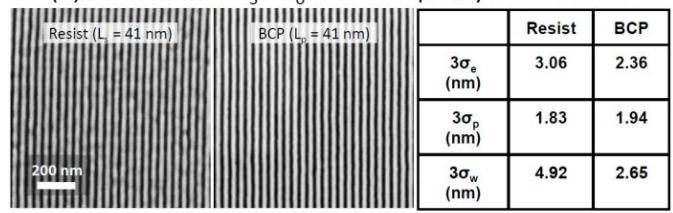

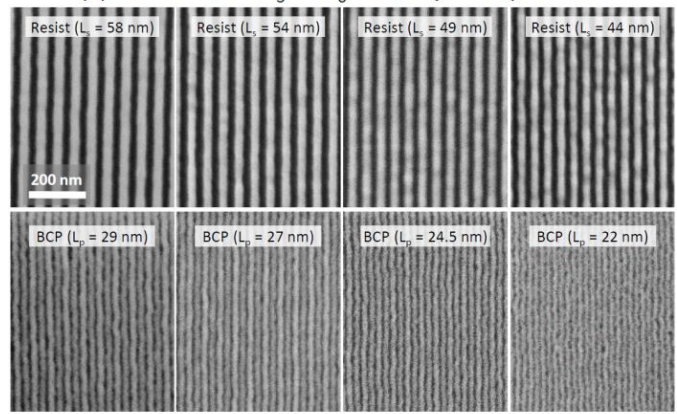

Fig. 15. (a) Schematic of DSA flow. (b) Top-down SEM images of e-beam resist pattern and BCP pattern (after PMMA removal) with a pitch of 41 nm. The table shows typical LER, LPR, and LWR values for resist and BCP patterns. (c) Top row: top-down SEM images of e-beam resist pattern with $L_s$ = 58, 54, 49, and 44nm; Bottom row: top-down SEM images of DSA patterns with $L_p$ = ½ $L_s$.

### 2) Patterning with PS-PMMA from 41 to 22 nm Pitch

The schematics in Fig. 15(a) illustrate the process flow used to achieve DSA in BCP thin films [53]. Chemical contrast patterns are created on silicon or quartz substrates by first depositing a ~7-8 nm thick film of cross-linkable polystyrene (XPS) mat. Cross-linking is attained by a thermal treatment after film deposition. Then a layer of ZEP 520A e-beam resist is applied and a series of line/space patterns arranged in either radial or circumferential arrays are exposed with a rotary stage e-beam (see Fig. 13). The pitch of the e-beam exposed lines, $L_s$, is roughly commensurate to the pitch of the target BCP, $L_o$, such that $L_s \approx nL_o$ where $n$ is an integer designating a density multiplication factor. The developed line/space resist patterns are further trimmed using $O_2$ plasma narrowing the resist lines to a width comparable to that of the PS domain in the BCP to be used during DSA. After thoroughly stripping the resist, a hydroxyl-terminated (-OH) polymer brush is selectively grafted in the interspatial regions between the remaining XPS stripes. The choice of polymer brush depends on the density multiplication factor used. When n = 1, PMMA-OH is used as the backfilling brush. When n ≥ 2, the selected brush is a random copolymer PS-r-PMMA-OH containing ~40-50% styrene [53]. While there is no density advantage when n=1, the use of DSA is still preferred due to its superior line width roughness as shown on Fig. 15(b). When the resolution of the e-beam resist is not sufficient to sustain n = 1, a higher value of n is chosen. Fig. 15(c) shows a series of DSA results with n = 2 at various pitches down to 22 nm.

### 3) Patterning below 22 nm Pitch: BCP + SADP

Any of the above line/space patterns formed by PS-b-PMMA BCPs could be used with SADP for pitch division. Details of SADP patterning will be covered later in this manuscript; here, we only note that the critical dimension deserves special attention when considering BCPs for SADP. Lamellar phase block copolymers tend to form line/spaces of roughly the same width (about ½ of the pitch) while in SADP, the width of the mandrel needs to be about ¼ of the pitch value [59]. So the formation of mandrels from a BCP lamellae needs some method to trim the width to about half of its original value. This can be achieved by trimming the mandrel itself or by using selective infiltration by atomic layer deposition where the final BCP-derived mask width is thinner than the original pattern as will be discussed later.

### 4) Patterning below 22 nm Pitch: "higher- χ" BCPs

In block copolymer lithography, a BCP film replaces the conventional resist. The best way to mimic the performance of a resist is when the BCP domains are oriented perpendicular to the substrate spanning a uniform cross section from top to bottom of the film for the reasons explained earlier. However, orienting BCP domains perpendicular to the plane of the substrate requires controlling interfacial energies at both interfaces surrounding the film such that neither domain preferentially wets either. The substrate interface ($I_1$ in Fig. 16), is typically straightforward to modify. Methods for creating neutral interfaces using an appropriately-designed cross-linked polymer, grafted polymer brush, or other surface modification techniques are well-documented [53], [60]. The BCP can then be directly spin-coated on this modified substrate. Modification of the free surface interface ($I_2$ in Fig. 16) is more challenging since a surface modification film cannot be simply spin coated on top of the BCP unless it is spin coated out of a solvent that will not dissolve the polymer film. Typically, a polymer with solubility opposite that of the BCP will not have the correct chemistry to provide a neutral interface.

PS-b-PMMA turned out to be a fortunate model system for BCP patterning since neither domain preferentially wets the free surface interface when it is thermally annealed in air or vacuum making perpendicular orientation at the top interface appear almost natural. However, due to its relatively low segregation strength, PS-b-PMMA does not phase separate at a pitch smaller than ~20 nm. Achieving high AD patterns without the use of line doubling, which has many fabrication-related challenges, requires selection and orientation of a different BCP system with a higher segregation strength quantified by its interaction parameter (χ). Many higher-χ BCP systems exist in the literature, some containing silicon or a metal component that also improves the inherent etch contrast between the domains upon oxidation during reactive ion etching [49], [61]–[64]. However, these BCP domains are notoriously difficult to orient perpendicular to the plane of the substrate due to the low surface energy of the silicon- or metal-containing domain, which almost always preferentially wets the free surface interface and causes the BCP domains to lie in the plane of the substrate. Other purely organic block copolymer systems that exhibit a higher-χ are not only challenging to orient perpendicular to the substrate, but may also lack etch contrast in any of the common RIE chemistries for pattern transfer [57], [65], [66]. While selecting a BCP within the class of higher-χ materials is under investigation for high AD patterning, most demonstrations of media fabrication have been done with PS-b-PMMA due to its predictability and ease of orientation control.

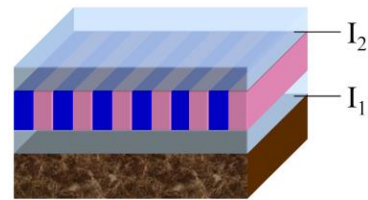

Fig. 16. Vertical orientation of BCP domains requires modification of the substrate interface ($I_1$) using a cross-linked surface treatment, grafted brush, or other surface modification method. The free surface interface ($I_2$) can be air with appropriate BCP chemistry. Otherwise this interface requires modification using a top coat surface treatment or a solvent vapor atmosphere.

Three solutions to the challenge of orienting higher-χ BCPs have been proposed in the literature. The most common solution involves modifying the atmosphere above the polymer film with a solvent vapor which also swells the polymer film, allowing it to reassemble [63], [67], [68]. However, this technique relies on finding a solvent that has sufficiently similar solubility to both blocks to avoid morphology changes, provides control of the interfacial energy to promote perpendicular orientation and depresses the glass transition temperature to give enough mobility to the BCP molecules all at the same time. A second option is to control the free surface interface by a top coat. Use of top coats presents chemistry challenges depending on how the top coat is used. If the top coat is spin coated on top of the BCP, this requires use of orthogonal solvents to prevent the original BCP film from being re-dissolved, but in general, the polarity of a "neutral" material may not be compatible with orthogonal solvents. A group at University of Texas resolved the conflict by using polarity-switching top coat materials [69]. Alternatively, the top coat material may be embedded in the original block copolymer solution, but in this case, it requires a low surface energy component capable of diffusing to the top of the film while also providing a neutral top interface. Another possibility is to float a top coat film on top, however it is generally difficult to contact a floated film over large areas of the BCP, resulting in gross defects [70]. The third solution to perpendicular orientation at the top interface is to chemically modify the formulation of the BCP itself such that neither wets the free surface interface upon thermal annealing in air while maintaining a higher $\chi$ [71]. This last solution, while ideal from the processing point of view, also presents significant challenges to be able to find the right chemistries to accomplish this goal.

### 5) Design Requirements for Self-Assembled Patterns

As mentioned earlier, BCP lithography has been a natural part of the BPM patterning solution not only because BCPs are capable of reaching the small dimensions needed for BPM, but also because of their flexibility to adapt to other specifications that are unique to BPM technology. Here we review some of these specifications and compare them to the current state of the art patterning with lamellae-forming PS-b-PMMA BCPs. We note that the specifications discussed here are exclusive to BPM technology and come in addition to those previously suggested for BCP lithography in general [47]. Table 1 provides a quick reference for some of these values.

TABLE 1. COMMON PATTERNING SPECIFICATIONS REQUIRED FOR BPM AND CURRENT STATE OF THE ART PATTERNING WITH PS-B-PMMA LAMELLAR BLOCK COPOLYMERS

|  | Defect density | Stretching/ compressing | Skew angle | Line roughness |
|---|---|---|---|---|
| Spec | $<10^{-5}$ | $\pm 4\%$ | $\pm 15^{\circ}$ | $\sigma_p < 0.05\,L_o$ |
| PS-b-PMMA | $<10^{-5}$ | $+3\%\,/-5\%$ | any | $\sigma_p < 0.03\,L_o$ |

a) Pitch flexibility. In order to optimize AD, HDDs are divided into annular zones (see Fig. 17). Within each zone, data is written or read at a constant clock rate. Because the disk rotational speed is constant, bits are placed at constant angular pitch around each track, with the result that within a zone, the bit spacing is not constant in every track, but is proportional to the track radius. In BPM templates, the bit pitch is determined by the pitch of the radial spokes in the radial sub-master template described in Fig. 11. Typical zone widths are 0.5 to 1 mm, which results in a down-track pitch variation of up to +/- 4% with respect to the value of the pitch at the center of the zone. To create these patterns, the BCP must accommodate this continuous pitch variation by compressing or stretching the pitch from its equilibrium value without introducing defects or disrupting the overall quality of the pattern. We have found that lamellae-forming PS-b-PMMA block copolymers can be compressed about 5% of its equilibrium value and stretched about 3%, which is sufficient to meet this requirement. If a BCP material cannot support this range of pitch variation, a larger number of narrower zones can be used.

b) Defect density. To achieve the required raw data error rate of $10^{-2}$ (see Sections II and V), the defect of rate of finished media needs to be well below $10^{-2}$. Because sequential fabrication steps have successively higher defect rates, the finished working template for nanoimprint should have a defect rate below $\sim 10^{-3}$ which in turn implies that for the patterns that define the sub-master templates at the self-assembly level, defect rates are preferably $<10^{-5}$. Owing to the lack of a proper inspection tool, we estimated defect rates by sampling SEM images of PS lines after PMMA removal at random locations.

On one wafer containing BCP patterns with $L_0$ of 24.5 nm, we took over 100 SEM images and found only one foreign particle on an area of $\sim 1700\ \mu m^2$. On another similar wafer, no defect was found in over 150 SEM images with a total area of $\sim 3400\ \mu m^2$. Assuming a bit cell of about $5.4 \times 10^{-4}\ \mu m^2$, we can estimate defect densities below $10^{-6}$. More systematic reports from the semiconductor industry [72], [73] further confirm that self-assembly is not the limiting element for low defect rates. In our own experiments we consistently observe that pattern transfer is the main source of defects. We have found that when an advanced dry lift-off method is used, the defectivity of the sub-master template is good enough to support a $\sim 10^{-3}$ defect rate in the final template.

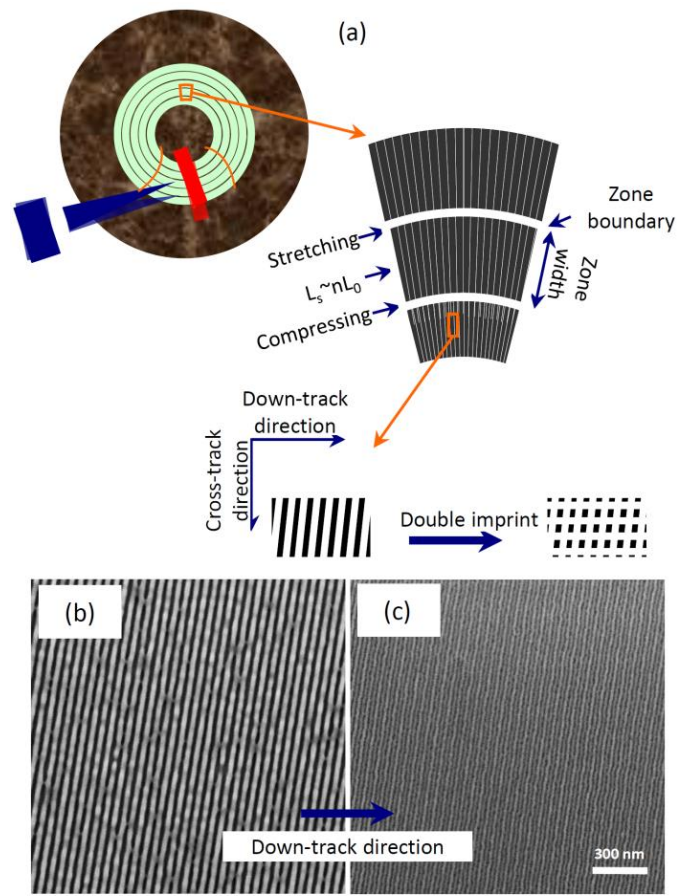

Fig. 17. (a) Cartoon of skew and zoning. (b & c) Top-down SEM image of skewed resist pattern and self-assembled BCP pattern respectively.

c) Skew angle. The head assembly is controlled by a rotary actuator that has its center outside the disk stack. As a result, the head moves in an arcuate trajectory from the outer to the inner parts of the disk as it scans over all tracks. These arcuate trajectories also imply that the orientation of the head with respect to the track line changes by a few degrees, which is referred to as the skew angle. A typical skew angle range is on the order of +/- 15°.

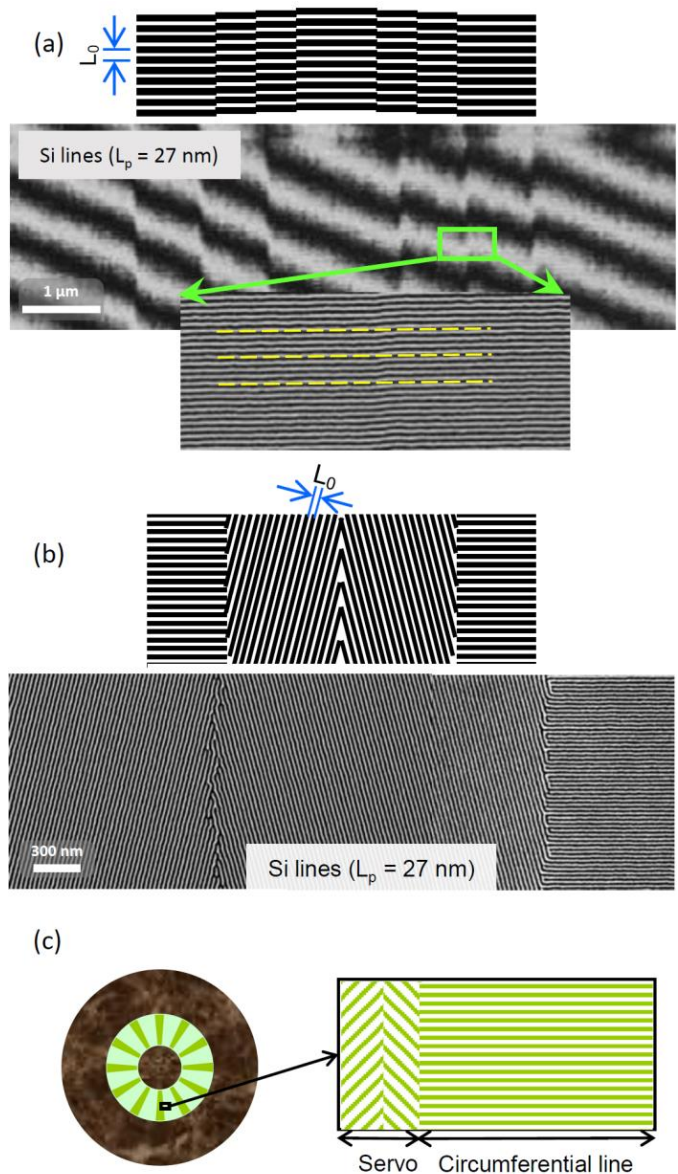

Fig. 18. (a) Top: cartoon of offset burst pattern; Middle: Moiré SEM image of offset burst pattern on a Si template; Bottom: A zoom-in top-down SEM image. (b) Top: cartoon of chevron pattern; Bottom: Top-down SEM image of part of a chevron pattern on a Si template. (c) Sector format of sub-master.

The implication of skew for BPM fabrication, assuming the head assembly and the rotary actuator remain without major design changes for BPM, is that the individual bits need to follow the arcuate patterns of the head trajectory. We note that this is one of the main advantages of using lamellae-forming block copolymers. Provided that the chemical contrast patterns, as defined by e-beam lithography, that form the radial lines are made in such a way as to mimic the head trajectory, the block copolymer lines can assemble without disruption. Additionally, because the fabrication of the radial lines is independent from the circumferential lines, any skew angle with respect to the track is possible. This contrasts with hexagonally packed arrays of dots that are also considered as an option for BPM. In hexagonally packed arrays, the track density is higher than the bit linear density and since the reader width is unlikely to scale accordingly, a hypertrack approach may be needed as shown in Fig. 4(b). The hypertrack architecture allows a relatively wide head to access a pair of tracks simultaneously, relying on the staggered formation of adjacent tracks when a hexagonal formation is used. However this staggered geometry would need to distort along with the head skew angle, a level of distortion that block copolymers cannot tolerate [74]. Not to mention that the increased track density of hexagonal patterns would add considerable challenges to the servo system for adequate track following.

d) Compatibility with servo patterns. In comparison to conventional lithographic methods, one obvious limitation of BCP lithography is that it is restricted to only a few geometries. Therefore, forming arbitrary non-regular patterns, such as servo patterns, is challenging for BCP lithography. However, two types of servo structures, offset field and chevron, were proposed, which can be generated by BCPs along with circumferential lines to encode the position error signal (PES) for track following [75]. Offset patterns are a series of sections of circumferential lines that shift along the cross-track direction. Chevron patterns are sets of stripes in "Λ" shape. As shown in Fig. 18(a,b), rotary stage e-beam was used to first generate sparse chemical patterns containing offset or chevron patterns along with circumferential lines, with the periods (line-to-line distance) equal to $nL_0$. Self-assembly on the chemical patterns produces BCP patterns with higher resolution. A subsequent pattern transfer step was carried out to transfer the patterns into the silicon substrates. As illustrated in Fig. 11(c), the finished sub-master templates will have many sectors, and each sector includes both circumferential lines and servo patterns. Other options include the ability to incorporate independent lower density patterns fabricated by conventional lithography and by masking the DSA patterns out of the servo sectors, as shown in Section IV-D and [76].

e) Line Roughness. Because the rectangular islands are made from two sets of line/space patterns, the line roughness of such patterns determines island size variation (through line width roughness) and placement jitter (through line placement roughness). Sources of line roughness in block copolymer films differ in nature from those in conventional resist due to the different mechanisms that contribute to form the line edges. We use standard definitions to measure line and width roughness [77]. Fig. 19 shows a representation of a line. Line edges are fitted to straight lines and the variance of the residuals is given by

$$\sigma_{\varepsilon 1}^2 = \frac{1}{N-1}\sum_j\left(\varepsilon 1_j - \overline{\varepsilon 1}\right)^2; \quad \sigma_{\varepsilon 2}^2 = \frac{1}{N-1}\sum_j\left(\varepsilon 2_j - \overline{\varepsilon 2}\right)^2$$

$$c = \frac{Cov[\sigma_{\varepsilon 1},\sigma_{\varepsilon 2}]}{\sigma_{\varepsilon 1}\cdot\sigma_{\varepsilon 2}} = \frac{\sum\left(\varepsilon 1_j - \overline{\varepsilon 1}\right)\left(\varepsilon 2_j - \overline{\varepsilon 2}\right)}{\sqrt{\sum\left(\varepsilon 1_j - \overline{\varepsilon 1}\right)^2\sum\left(\varepsilon 2_j - \overline{\varepsilon 2}\right)^2}} \quad (5)$$

where $\varepsilon 1$ and $\varepsilon 2$ stand for the points representing the first and second edges forming the line, N is the number of sampling points and c is the linear correlation coefficient. For self-similar lines and in the limit of large N, the variance of the width is given by

$$\sigma_w^2 = \frac{1}{N-1}\sum_j\left(w_j - \overline{w}\right)^2 = 2\sigma_\varepsilon^2 - 2c\sigma_\varepsilon^2 \quad , \quad (6)$$

where $\sigma_\varepsilon^2$ is the edge variance. Additionally, we introduce a measurement of the centroid line which we call the placement line with a variance given by

$$\sigma_p^2 = \frac{1}{2}\sigma_\varepsilon^2 + \frac{1}{2}c\sigma_\varepsilon^2 \qquad . \tag{7}$$

In measuring line edge, width and placement roughness (LER, LWR and LPR), we use the standard 3-$\sigma$ definitions where LER = $3\sigma_\varepsilon$, etc. Unless otherwise noted, all of our roughness measurements are performed on square SEM images acquired with a pixel resolution of 0.9 nm/pixel and a lateral size of 1.8 µm giving cutoff wavelengths of 1.8 nm and 1.8 µm for the roughness measurements. In making rectangular arrays with two orthogonal sets of orthogonal lines, we note that the width roughness will impact the feature size uniformity and the placement roughness will impact placement accuracy. Our current specification is to keep $\sigma_w < 0.05\,\overline{w}$ and $\sigma_p < 0.05\,\overline{L_o}$ in the final structures after pattern transfer. We also note here that the line roughness of the block copolymer pattern evolves during the various steps involved in pattern transfer.

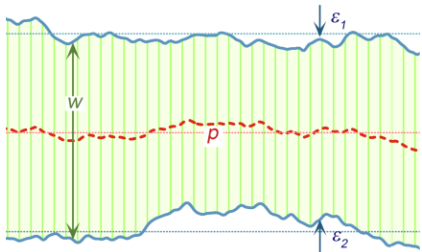

Fig. 19. Line roughness definitions.

### 6) Pattern Transfer of BCP DSA Patterns

To fabricate imprint templates, a pattern transfer step from the block copolymer film is needed. We have investigated various pattern transfer methods from lamellae-forming PS-b-PMMA block copolymers of various dimensions. The results are summarized in Table 2. Except for the sequential infiltration synthesis (SIS) method, selective removal of the PMMA blocks by an $O_2$ RIE has to occur first. After PMMA removal, the resulting PS lines have a typical aspect ratio of ~1-1.3, which means <16 nm in height for BCPs with $L_0$<25 nm. A conventional Cr lift-off method [78] fails to provide clean lift-off for a BCP with $L_0 \leq 27$ nm mainly due to the low aspect ratio of the PS lines. Since PS lines are carbon-based polymer, and erode quickly in plasma, direct etching using PS lines as masks [79] results in high roughness. The Mo etchback method (discussed later) has been demonstrated for generating high quality magnetic dots from imprint resist patterns; however, it generated rough lines from PS lines.

For PS-b-PMMA with $L_0$>25 nm, the SIS method (discussed below) can be used to produce lines with low roughness and small critical dimensions, which are desired for generating mandrel lines for self-aligned double patterning (discussed later). However, it should be noted that SIS can exacerbate pitch walking (discussed below in Fig. 21) if the BCP line pattern was generated using a density multiplication approach (e.g., the 27 and 29 nm cases shown in Table 2). Using the SIS method for BCPs with smaller $L_0$ generally results in either broken or merged lines.

Two novel dry lift-off methods have been developed for high-fidelity pattern transfer from BCP patterns. In dry lift-off method #1 [80], a spin-on glass material was used to fill the trenches of the PS lines and then partially etched back to form lines of spin-on glass as an etch mask. This method is effective for BCPs with $L_0$ down to 22 nm, but not robust enough to extend further due to the high organic content in spin-on glass. Therefore, another method (dry lift-off #2) was implemented, in which pure inorganic material ($AlO_x$ deposited by ALD) was used to replace the spin-on glass. The replacement leads to a much wider processing window for etching back the $AlO_x$ and etching into the underlying substrate.

TABLE 2. SUMMARY OF PATTERN TRANSFER FROM PS-B-PMMA FILMS.

| Method / $L_0$ | Cr lift-off | Direct etch | SIS | Mo etchback | Dry lift-off #1 | Dry lift-off #2 |
|---|---|---|---|---|---|---|
| 41 nm | | | Good in LER, small critical dimension | | | |
| 37 nm | | | | | | |
| 29 nm | | | Sometime pitch walking | | | |
| 27 nm | Debris | | | | Good in roughness, no pitch walking | Robust, good in LER, large process window |
| 24.5 nm | Not working | Rough lines | No process window | Rough lines | | |
| 22 nm | | | | | Little process window | |

As the demand for lithographic dimensions squeezes into the single-digit nm scale, pairing block copolymer materials that segregate at these dimensions with adequate pattern transfer techniques is increasingly difficult. Innovations for pattern transfer in BCP lithography below 20 nm pitch have to focus on the following properties: a) high etch contrast between the two blocks so that one can be selectively removed without losing much of the remaining block. b) Size tuning of the critical dimension (especially useful for mandrel fabrication in SADP). c) Maintaining low values for LPR and LWR as dimensions scale down. d) Control over the mechanical properties of the mask materials.

Recent demonstrations that combined sequential vapor infiltration [68] in an atomic layer deposition (ALD) reactor [81] with block copolymer films for selective synthesis of metal oxides inside one of the two polymer blocks [82], [83] hold some promise to address the challenges in etch contrast, size control, line roughness and potentially even tuning mechanical properties. In sequential infiltration synthesis (SIS), a block copolymer film is exposed to an organometallic vapor precursor inside an ALD reactor for a given time. The vapor diffuses inside the polymer and selectively binds to target groups in one of the two polymeric domains, e.g. the carbonyl groups in PMMA [82], [84] as shown in Fig. 20. The excess precursor vapor is pumped away and a co-reactant vapor is introduced to finish the reaction forming a metal oxide only where the first precursor was located. Most commonly, this process is done using trimethyl aluminum (TMA) and water to synthesize aluminum oxide. The full cycle can be repeated a number of times to increase the amount of deposited material. A subsequent plasma treatment removes the polymer matrix and densifies the oxide material leaving behind an inorganic pattern that resembles that of the original block copolymer pattern which in turn can be used as an etch mask to transfer the pattern into the substrate [85].

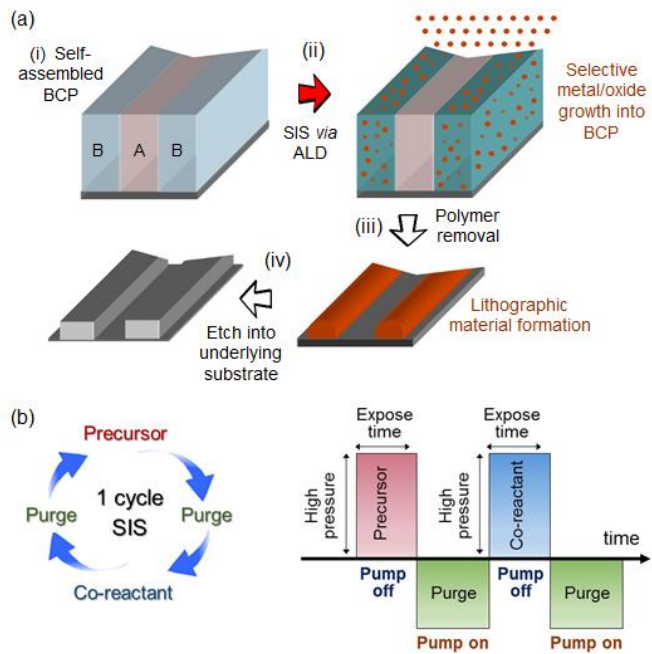

Fig. 20. (a) SIS process flow: (i) A-block-B BCP; (ii) infiltration and selective synthesis of precursor and co-reactant using ALD; (iii) polymer removal treatment to densify metallic nanostructure and form lithographic material; (iv) transferring patterns into underlying substrate. (b) ALD conditions for 1 cycle of SIS (precursor/purge/co-reactant/purge): first a precursor is diffused at high pressure and exposed for a long time without evacuation (vacuum pump off), following by a purge sequence (vacuum pump on) to remove unreacted molecules and byproducts, then co-reactant is diffused and exposed in similar manner similar to oxidize the inorganic materials before purging again the chamber reactor.

While some of the chemical and physical mechanisms of SIS are still unknown, our results on PS-b-PMMA and poly(2-vinyl pyridine-block-styrene-block-2-vinyl-pyridine) (P2VP-PS-P2VP) block copolymers indicate that SIS may be promising in that, for some polymers, the LPR may be better than that obtained from other conventional pattern transfer techniques. Additionally, with SIS the width of the lines can be tuned proportionally to the amount of infiltrated material which is controlled with the number of ALD cycles within a certain range [82], [85]. This is particularly useful for the fabrication of mandrel features for SADP that require a narrow CD of about ¼ of the mandrel pitch while maintaining a low LER. And last, but not least, SIS brings a good method to create etch contrast in some higher-$\chi$ block copolymers that naturally lack etch selectivity such as PS-b-P2VP and poly(styrene-block-ethylene oxide) (PS-b-PEO).

To evaluate the line roughness and demonstrate a CD < ½ $L_o$, we took a DSA film made of PS-b-PMMA with $L_0 = 29$ nm as shown in Fig. 21(a). $AlO_x$ was selectively synthesized in the PMMA domains by sequentially exposing the film to TMA and water vapors for a total of three cycles as shown in Fig. 21(b). The exposure and purge times were kept constant at 300 s for both precursors. The pressure during exposure was 1.5-2.5 Torr. After SIS, the BCP film was treated in oxygen plasma to remove the polymer and form $AlO_x$ lines where the PMMA domains were as shown in Fig. 21(c). The width of the $AlO_x$ lines has a duty cycle (CD/$L_0$) of 0.35 as shown in the inset of Fig. 21(c). The $AlO_x$ lines were next transferred into the Si substrate using fluorine plasma as shown in Fig. 21(d) and cross section inset.

Table 3 shows the measured LER, LPR and LWR values for the $AlO_x$ lines and for the transferred features. These values are generally lower than those obtained by other pattern transfer methods. In particular, LPR tends to be better when using SIS. It may also be observed that the roughness appears lower in Fig. 21(c) and (d) than it is in Fig. 21(b). Note that Fig. 21(b) shows the block copolymer film right after SIS infiltration. The observed roughness in Fig. 21(b) is at the interface between PS and PMMA at the top of the film, where roughness is always relatively high, plus the added granularity coming from the $AlO_x$ infiltrated in the PMMA which at that point is not very dense [84]. Fig. 21(c) shows the remaining, densified $AlO_x$ after both PS and PMMA have been removed. It is also speculated that during polymer removal and the consequent densification of $AlO_x$, tensile stress in the material may create tension in the remaining lines which may tend to straighten them.

We also note, however, that SIS seems to be particularly sensitive to subtle mismatches in the chemical contrast pattern used for DSA. If the width of the guiding line does not match that of the BCP, the distortion at the bottom of the film is replicated and accentuated by the $AlO_x$ lines in the form of a "pitch walking" whereby the pitch oscillates between consecutive lines as observed in Fig. 21(c,d).

TABLE 3. LINE ROUGHNESS

| BCP Full pitch | PS-PMMA $L_0 = 29$ nm | |
|---|---|---|
| Line pattern type | $AlO_x$ | Hard carbon *via* $SiO_2$ |
| LER ($3\sigma_L$) | 2.08 nm | 2.57 nm |
| LWR ($3\sigma_W$) | 1.80 nm | 2.04 nm |
| LPR ($3\sigma_P$) | 1.88 nm | 2.35 nm |

A particularly positive benefit of SIS is its potential to bring etch contrast in some of the higher-$\chi$ materials that naturally lack any contrast. Lamellae-forming P2VP-b-PS-b-P2VP triblock copolymers are a prominent example of a material that could be amenable to sub-10 nm lithographic applications in conjunction with SIS. Similarly, PS-b-PEO and poly(styrene-block-lactide) (PS-b-PLA) block copolymers are other examples that could benefit from the etch contrast generated by SIS.

### C. Self-Aligned Double Patterning (SADP)

SADP is a technique used by the semiconductor industry to double the density of an existing prepattern [86], [87]. There are three major unique features in our SADP for BPM: (1) Our prepattern is created using DSA of BCP (2) SADP is performed on circumferential or radial line arrays instead of straight lines; (3) SADP is used to pattern a nanoimprint template instead of active devices.

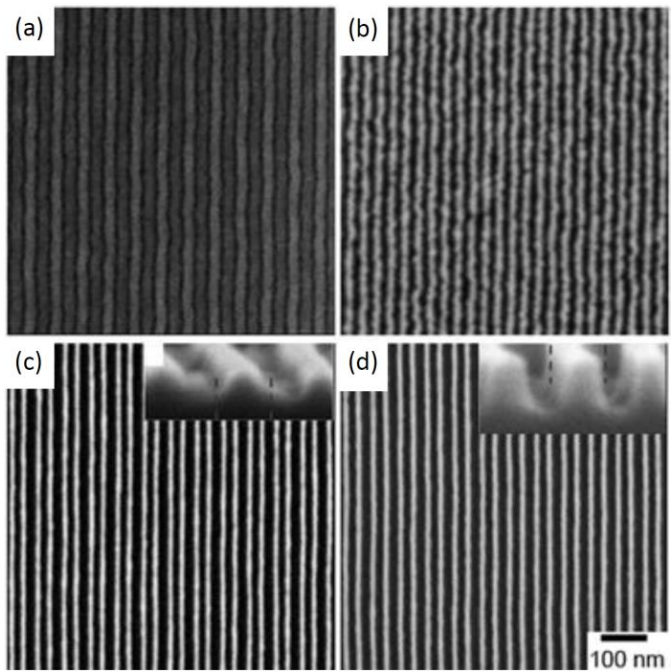

Fig. 21. Pattern transfer steps using SIS on DSA of PS-b-PMMA at 29nm pitch. (a) Striped patterns from PS-b-PMMA after DSA. (b) Pattern after selective AlOx synthesis in the PMMA domains by 3 cycles of SIS. (c) AlOx patterns after selective removal of polymeric material by O2 plasma reactive ion etching. (d) Pattern transferred to Si by reactive ion etching using the AlOx lines on (c) as the etch mask.

Starting with Fig. 22(a), arrays of circumferential or radial lines are first patterned using DSA of a lamella forming BCP (e.g. PS-b-PMMA) on a stack which at least consists of a substrate, a sacrificial material layer known as the mandrel layer, and a hard mask on top of it. The line patterns are subsequently transferred through the hard mask into the sacrificial layer (Fig. 22(b)). A lateral trimming step may be needed to achieve the desired mandrel line width, which is typically close to ¼ of the BCP full pitch. Another layer of material, known as the spacer, is conformally deposited over the mandrel lines (Fig. 22(c)). An anisotropic etch is then performed to remove the portions of the spacer layer on top of the mandrel and the substrate (Fig. 22(d)), followed by removal of the mandrel, leaving only the sidewalls of the spacer layer (Fig. 22(e)). If necessary, the spacer line pattern, which has a pitch that is half of the BCP pitch, can be further transferred into the substrate (Fig. 22(f)).

Our mandrel material of choice is diamond-like carbon (DLC), deposited using sputtering or ion beam deposition. DLC can be reliably patterned using oxygen based ($O_2$, $CO_2$, CO) reactive ion etching (RIE), resulting in a nearly rectangular profile with sidewall angle (SWA) larger than 80°. Fig. 23(a) shows an example of a cross-sectional SEM view of 41 nm full pitch DLC mandrel lines.

Atomic layer deposition (ALD) is employed to achieve uniform conformal coating of the spacer material on mandrel surfaces. The process should not damage the mandrel, and the spacer material should be dry etchable in an anisotropic manner. One example of spacer material we use is $TiO_2$ deposited using thermal ALD without plasma assistance. Fig.

23(b) shows an example of the conformal $TiO_2$ spacer coverage over DLC mandrel of 41 nm full pitch. $TiO_2$ can be etched using fluorine based (e.g. $CF_4$, $CFH_3$, $SF_6$) inductively coupled plasma RIE (ICP-RIE) processes.

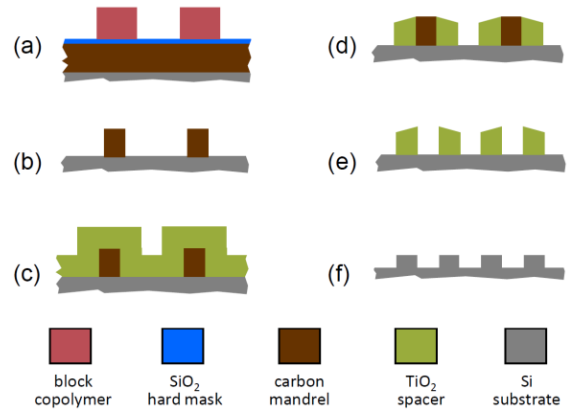

Fig. 22. Line doubling process flow with DSA (a) DSA BCP lines; (b) etched carbon lines with line width control; (c) conformal spacer deposition; (d) anisotropic spacer etch back; (e) mandrel removal; (f) transfer into substrate.

The roughness of the mandrel lines is critical to the overall quality of the final spacer lines. During the patterning of DLC mandrel lines, instead of direct pattern transfer from the BCP lines through a hard mask, we can employ alternative pattern transfer method such as the aforementioned SIS method to minimize the mandrel line roughness. Fig. 24(a) and (b) show 29 nm pitch DLC mandrel lines fabricated using 29 nm pitch DSA of PS-b-PMMA along with SIS pattern transfer into DLC. Edge, width, and placement roughnesses for mandrels have been controlled to within 2.5 nm (3σ). After thermal ALD of the $TiO_2$ spacer, fluorine based ICP-RIE etch back, and mandrel removal using $O_2$ RIE, 14.5 nm full pitch $TiO_2$ lines are produced (Fig. 24(c) and (d)). The roughnesses of the $TiO_2$ spacer lines are also within 2.5 nm (3σ).

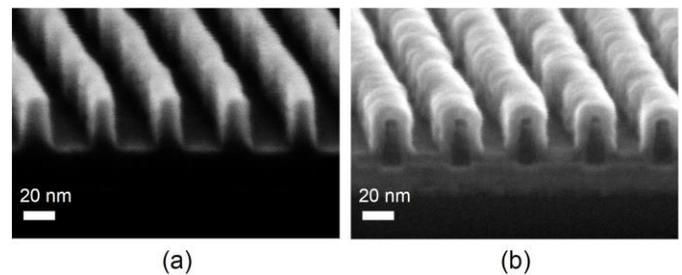

Fig. 23. (a) 41 nm pitch DLC mandrel cross section SEM (b) conformal thermal ALD $TiO_2$ spacer deposition over 41 nm pitch DLC mandrel.

While it is typically preferred to transfer the spacer patterns into the substrate, especially for active device fabrication, $TiO_2$ spacer lines can directly serve as template features for nanoimprinting. Fig. 25(a) shows an SEM image of 20.5 nm pitch $TiO_2$ spacer lines on a Cr coated quartz template. The template is used to imprint directly into a nanoimprint resist film, and the resulting 20.5 nm pitch resist lines are shown in Fig. 25(b). This allows us to bypass the spacer line pattern transfer into the substrate, and replicate the spacer lines into a more desired stack. More on nanoimprint is discussed in later sections.

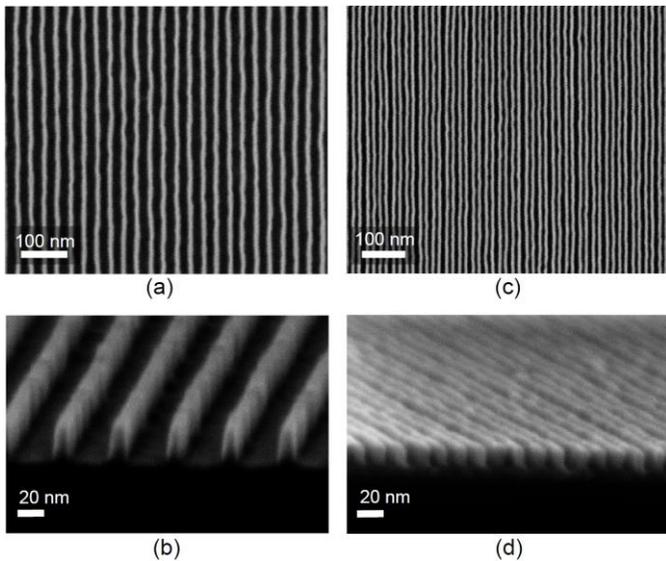

Fig. 24. (a) 29 nm pitch DLC mandrel top view SEM image (b) 29 nm pitch DLC mandrel cross-sectional SEM image (c) top view SEM image of 14.5 nm pitch TiO₂ lines after SADP; (d) cross-sectional view SEM image of 14.5 nm pitch TiO₂ lines after SADP.

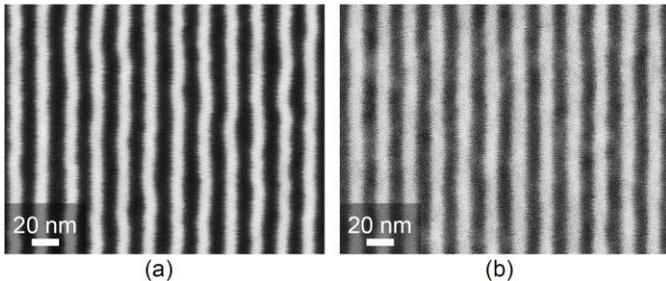

Fig. 25. SEM images of (a) TiO₂ spacers lines and (b) imprinted resist lines at 20.5 nm pitch. Comparing spaces between TiO₂ lines and the resist lines, the LERs are similar, while there is a slight improvement in LPR for the latter, which may be related to tension in the lines caused by the resist shrinkage during UV curing.

### D. Integration of Servo Patterns

When a disk drive reads or writes data on a recording medium (including BPM), it needs position information in both the disk radial and circumferential directions. This is provided by servo patterns on the disk. A recording head reads the servo patterns and determines the position of the head with nm-scale accuracy. Servo patterns consist of several magnetic sequences, including track number, sector number, and sub-track position reference pattern (usually called a "burst" pattern). For an HDD with conventional granular media, servo patterns are recorded on the media with magnetic write heads in a time-consuming track-by-track process. BPM offers the possibility of eliminating conventional servowriting by creating physical servo patterns of magnetic islands along with data islands. One way to accomplish this is to fabricate a master template that contains both servo and data patterns, and imprint this on each disk.

Since servo patterns are inherently complex, with unique track and sector identifiers for different locations on the disk, such patterns are generally not amenable to self-assembly, and must be produced by e-beam lithography, with its arbitrary pattern capability. Ideally, a single e-beam writing session generates both the servo patterns and the guiding patterns for self-assembly of data island patterns [64], [88], since this results in servo and data patterns that are self-registered (see Section V). Since servo and data patterns are fabricated by very different processes (servo pattern by e-beam alone; data patterns by DSA), a special integrated process must be developed to combine these two different processes on a single master wafer.

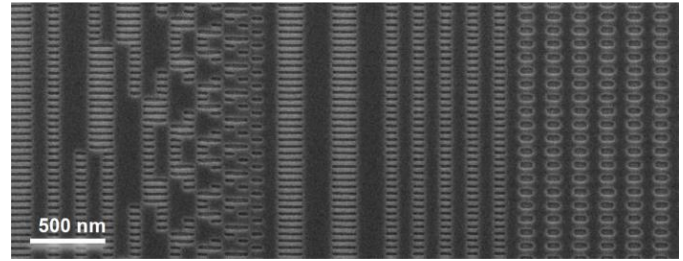

Fig. 26. Servo pattern written on circumferential sub-master.

One option for such an integrated process is as follows: a Si wafer is first coated with cross-linked PS mat (XPS-mat) and baked. The E-beam resist is then coated on the wafer and servo patterns are written on the circumferential sub-master along with the circumferential DSA guiding lines for data tracks as shown in Fig. 26. From here the servo and data patterns are processed differently, using the following sequence: 1) data area protection using photolithography, 2) servo pattern tone reversal and transfer to substrate using conformal ALD deposition, 3) servo protection using protective layer, 4) data pattern retrieval and DSA, and 5) data pattern transfer to substrate. The steps required to achieve this are shown in Fig. 27.

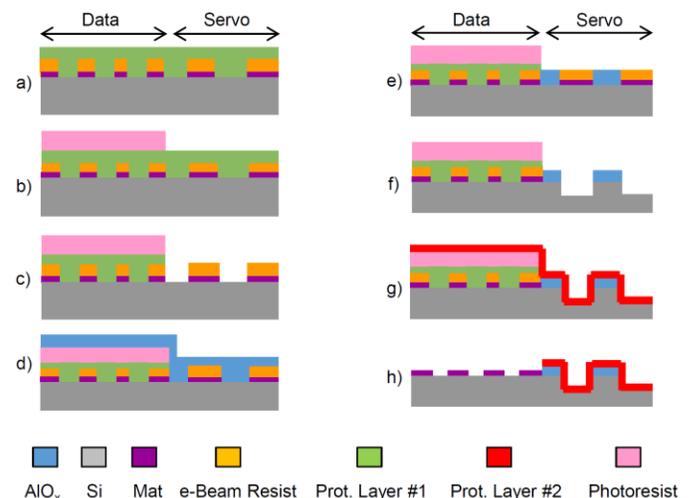

Fig. 27. Servo/data integration process: a) Circumferential sub-master after protective layer #1 deposition. b) Data area protected with photoresist. c) Servo patterns exposed using wet etch. d) Alumina deposition by ALD. e) Alumina etched to produce tone-reversed servo pattern. f) Servo pattern etched into silicon. g) Deposition of protective layer #2. h) Data area DSA guiding patterns retrieved by lift off and wet etch.

The patterned wafer is first etched in oxygen plasma to remove XPS mat from the required regions in both the servo and data areas. A protective barrier layer is deposited, followed by photoresist, which is then exposed and developed to open up the servo areas while protecting the data areas. The servo pattern is then transferred via a conformal deposition of alumina by ALD, followed by an etch back that has the effect of creating a tone-reversed hard mask pattern. The servo pattern is then etched into the Si substrate and the alumina is stripped. A second protective layer is then deposited to cover the servo area, and various etches are used to remove the layers covering the XPS guiding pattern in the data area. BCP DSA then takes place in the data area to form circumferential lines at the final track pitch, and this pattern is then transferred into the Si substrate using a suitable pattern transfer process such as ALD SIS.

Fig. 28 shows patterns in the servo area of a combined sub-master with servo patterns and circumferential arcs. Note that the servo patterns have been cut into islands of limited size, both to facilitate reverse-tone pattern transfer for template replication and disk patterning, and to limit the size of magnetic features on the disk to prevent undesired domain wall motion.

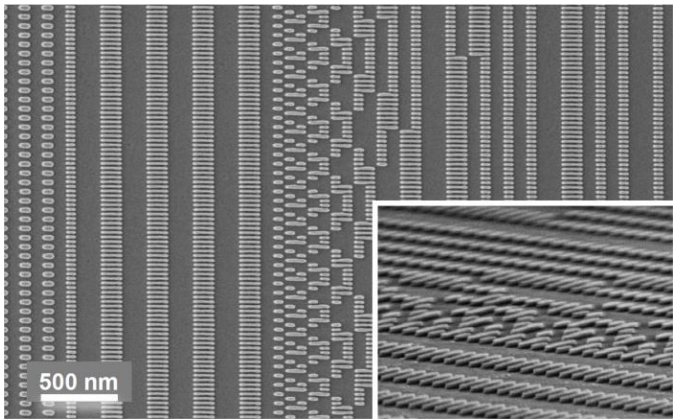

Fig. 28. Servo pattern etched into silicon circumferential sub-master. The inset shows an oblique view of the servo pattern.

### E. Pattern Transfer in Fabrication of Imprint Templates

High volume manufacturing of bit patterned media via nanoimprint lithography requires robust master or sub-master templates usable thousands of times without degradation while withstanding occasional cleaning to remove built-up organic residues or contaminants. One route to achieving this is to transfer patterns directly into the substrate onto which the patterns are primarily formed. This is, however, a formidable challenge at the dimensions required for bit patterned media, especially in the case of SADP. At this small scale, complications to reactive ion etch processes like aspect ratio dependent etching [89] and etch byproduct redeposition [90] render it very difficult to clear the spacer material deposited by ALD at the bottom of trenches. Finding redeposition to be principal issue in our etch processes, successful transfer of spatial frequency doubled patterns formed by SADP was enabled by lowering the parallel plate RF power to increase the likelihood for completed chemical reactions and to reduce the amount of undesirable sputtering.

An example of $TiO_x$ spacer lines with an 18.5 nm full pitch (after mandrel removal using an $O_2$ plasma etch) produced through SADP using this lower power etching is shown in Fig. 29(a). A longer etching time is required to etch through $TiO_x$ at the bottom of trenches as opposed to the top, and so the line height is only ~9 nm. This is not tall enough if the spacer lines are expected to be used directly as the structural template material, but is sufficient to transfer the pattern into the underlying Cr etch stop/hard mask layer by RIE using $Cl_2$ and $O_2$ chemistry that exhibits high etch selectivity for Cr over $TiO_x$. Though only ~4 nm in height, the transferred Cr lines after a wet chemical etch of the $TiO_x$ spacers in a mixture of $H_2SO_4$ and $H_2O_2$ commonly called "piranha" (Fig. 29(b)) are an excellent hard mask for further transfer to Si or quartz, as shown by an SEM image of the final Si lines after etching and wet chemical stripping of the Cr hard mask in Fig. 29(c). This topographical Si line pattern may be used directly in a sub-master template to create high BAR media according to a double imprint scheme. Alternatively, the minimal topography of the Cr lines introduces little complication to the patterning of orthogonally-oriented lines of another material above them. These orthogonal lines can then be used as a mask to create rectangular Cr dots using RIE, which in turn are used as a hard mask to generate pillar-tone master templates directly. This pattern transfer method appears extendible to even smaller pitches, as shown by Si lines with a 14.5 nm full pitch shown in Fig. 29(d). In this case, $CrN_x$ was used in place of elemental Cr for its smaller grain structure more amenable to patterning at smaller dimensions.

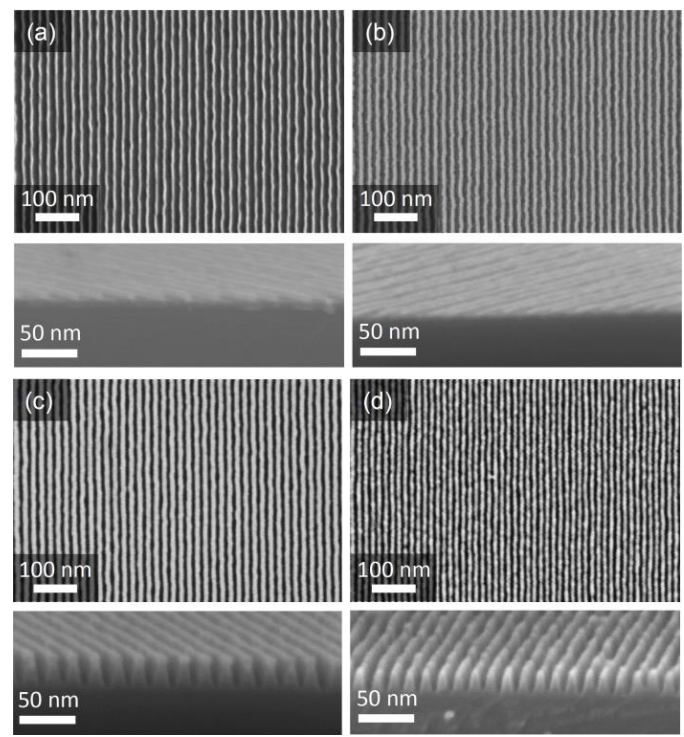

Fig. 29. Pattern transfer of lines created through self-aligned double patterning into Si for sub-master template preparation. (a-c) Plan view and cross-section SEM images of lines at 18.5 nm full pitch after key process steps including (a) $TiO_x$ spacer lines after etch back and mandrel removal, (b) Cr lines after $Cl_2/O_2$ RIE and $TiO_x$ wet strip, and (c) Si lines after $CF_4$-based RIE and Cr wet strip. (d) Plan view and cross-section SEM images of lines at 14.5 nm full pitch transferred into Si using the same process after $CrN_x$ wet strip.

Our nanoimprint template fabrication strategy has undergone two major transitions: (1) hexagonal island arrays to rectangular, or high bit aspect ratio (BAR) arrays, which has been discussed in detail in earlier sections, and (2) hole-tone to pillar tone. Nanoimprinting, by virtue of its nature as a "molding" process, results in an image tone reversal between the template pattern and the imprinted resist. While this tone reversal can in principal be taken into account and the overall process (from basic pattern generation all the way to finished BPM disk, for which there may be multiple tone reversals), the preferred approach is to transfer patterns from imprint resist using a tone-reversal process, so that the tone reversals of the imprint and pattern transfer result in a positive tone image transfer. Using this strategy, we use pillar-tone ("feature proud") patterns on our templates at every imprint step. Advantages of this approach include better resist flow, elimination of resist pattern collapse, superior etch selectivity in pattern transfer, and reduced sensitivity to residual layer thickness of imprinted patterns.

The advantages of pillar tone templates become significant for areal densities beyond 1 Td/in$^2$. For hexagonal patterns, a pillar tone template can be achieved using one of the aforementioned wet or dry liftoff type BCP pattern transfer methods followed by etch into a substrate such as Si or fused silica. The fabrication of high BAR templates involves cross cutting of the circumferential and radial lines.

The preferred cross line cutting option makes use of sequential imprints from two separate sub-master grating patterns, one with circumferential lines and the other radial lines to create high BAR patterns on a suitable template substrate, or to directly pattern disks. The first imprint is used to transfer one set of line patterns (e.g. radial lines) to a hard mask, which is in turn cut by transfer of circumferential line patterns from the second imprint. Fig. 30(a) shows the top view SEM image of a 1.6 Td/in$^2$ DLC dot array on a disk master, with 22 nm track pitch and 18.5 nm bit pitch, fabricated using this option. An advantage of using this dual-imprint approach is that both the radial and circumferential sub-master patterns are reusable; it is not necessary to create new grating patterns to make additional copies of working templates or disks. However, imprinting of high density parallel lines is challenging. Some sub-10 nm polymer resist lines with more than 10 nm in height may not be mechanically strong enough to survive the template and substrate separation, resulting in collapsed or delaminated patterns.

The second option removes one nanoimprint step for high BAR template fabrication. Because bit pitch is smaller than track pitch for high BAR patterns, the radial lines are patterned using BCP DSA and SADP, with an optional transfer into a hard mask. Nanoimprint of circumferential lines will then follow to cut the existing radial lines. Fig. 30(b) shows the top view SEM image of a 1.6 Td/in$^2$ silicon master pattern, with 22 nm track pitch and 18.5 nm bit pitch, fabricated using this option.

When AD further increases, even the circumferential lines with larger pitch will face nanoimprint challenges. Therefore a third option may be needed, which eliminates any nanoimprinting steps during master template making. One possible process is to pattern the radial lines using DSA and SADP, and then perform a second DSA of circumferential lines (and optional SADP) to cut the underlying spacer lines.

Fig. 30(c) shows the top view SEM image of a 1.65 Td/in$^2$ TiO$_2$ dot array on a silicon substrate, with 27 nm track pitch and 14.5 nm bit pitch, fabricated using this option.

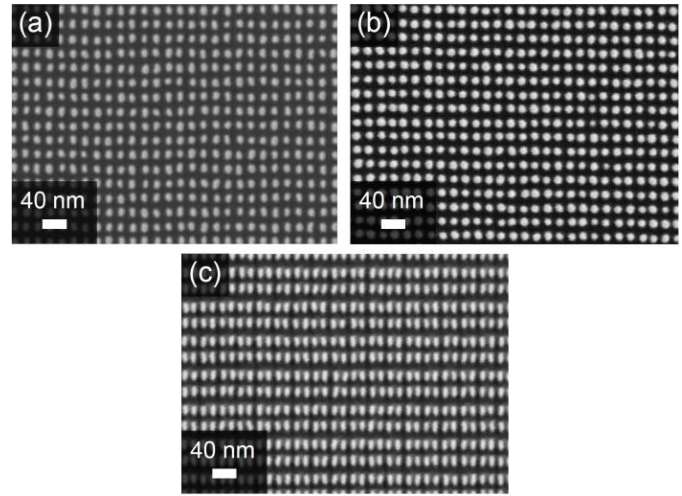

Fig. 30. Top view SEM images of (a) 1.6 Td/in$^2$ DLC dot array on a disk master, 18.5 nm x 22 nm, two nanoimprint steps; (b) 1.6 Td/in$^2$ silicon master pattern, 18.5 nm x 22 nm, DSA, SADP and one nanoimprint step; (c) 1.65 Td/in$^2$ TiO$_2$ dot array, 14.5 nm x 27 nm, DSA, SADP and 2nd DSA, no imprint.

After the fabrication of a BPM master template, it needs to be replicated to multiple "working templates" for high volume manufacturing of disks using nanoimprinting. The template replication is done by imprinting the master template pattern into nanoimprint resist on working template substrates, which is further transferred into these substrates using one of the aforementioned tone reversal pattern transfer techniques. The degradations in the roughness numbers during the template replication process are typically less than 10% of those on the master template. As template replication does not have as high a throughput requirement as disk manufacturing, some steps, such as nanoimprinting, can take longer, which potentially reduces the chance of defect formation. More on defectivity is discussed in the nanoimprint section below.

*F. Nanoimprint Lithography*

Nanoimprint lithography [91], [92], [93] is the preferred technology for low cost high volume replication of BPM patterns because of its extendibility to single digit nm-scale feature size and its ability to replicate full disk patterns without stitching. Nanoimprinting was first developed specifically as a solution for fabrication of BPM [94].

HGST has collaborated with Molecular Imprints, Inc. since 2005 to develop equipment and processes for double sided full disk imprinting. Three generations of nanoimprinters (the Imprio$^{TM}$ 1100 single side imprinter, and the Imprio$^{TM}$ 2200 and NuTera$^{TM}$ HD7000 multiple head double sided imprinters) have been developed and extensively exercised in BPM template and media fabrication, with over 50,000 imprints performed at HGST. Concurrent with equipment innovations there has been extensive work in process development, including new resist materials, refinement of template release strategies (including release agents incorporated into resist), optimization of the adhesion promoter (vapor deposited on disks prior to imprinting), improved template/disk mechanical contact and release methods.

### 1) UV Cure Ink Jet Dispense Nanoimprinting Process

The core imprinting steps include resist dispensing, bringing the template and resist-coated disk into conformal contact, UV exposure and template/disk separation as illustrated in Fig. 31. Resist may be applied either by droplet dispensing from a linear array of ink jet nozzles under which the disk passes, or by spin coating. Ink jet offers the advantages of rapid in-line dispensing (immediately prior to imprinting) and localized resist volume control. However, resist dispensed as droplets may take more "spreading time" to form a uniform continuous layer and conform to template pattern features. Spin coating may reduce spreading time, but requires a dedicated spin coating system. The UV curing step requires as little as a few seconds if there is sufficient UV intensity and resist sensitivity. Separation of the disk and template can also be very rapid (<1 sec) but requires a well-engineered process, both in terms of a release agent to prevent adhesion and a mechanical system that avoids damage to the imprinted resist patterns caused by mechanical strain during release. Conversely, ensuring good adhesion of the resist to the disk requires ~2 nm of an adhesion promoter on the disk surface.

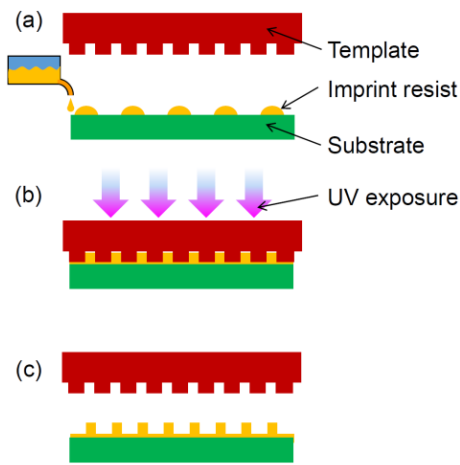

Fig. 31. UV-cure nanoimprinting process steps: (a) resist dispensing, (b) conformal contact of the disk and template, resist spreading, and UV curing, and (c) separation of the disk and template.

With droplet dispensing, the micro-fluidic spreading by which droplets coalesce into a continuous film and conform to the template features is a critical process. Resist flow in the rapidly decreasing channel width as the template is brought into contact with the resist-coated disk is subject to subtle effects governed by surface chemistry, pattern feature properties, and the mechanical procedure for bringing the two bodies into contact. Contact occurs first at the inner diameter of the disk and propagates outward from there. In addition to the time needed to initiate and propagate contact, there can be considerable additional "spreading" time to allow the resist film thickness to become uniform and for the resist to conform to nanoscale topographic features on the template. Spreading time ends up being the rate limiting factor for nanoimprinting on BPM with ink jet dispense, requiring >10 sec per imprint. Our current minimum droplet size is 3 pl; experiments showing the feasibility of dispensing 1 pl droplets suggest that a denser array of smaller droplets could be a route to reduced spreading time.

UV curing has been successfully implemented with either a mercury lamp or LED light source. While thermal curing [91] can be an alternative nanoimprinting strategy, avoiding thermal expansion issues is thought to be an advantage of UV curing.

The final step is separation of the disk and template, leaving behind on the disk a solid resist film with an inverse replica of the topographic pattern on the surface of the template. The separation process is roughly the reverse of the process by which the template was brought into contact with the disk; initial separation occurs at the outer diameter of the disk and propagates toward the disk center.

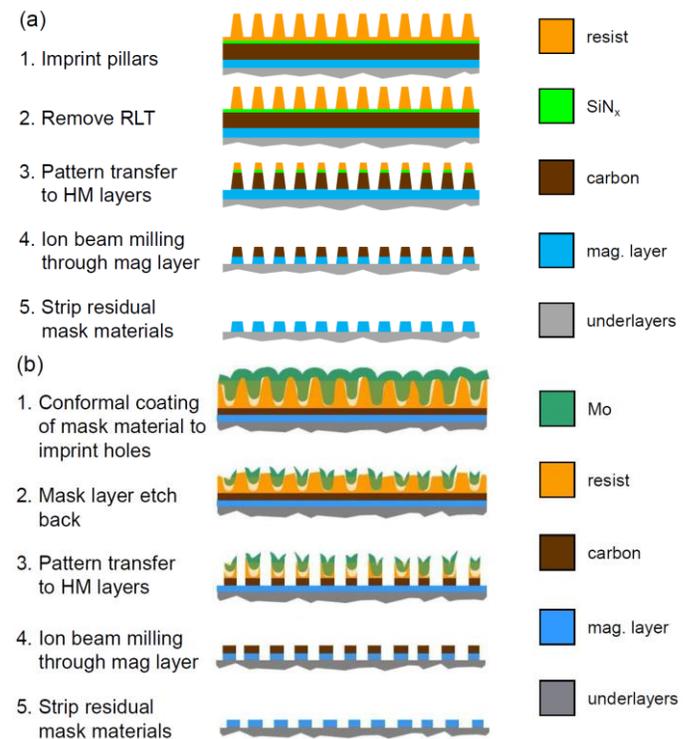

Fig. 32. (a) Normal tone and (b) reverse tone nanoimprinting and pattern transfer processes.

### 2) Normal Tone vs. Reverse Tone Imprinting

Nanoimprinting is similar to a molding process in that the topographic pattern in the resist is a negative tone image of the template pattern. To make use of the resist image for pattern transfer to another layer (such as the magnetic layer of BPM) there are two choices: a *normal tone* process, in which the resist itself is used as an etch mask, and a *reverse tone* process, in which depressions in the resist are backfilled with another material which serves as the etch mask for pattern transfer (see Fig. 32) [93]. Although the terminology can be confusing, it is important to note that *normal tone* produces a negative image of the template features, and *reverse tone* produces a positive image.

While the normal tone process is conceptually simpler, it has important drawbacks. Nanoimprint lithography, unlike photolithography, does not produce a pattern of resist areas and clear areas. Instead of clear areas, nanoimprinting leaves behind a thin residual layer, so the etch contrast of the resist mask in a normal tone process is between relatively thick vs. relatively thin areas, as shown in Fig. 33. The residual layer thickness (RLT) can be a substantial fraction of the main feature height, and can also be nonuniform and difficult to precisely control. In pattern transfer it is necessary to clear the residual layer before etching into the material below; the etch used to clear the RLT can result in critical dimension (CD) shifts and degrade overall image fidelity. A further disadvantage of a normal tone process for BPM patterns is that a normal tone (hole tone) template does not leave open channels for resist flow during spreading; thus resist is forced to spread via the thin residual layer, which results in slower spreading.

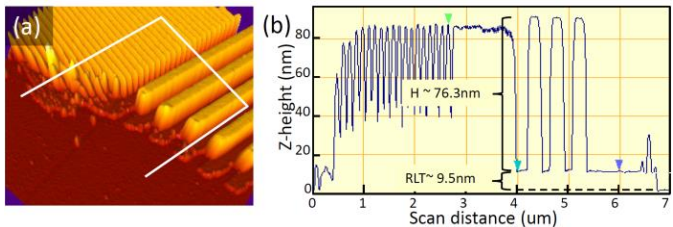

Fig. 33. AFM image of a nanoimprint resist pattern showing the height of features and the thickness of the residual layer between the imprinted lines. Observation of the residual layer and its thickness is facilitated by scratching away part of the resist pattern (foreground), leaving resist edges exposed.

A reverse tone process, while requiring more process steps, offers the advantages of open channels for better resist spreading on BPM patterns, and superior etch contrast for pattern transfer (depending on the backfill material). Since the resist itself is not the etch mask, the reverse tone process is significantly less sensitive to RLT and RLT variation. In addition, the reverse tone resist image is an interconnected network which is far more mechanically robust than a normal tone pillar pattern and thereby eliminates pattern collapse.

### 3) Defects in Nanoimprinting

Fig. 34 illustrates a variety of common imprinting defects which can be mitigated by appropriate methods.

#### a) Non-Fill Defects.

Fig. 34(a) shows an AFM image of non-filled resist pillars in a normal tone BPM process caused by blockage of template holes by foreign material or insufficient resist volume. The pillars present are only 8-10 nm in height, compared to template holes that are 42 nm deep. Some pillars are completely missing. Cleanliness and optimization of resist dispense volume can prevent these defects. Fig. 34(b) shows a partially filled network in which the resist failed to completely fill the narrow valleys around pillars on a pillar tone template.

#### b) Resist Fracture / Clogged Template.

Fig. 34(c) shows defects caused by insufficient release treatment. Most pillars were fractured during disk-template separation because the adhesion between template holes and cured resist pillars exceeded the mechanical strength of the pillars. In Fig. 34(d), an AFM image of the corresponding template confirms the failure mechanism in that the broken resist pillars lodged in the template holes are visible.

Subsequent imprints with this template may be affected by the clogged holes. Optimization of resist mechanical properties and template feature aspect ratio as well as sidewall slope can eliminate broken features. Use of a reverse tone process provides an even more definitive method to prevent such defects.

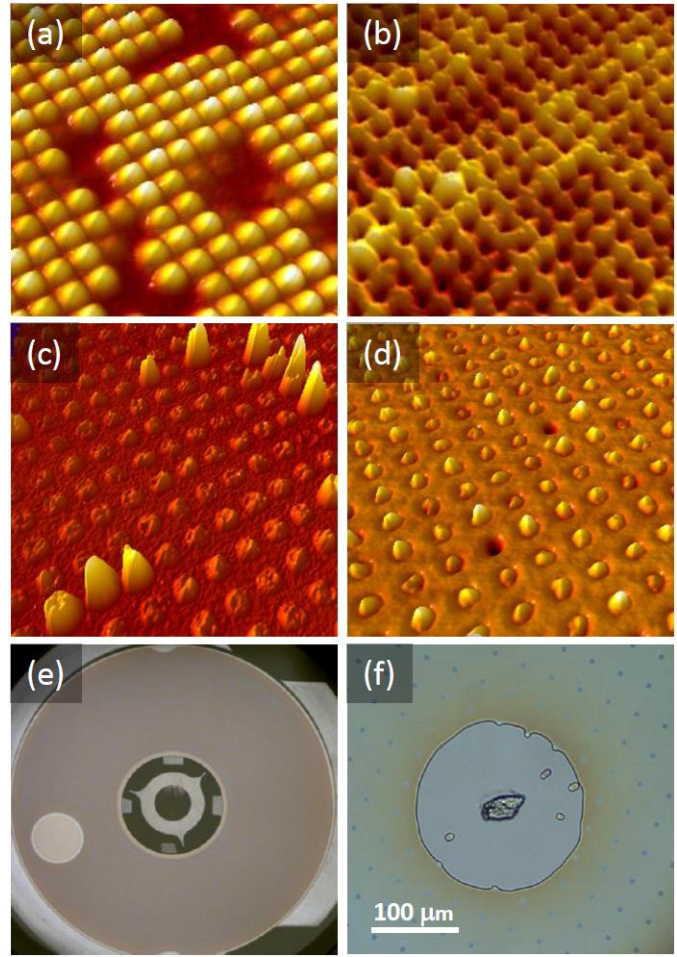

Fig. 34. Typical defects in nanoimprinting: (a) Non-filled features in a 300 Gd/in² pattern with a hole tone template. (b) Non-filled features in a 300 Gd/in² pattern with a pillar tone template. (c) Fractured pillars. (d) Fractured pillars lodged in 300 Gd/in² template holes. (e) Full disk view with large cm-size tent defect caused by large air-born particle. (f) Tent defect with particle visible at center of tent.

#### c) Particles / Tenting.

Fig. 34(e) is a full-disk optical image which shows an example of a large (cm-size) defect caused by a particle. Particles lodged between the template and disk surfaces prevent proper conformal contact between the two, leaving regions of thick and/or incompletely filled resist patterns. Fig. 34(f) shows a smaller tent defect introduced by an air-born particle; the particle which caused the tent is visible in the center. Depending on the size and type of particle, these types of defects can range in lateral extent from <1μm to >10mm. Clean templates, disks, and environment are critical to preventing tenting defects.

Although the above types of defects can be encountered in nanoimprinting, appropriate measures as indicated above can easily keep the defect rate below $10^{-3}$, which is sufficient for BPR to achieve a typical raw error rate target of $10^{-2}$.

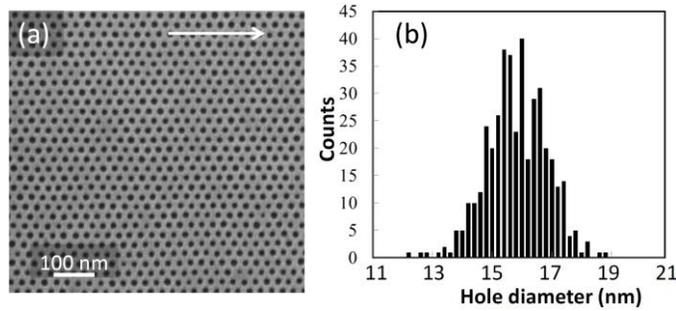

Fig. 35. (a) SEM image of a reverse tone, 1 Td/in² imprint. The imprint features are holes, patterned in a hexagonal array with a pitch of 27.3 nm. The arrow indicates the circumferential, or down-track direction. (b) Histogram of the imprint hole diameter. The mean diameter is 16.38 nm and the sigma is 0.94 nm. The down-track and cross-track placement accuracy sigmas are 1.19 nm and 1.08 nm, respectively.

### 4) Imprint Metrology and Image Quality Analysis

As explained in Section II, island uniformity and placement tolerance are key factors for successful high density BPR. Image analysis of top-down SEM micrographs can provide statistical metrics of key parameters. Fig. 35(a) shows an SEM image of a reverse tone imprint of a 1 Td/in² hexagonal hole pattern with 27.3 nm pitch. Fig. 35(b) shows a histogram of the diameter in which the mean is 16.38 nm and the 1-sigma is 0.94 nm. Analysis of hole placement tolerance (deviation of hole centroids from ideal lattice points) indicates a down-track 1σ placement tolerance of 1.19 nm and 1.08 nm cross-track. These tolerances are less than 5% of pattern pitch, which generally indicates sufficiently tight tolerance for good BPR performance.

### 5) Template Lifetime

Although we have performed thousands of imprints on some of our templates, no detailed statistical analysis of defect generation or CD erosion has been performed. Others have reported anticipated template lifetime exceeding 10,000 imprints [95]. A key factor affecting template longevity is the ability to clean templates without feature erosion. A cleaning procedure based on ozone water [96] has been reported to reduce feature erosion compared to Piranha ($H_2SO_4 + H_2O_2$) cleaning, which we have used routinely without any obvious erosion problems.

Of particular concern for template lifetime is damage caused by particles that lodge between the template and disk. Such particles can cause tent defects, and if the particle adheres to the template, such defects will repeat on subsequent disks. An effective strategy for removing particles that have adhered to the template is to perform an imprint on a dummy disk with an increased resist amount (e.g., sufficient for a RLT of ~100 nm). In most cases, this will result in the particles adhering to the dummy disk instead of the template, with the result that the template is successfully cleaned in situ and can continue with normal imprinting on subsequent disks. In the event this procedure fails, the template can be cleaned ex situ.

Optical inspection of disks just prior to imprinting and rejection of disks with incoming particles is a key factor in managing yield and preventing template contamination. Optical inspection during resist spreading or immediately after imprint can easily detect tent defects, allowing automatic execution of in situ or ex situ cleaning procedures.

While particles, even if removed, can cause permanent template damage (e.g., fracture of template features), as long as the area of damage is limited, a template can continue to be used for BPM, since a BPR system can tolerate defect rates as high as $10^{-3}$.

### 6) Extendibility of Nanoimprinting

UV-cure nanoimprinting appears to be extendible from a basic image resolution standpoint to at least 4 Td/in². We have imprinted ~5.5 nm features with ~12 nm pitch using a template created from a random nanoparticle pattern; the size and density of these features is consistent with what would be expected on a 4 Td/in² BPM pattern. Improved resists could extend nanoimprinting even further. In general, demonstrations of nanoimprinting for higher density patterns is limited mainly by the ability to produce a high quality template rather than by inherent resolution limits of the imprinting process or resist.

Although sensitivity to RLT and RLT variation is greatly reduced when using a reverse tone process, extending pattern density toward 4 Td/in² is likely to result in renewed emphasis on thin and uniform RLT. Reducing droplet size to ~1 pl and using a more dense droplet array may provide a route for better RLT control as well as reduced spreading time.

### G. Media Pattern Transfer

### 1) Media Etch Mask Preparation

For successful employment of the reverse tone approach to pattern transfer, the deposited fill mask (FM) material must selectively amass within depressions in the imprinted resist to an extent in which it provides sufficient masking power. Two basic phenomena have been exploited to achieve this: (1) Inefficient trench etching, where factors like Knudsen transport, etch byproduct redeposition, and shadowing by protruding resist features result in substantially lower relative etch rates for FM material within imprint resist depressions, and (2) conformal/liquid-phase HM deposition, where an excess of FM material accumulates within the resist depressions at the FM deposition stage. FM materials for reverse tone pattern transfer include Ta, Ti, Cr, Mo, $Al_2O_3$, $TiO_2$, $SiO_2$ and spin-on glass deposited by various methods such as e-beam evaporation, sputtering, ion beam deposition, atomic layer deposition, or spin-coating.

Despite the advantages of the reverse tone approach discussed earlier, it still suffers from microloading effects, where the FM material is etched back faster in larger holes than in smaller ones. This is a potential problem for creating servo features, which can be substantially larger (~5X) than data features, and recommends servo designs where features are approximately equal in size to data features.

### 2) Magnetic Dot Pattern Transfer by Etching

Transfer of the pattern formed in imprint resist by either the normal tone or reverse tone approaches into magnetic media is facilitated by the inclusion of a hard mask layer(s) under the imprint resist. Successful hard mask layers must be designed to meet several requirements. Firstly, the hard mask needs to be free of particles that affect imprint quality and damage the imprint template. Chemical/mechanical washing is generally used to remove particle contamination, and a useful hard mask must be resistant to damage from this treatment. Secondly, the hard mask must have good adhesion to both the media substrate (capping layer) and the imprint resist. Poor adhesion with imprint resist can prevent resist spreading uniformly (due to unfavorable surface chemistry) and resist adhesion failure during template separation at the end of the imprinting process. Poor adhesion to the media substrate can result in hard mask flaking, increased particulate contamination, and non-uniformity in media etch. Thirdly, the hard mask must have enough masking power (selectivity) for media etch to the required depth, primarily by an ion milling process. Finally, the remaining hard mask after media etch must be stripped in a way that does not damage the media or promote media corrosion, prohibiting the use of fluorine- or chlorine-based RIE for this step.

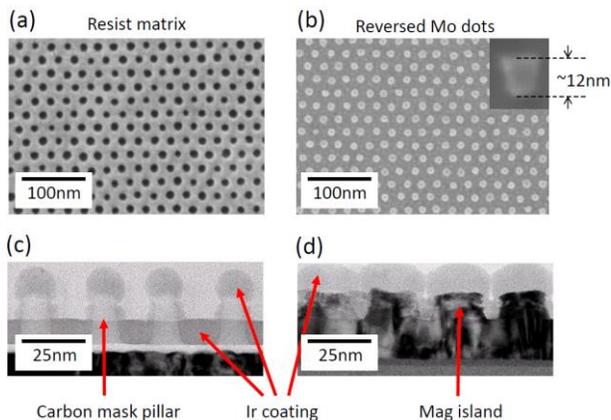

Fig. 36. (a) SEM image of a hole-tone imprint resist pattern at 1.0 Td/in$^2$. (b) SEM of Mo dots produced by tone-reversal of a resist pattern like that shown in (a). (c) TEM image of carbon hard mask pillars transferred from a Mo dot pattern. (d) TEM image of magnetic media islands after media etch and hard mask strip.

To satisfy these specifications, we employ a dual-layer hard mask. A robust amorphous or diamond-like carbon (DLC) layer, deposited either by plasma-enhanced chemical vapor deposition or facing target sputtering [97], is deposited on top of the magnetic media layers as the primary masking layer due to its low Ar sputter yield and facile stripping by oxygen-based plasma etching. Because carbon exhibits poor adhesion to imprint resist and is etched using the same chemistry, a secondary SiN$_x$ or SiO$_x$ hard mask layer (< ~3 nm) is added on top of the carbon layer. Fig. 36(a) shows an SEM image of a hole-tone imprint resist pattern at 1.0 Td/in$^2$ and Fig. 36(b) shows free standing reverse tone Mo dots on top of a SiN$_x$ hard mask after Mo deposition, Mo etch back and resist etch. After tone reversal, the pattern is transferred from Mo dots to SiN$_x$, using a very short low power fluorine-based etch. To etch through the primary carbon layer, a CO$_2$-based RIE is used instead of an O$_2$ etch because we have found it etches carbon more anisotropically. Fig. 36(c) shows a TEM cross section of carbon hard mask pillars.

Currently, Ar plasma milling is our preferred approach to etch media without damaging the magnetic properties and inducing corrosion. There are different sources to generate Ar plasma. Ion Beam Etching (IBE) is commonly employed for research purposes due to the high etch anisotropy provided by the use of a highly collimated ion beam in an ultra-high vacuum environment (<10$^{-5}$ Torr). Fig. 36(d) shows a TEM cross section of IBE etched media islands.

Due to the low plasma density, the etch rate of IBE is too low for high-volume manufacturing, leading to the introduction of a plasma media etching process for pilot line production. Capacitively coupled plasma (CCP) etching does not show the same anisotropy and selectivity as IBE and can only etch one side of each disk. On the other hand, double-sided etching is made possible using an inductively coupled plasma (ICP) process where bias is applied through the carrier to the disk. While DC or RF bias leads to dramatic over-heating during ICP etch, a pulsed DC bias was found to mitigate this effect. Then, by applying low ICP power and high bias voltage, a media etch rate 20 times faster than IBE with similar selectivity can be achieved. Table 4 shows the current pilot line process for BPM. The maximum throughput is about 16 sec per disk if each step is processed in a single station. With multiple stations, throughput can be improved to 5 sec per disk.

TABLE 4. PILOT LINE PROCESS FOR PATTERNED MEDIA

| Step | Process | Time (sec/disk) |
|---|---|---|
| Mo etch back | Fluorine | 16 |
| Resist etch | H$_2$/Ar | 9 |
| SiNx HM etch | Fluorine | 2 |
| NCT HM etch | CO$_2$ | 15 |
| Media etch | Ar | 10 |
| HM strip | O$_2$ | 15 |
| Carbon dep. | C$_2$H$_2$ | 2 |

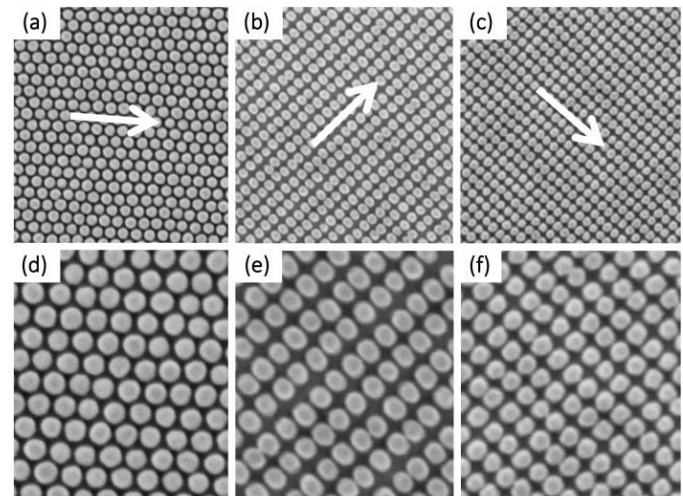

Fig. 37. SEM images of BPM magnetic dots with an AD of (a) 1.0 Td/in$^2$ in a 27.3 nm hexagonal array, (b) 1.2 Td/in$^2$ in a 27 nm Circumferential x 20.5 nm Radial rectangular array, and (c) 1.6 Td/in$^2$ in a 22 nm circumferential x 18.5 nm radial rectangular array. Arrows indicate the down-track direction. (d) - (f) Higher magnification images of the same samples as (a) – (c).

Fig. 37(a-c) show SEM images of BPM media at 1.0 Td/in$^2$ (hexagonal array), 1.2 Td/in$^2$ (rectangular array, BAR = 1.3) and 1.6 Td/in$^2$ (rectangular array, BAR = 1.2), respectively. Fig. 37(d-f) are higher magnification images of a-c, respectively, and Table 5 lists the CD and placement accuracy (all in units of nm) of the magnetic dot arrays. Advancements in BPM fabrication have improved the overall quality of the 1.6 Td/in$^2$ rectangular patterned media (built in 2014) in comparison to the 1.0 Td/in$^2$ hexagonal patterned media (built in 2012).

TABLE 5. CD AND PLACEMENT ACCURACY OF BPM MAGNETIC DOTS

| Areal density (Tdots/in$^2$) | 1.0 | 1.2 | 1.6 |
|---|---|---|---|
| Dot cross-track mean | | 22.44 | 19.10 |
| Dot cross-track 1σ | | 1.14 | 1.34 |
| Dot down-track mean | 21.03 | 18.50 | 17.58 |
| Dot down-track 1σ | 0.88 | 1.20 | 1.05 |
| Dot aspect ratio | 1.00 | 1.24 | 1.10 |
| | | | |
| Placement down-track 1σ | 1.15 | 1.28 | 0.94 |
| Placement cross-track 1σ | 1.22 | 1.35 | 0.91 |
| Total placement 1σ | 1.68 | 1.86 | 1.31 |

In part due to ion scattering off the carbon hard mask sidewalls during the Ar milling process, the etched media islands develop sloped sidewalls and a fraction of the sputtered media is redeposited onto the carbon hard mask forming a media "fence" structure that effectively becomes part of the hard mask as shown in Fig. 38(a). These aspects of media etching contribute to a substantial growth in media island diameter compared to the hard mask islands, in some cases by more than 100%. Angled IBE may minimize fencing but cannot match the throughput of an ICP etch process. Although using thinner carbon hard mask and magnetic media layers may also reduce media island growth from the Ar mill process, this issue poses a major challenge for scaling BPM fabrication beyond 2 Td/in$^2$.

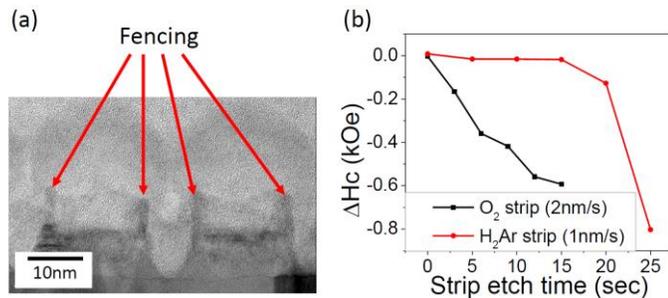

Fig. 38. (a) TEM of media islands before strip. (b) Coercivity drop of O$_2$ strip and H$_2$Ar strip.

After media etch, the carbon and redeposited magnetic material (including fencing) must be removed with minimal damage to the magnetic media. As shown in Fig. 38(b), pure O$_2$ plasma etching causes a severe drop in coercivity as a result of media oxide growth. On the other hand, an H$_2$ + Ar (10% H$_2$) strip has a minimal impact on the magnetic media quality while the protective capping layer remains. Combining these, our process employs an O$_2$ plasma etch to remove most of the carbon, followed by an H$_2$/Ar etch to remove the remaining carbon, fencing and residue. By controlling the strip time, coercivity drop can be controlled within 500 Oe. The dash line in Fig. 38(b) shows optimal process time for the combined strip.

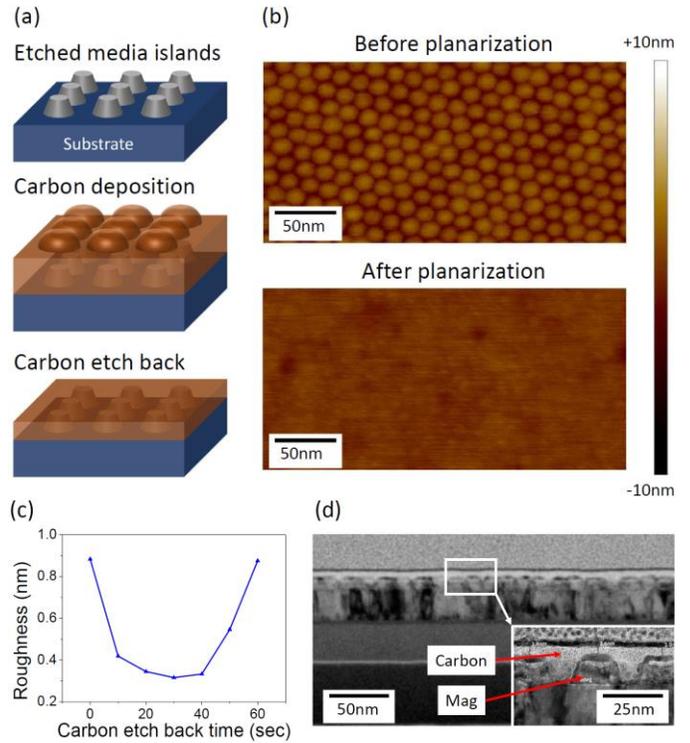

Fig. 39. (a) Planarization diagram. (b) AFM of patterned media surface before and after planarization etch back. (c) Surface roughness vs planarization etch back time. (d) TEM of large area of planarized media islands. Inset is zoom-in.

### 3) Vacuum Planarization

Patterning BPM via etching from a full magnetic film introduces intrinsic roughness to the disk surface. To improve the flyability of patterned disks, a planarization process may be used to flatten the disk surface seen by the head. In vacuum planarization, carbon is first deposited to fully cover the patterned media surface after hard mask strip. An Ar plasma with high ICP power and low bias is applied to redistribute the carbon from protrusions to valleys as shown in Fig. 39(a). Fig. 39(b) shows AFM images of the surface before and after planarization etch back. As the planarization etch back time is increased, the surface initially becomes smoother. When the planarization etch removes all the carbon on top of the media islands, it begins to etch the media faster than the carbon remaining in the grooves, which results in increasing roughness with overetching as shown in Fig. 39(c). With optimal Ar etch back time, a well-planarized surface is achieved as shown in Fig. 39(d). The inset indicates the presence of ~3 nm carbon on top of the media from an additional carbon overcoat deposited after planarization. In addition to providing a smoother head-disk interface for the flying magnetic head, planarization can also improve corrosion resistance of BPM by improving island sidewall coverage compared to simple overcoating of etched islands.

### H. Alternative Pattern Transfer: Templated Growth

As previously introduced in Section III-D, templated growth (TG) is an alternative process for fabricating magnetic islands for BPM. Instead of etching islands out of a continuous film, segregated islands are grown on specially prepatterned nucleation sites. To create epitaxial templated growth (eTG)

BPM, an array of topographic nucleation sites is created on the surface of a Pt(111) seed layer, using a fabrication scheme similar to that used to etch islands out of full film media (as discussed above). After imprinting, a tone reversal process is performed with deposition and etchback of Mo. The Mo dot pattern is transferred into a carbon hard mask layer and then into the Pt seed layer using IBE. A key requirement is to preserve the crystalline structure and orientation at the surface of the Pt seed layer after pattern transfer. The media etch fabrication scheme, when employed for patterning Pt seed layer for eTG, causes similar challenges as in media etching (fencing and rounding of the top surface). These are critical issues for eTG as the top surface texture and morphology affect the growth of the magnetic layer on these features. Use of typical dry etching methods to etch Pt, strip the hard mask, and remove fencing results in rounded morphology as shown in Fig. 40(a). An improved process incorporating a sacrificial wet-strippable layer under the hard mask produces a more ideal morphology as shown in Fig. 40(b,c). Upon completion of the patterned array of nucleation sites on Pt(111), epitaxial growth of various underlayers and the co-sputtered magnetic alloy and segregant proceeds as discussed in Section III-D.

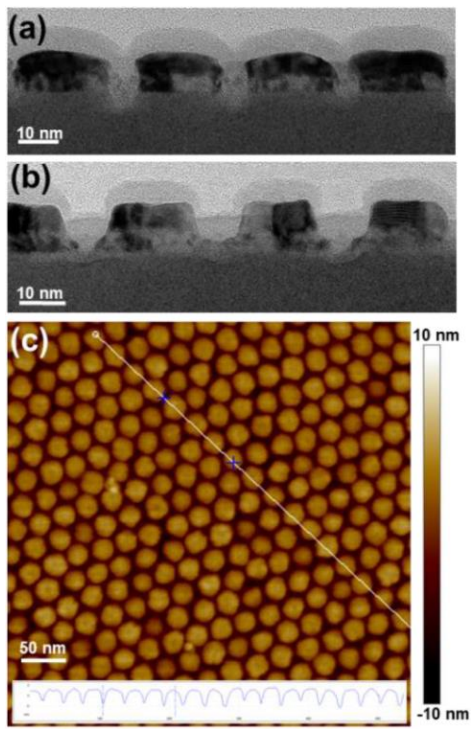

Fig. 40. Cross-section TEM of a disk showing Pt nanopillars for eTG (a) after pattern transfer using POR method for pattern transfer, (b) after modified fabrication process i.e. removal of fencing using angled IBE and wet etch to remove carbon hard mask, (c) AFM and section analysis showing top surface flatness of Pt nanopillars in (b).

### I. Alternative Pattern Transfer: Ion Implantation

Ion implantation magnetic patterning embeds high energy ions in the recording layer to locally modify the magnetic properties while introducing minimal topography [98]. The implant species and implant energies are chosen to yield high lateral resolution by minimizing lateral straggle and matching the implant depth (including longitudinal straggle) to the magnetic recording layer thickness. Low straggle is favored by implanting with heavier mass species such as P, As, or Sb. Requirements for the implant hard mask are more stringent than for etching – the mask has to be fabricated with sufficiently high aspect ratio (high height / narrow width) and have adequate robustness against sputter degradation during the high dose implantation. Diamond-like carbon is a good hard mask material candidate since it has one of the lowest sputter yields of any material and supports reactive-ion etch patterning of tall and narrow mask features with vertical walls.

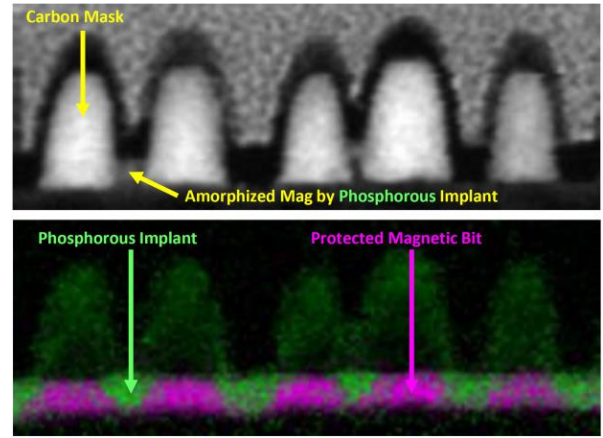

Fig. 41. Top: Bright-field TEM cross-section of carbon hard mask of sample after implanting P. Bottom: EELS image of the same sample. Green shows the location of P (in the mag layer and along the walls of the carbon hard mask after implantation. Magenta shows the protected magnetic alloy which are the magnetic bits providing the switching plotted in Fig. 42.

$Co_{70}Cr_7Pt_{25}$ media has been magnetically patterned at 1 Td/in$^2$ densities using high dose (2 x$10^{16}$ atoms/cm$^2$) of phosphorus (P) ions implanted at ~7keV energy. As shown in Fig. 41, the implanted P magnetically patterned by locally amorphizing the crystalline Co-alloy and reducing its magnetic moment relative to the protected, non-implanted regions of the recording layer, while introducing ≤1nm of vertical topography. Magnetic properties suitable for high AD and high thermal stability were achieved with H$_C$ >8kOe and K$_U$V/k$_B$T ~180 in the as-implanted patterned state, but switching field distributions do need tightening via future process refinements. Further increases of H$_C$ to >11kOe were obtained by post-implant thermal annealing as shown in Fig. 42.

As the BPM AD increases, the magnetic media thickness is reduced which reduces the required implant dose and will help increase disk manufacturing throughput. Implantation is compatible with all servo pattern designs, may offer higher quality flyability from reduced topography, and lower cost since it replaces the etch step and eliminates the need for additional process steps to planarize, which might otherwise be required with etch-based magnetic patterning approaches. When combined with etching, ion implantation may offer a path to even higher resolution than etch-based patterning alone, since it circumvents the resolution limit caused by side-wall re-deposition which constrains etch-based patterning.

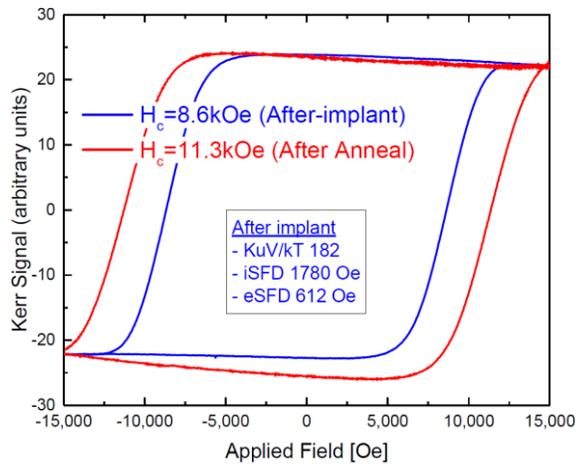

Fig. 42. Kerr Signal vs. field after P implantation of 7.5 nm thick $Co_{70}Cr_7Pt_{25}$ magnetic media masked with a 1 Td/in² hard mask. 8.6kOe (blue trace) was achieved from the initial implant and increased coercivity of 11.3kOe after annealing of the implanted media (red trace).

## V. RECORDING SYSTEM INTEGRATION

Many components work in tandem in the drive to store and retrieve data with high fidelity. On the component side there are the read/write head and the media components. Closely tied to these is the head-disk interface. On the system side there are the servo/mechanical, read channel, write synchronization, front-end electronics sub-systems, data architecture, and firmware. The integration of all of these systems into a functioning drive requires a careful awareness of the specifications of each system and the interactions between the various systems in order to arrive at a detailed balance that can meet customer requirements. This section will focus on several topics regarding the integration of BPM into an HDD. First we will discuss issues related to the head-disk interface followed by implementation of servo and write synchronization.

### A. Head-Disk Interface

Since the technology for fabricating BPM media generates a patterned disk topography, the resulting disk surface is inherently rougher than smooth continuous media in conventional PMR magnetic recording. This greater roughness can have a negative impact on the tribology and long-term reliability of BPM disk drives.

Some of the tribological challenges expected to be aggravated by the rougher BPM surface:

- Higher disk defect and asperity densities
- Poorer contact detection and clearance control for recording heads flying over the disk surface
- Higher average flying heights and spacing modulations of the recording head
- Poorer robustness of the BPM islands towards head-disk contacts
- And increased tendency for corrosion.

A potential way to solve the problems associated with the patterned topography of BPM is to planarize the BPM disk surface [99]–[101]. This may be accomplished by depositing an appropriate material to fill the grooves between the patterned bits, and then removing the excess fill material on the islands so that the final disk surface has a roughness comparable to conventional continuous media. (Using a vacuum etch process for planarization was discussed in Section IV-G-3.)

Implementing a planarization process, however, not only introduces additional fabrication steps (and associated costs) but may also introduce additional technology issues. A more economical approach would be to use unplanarized or partially planarized patterned media, combined with technology solutions for the tribology problems associated with the rougher BPM surface. An even more attractive solution would be to use templated growth BPM as discussed in Sections III-D and IV-H, which has the potential to produce BPM with a self-planarized surface.

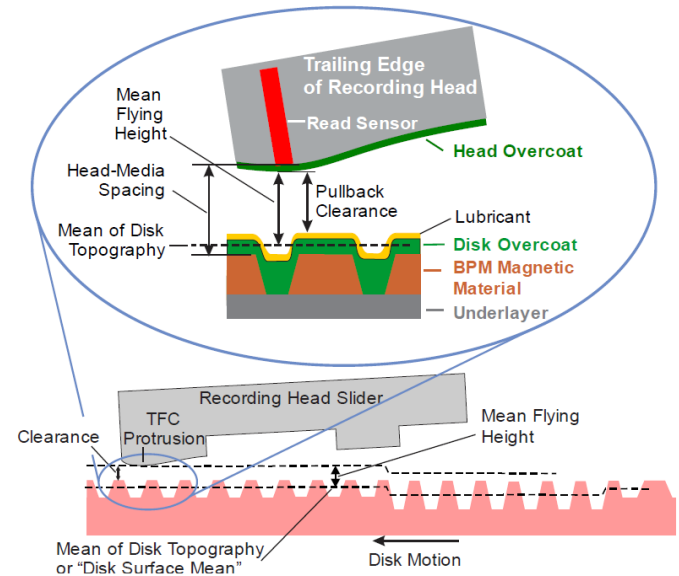

Fig. 43. The bottom schematic illustrates a recording head flying over a patterned disk surface. The mean flying height is defined as the separation from the mean of the disk surface to the closest part of the slider surface. The upper schematic provides an expanded view of the head-disk interface to illustrate the definition of head-media spacing (HMS) and some of the contributors to this spacing for BPM.

In this section, we discuss some of the impacts and tribological challenges that using unplanarized and partially planarized BPM disks can have on the performance of the head-disk interface (HDI). In particular, we address: 1) Head-media spacing, 2) Contact detection, 3) Flying height and spacing modulation.

#### 1) Head-media spacing (HMS)

One of the most important parameters for determining the final signal-to-noise ratio (SNR) performance of a magnetic recording system is the head-media spacing (HMS), typically defined as the distance from the top of the magnetic media to the bottom of the read and write elements in the recording head. (Fig. 43 illustrates how the BPM head-media spacing is defined, as well as some of the contributors to this spacing.)

Since the read and write processes in magnetic recording are a near field effect: as the bit AD goes up, the head-media spacing must go down. Historically for continuous media, the head-media spacing has been about 60% of the bit pitch length [102], and a similar head-media spacing criteria are expected to apply to BPM. For example, a 2.0 Td/in² BPM drive with BAR

= 1.5 and bit pitch = 14.5 nm is expected to require an HMS ≤ 8.7 nm.

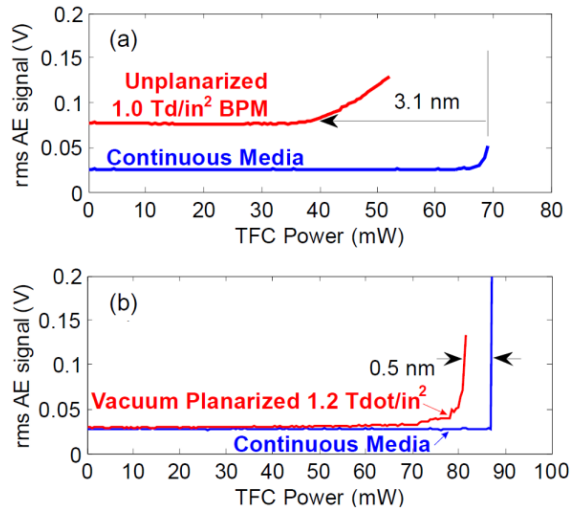

Fig. 44. Acoustic emission (AE) signal as a function of the power applied to the TFC heater for recording heads flying over BPM disks either (a) with a hexagonal array of unplanarized 1.0 Td/in² BPM islands or (b) with a rectangular array of 1.2 Td/in² BPM islands, partially planarized with the vacuum planarization etch process like that shown in Fig. 39. The AE vs. TFC touchdown curves on BPM are compared to ones done on conventional continuous PMR disks at the same conditions.

Even for continuous media, achieving such small head-media spacing is extremely challenging, as the thickness of the disk lubricant film and the head and disk overcoats are approaching the atomic and molecular limits of what can still provide adequate protection against corrosion and wear, while still allowing for a finite clearance between the head and disk surfaces, which move at speeds > 10 m/s relative to each other [102], [103]. For BPM, there is the added complication that all the islands need to be fabricated to have a very narrow height distribution so that HMS spacing does not vary substantially from bit to bit. For example, if it is desired to have a total HMS that varies < 10 %, then the three sigma width of the island height distribution for 2.0 Td/in² BPM will probably need to be << 0.9 nm to allow for additional contributions to variations in HMS, such as from variations in overcoat thickness, media land roughness, and flying height.

### 2) Contact detection on BPM

a.) Unplanarized media: To maintain the appropriate amount of clearance and HMS during read-write operations, modern HDDs use a small heater located near the recording elements to create thermal protrusion of the recording head toward the disk surface; this method is often referred to as thermal flying height control or TFC [104]. To set the proper clearance, the head is first protruded into contact with the disk and then backed off for a pullback clearance that ensures that excessive wear does not occur on the head and disk surfaces. Fig. 44(a) shows an example of contact detection during the TFC protrusion of a head towards an unplanarized 1.0 Td/in² BPM disk and towards a conventional continuous media disk. In these experiments, contact detection occurs when the acoustic emission (AE) signal rises above the background level, indicating that the head has started to contact the highest points on the disk surface. For conventional PMR, the AE signal rises sharply, as contact occurs on a smooth surface. For the unplanarized BPM surface, however, the AE signal rises only gradually. This more gradual rise occurs due to the smaller area of contact caused by the rougher patterned media surface, which generates a lower friction force and less friction induced bounce that would picked up by the AE sensor. Similar behavior has been previously reported by our lab for discrete-track media (DTM) [105].

b.) Partially planarized media: Fig. 44(b) shows an example of an AE vs. TFC power curve for contact detection for a head protruding into contact with a BPM disk partially planarized using a vacuum planarization etch process as discussed in Section IV. As can be seen in Fig. 44(b), the onset of contact is much sharper for this partially planarized surface than for the unplanarized BPM surface in Fig. 44(a), consistent with the partially planarized surface being substantially smoother than the unplanarized surface (but still not quite as smooth as conventional media).

### 3) Flying height and modulation of HMS

Another issue with flying a recording head over the rough, unplanarized pattern media is that the mean flying height can be much larger than for continuous media. (The definitions of clearance and mean flying height are illustrated in in Fig. 43.) One of the consequence of this is that the onset of contact occurs at much lower TFC power for on unplanarized BPM than for continuous media. For example, in Fig. 44(a), the onset of contact occurs 3.1 nm earlier for the unplanarized BPM surface, indicating that the disk surface mean is 3.1 nm lower than the maximum heights of the BPM islands compared to these maximum asperity heights on continuous media. For the partially planarized example in Fig. 44(b), however, the onset occurs only 0.5 nm earlier, demonstrating the effectiveness of this planarization process at reducing the mean groove depth.

Since the design of the appropriate air bearing surface on the recording head that results in the proper flying height is typically done using air bearing simulations, many groups have examined how having patterned disk topography influences the flying height performance [100], [105]–[109]. Fortunately, the consensus is that the presence of a patterned topography adds only a small and predictable correction to air bearing codes for heads flying over ideally smooth surfaces. This means that for BPM the mean flying height and other flying characteristics of a recording head can usually be well simulated using existing air bearing codes by substituting, at the disk surface mean height, a smooth surface for the patterned surface.

As a consequence, the flying height is expected to track, to first order, variations in the mean groove depths of the BPM disk surface. The bottom portion of Fig. 43 illustrates, however, a negative consequence of the flying height tracking the mean groove depth, in that a sudden change in head-media spacing can occur as the recording head flies over the transition between the data and servo zones, if these zones have different mean groove depths [110].

Measurements of such flying height modulation are shown in Fig. 45. For the disk in this example, the difference in surface means between the data and servo zones was ~1.5 nm. This led, however, to a flying height modulation that is almost 2× the height difference in the surface means. The enhancement of the magnitude of the flying height modulation compared the mean height difference between the data and servo zones is attributed to an air bearing pitch mode resonance being excited by the abrupt change in mean heights.

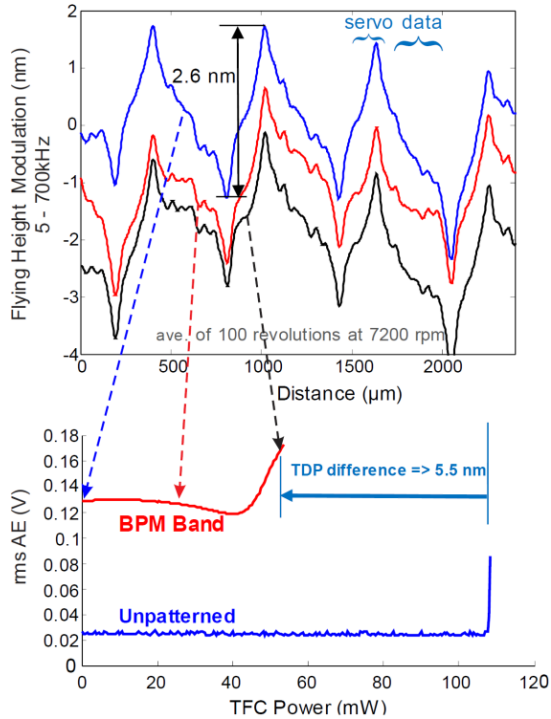

Fig. 45. Flying height modulation for a head flying over BPM. Top plot shows the flying height modulation measured at different TFC protrusion powers by laser Doppler vibrometry (LDV) for a recording head flying over disks patterned with data and chevron type servo zones with a difference in surface mean heights of 1.5 nm. Bottom plot shows the AE vs. TFC curve for this experiment. The magnitude of the flying height modulation was found to be independent of the amount of clearance.

## B. Servo and Servo Patterns

Integrating servo with BPM is challenging as the servo patterns have very different features than the data. This poses both fabrication and HDI concerns. The small BAR and resulting small track pitch places additional importance on the servo system, as the budget for track misregistration (TMR) becomes a larger fraction of the track pitch and a dominant limiter of the achievable AD. Nevertheless, we discuss in this section the servo requirements for 25 nm track pitch BPM and demonstrate that they are achievable today. We also discuss servo strategies that ease some of the HDI and fabrication risks.

### 1) Self-Registered Servo

BPM fabrication provides an opportunity to produce servo patterns as part of the patterning process, eliminating the need for servowriting during HDD manufacturing [64], [88]. Section IV-D presents one approach for creating such patterns. When servo patterns and data patterns originate from a common e-beam lithography exposure (as in Section IV-D), the data tracks and servo patterns are self-registered, which is the simplest case for servo system integration. Self-registered patterns can be conventional HDD patterns containing the usual sector header information such as track number, sector number, various sync fields, and a position burst (e.g., quad burst, null, or phase pattern). Bulk DC magnetization of such patterns yields a "ready to use" disk equivalent to a servowritten conventional disk. A limitation of this approach, however, is that the servo pattern resolution is limited to that of the e-beam lithography system, and thereby may not be written at the same resolution as self-assembled data track patterns. An alternative is to use self-assembled servo patterns, such as chevrons or offset fields as discussed in Section IV-B-5-c, which can also originate from a common e-beam guiding pattern and thereby be self-registered. These types of patterns, however, do not include prepatterned sector and track numbers, so an additional magnetic formatting step is generally required to provide this additional information in sector headers.

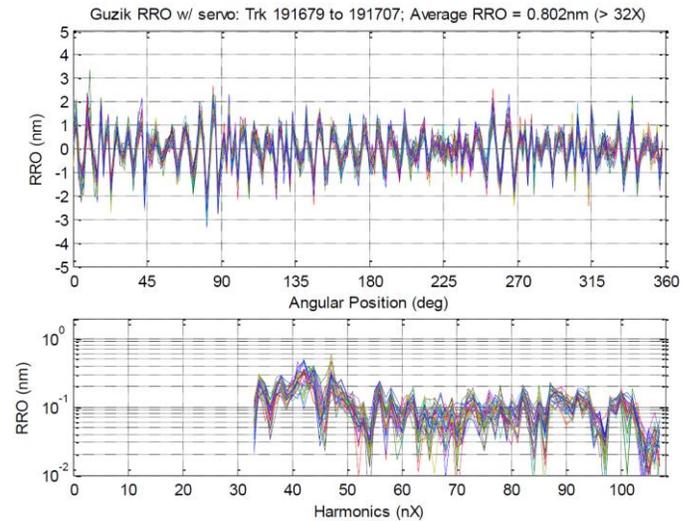

Fig. 46. High frequency (>32X) RRO measured for 200Gd/in² disk. Each color represents data from an individual revolution of the disk.

The performance of a BPMR servo system using e-beam patterned conventional servo patterns was evaluated on a spin stand. Through the analysis described later in this section, we estimated that the total position error signal (PES) requirement to meet 1000 KTPI AD is 1.17 nm. The total PES is sum of non-repetitive runout (NRRO), repetitive runout (RRO) and PES noise. The RRO itself can be divided into high frequency RRO (> 32X disk rotation frequency) and low frequency RRO. While the low frequency component of the repeatable run out can be compensated by servo track following, the high frequency RRO is difficult to compensate. In order to meet the requirements for 1000 KTPI, we need a high frequency RRO less than 0.82 nm. Since the RRO of BPM disks is generated by the e-beam writing step, a low density servo data template (200 Gd/in²) was written and disks made using nanoimprinting. The disk PES was measured on the spin stand and the high frequency RRO was calculated. Fig. 46 shows the high frequency RRO spectrum measured for the disk. The smallest high frequency RRO measured was 0.802 nm. The e-beam writer's capability to write master templates with high frequency RRO less than 0.82 nm was demonstrated and the overall servo system performance was found to be satisfactory.

*2) Nonregistered Servo*

At data densities above 1 Td/in², the strategy for scaling the servo pattern features must keep pace with the data. We have demonstrated the integration of servo and data on the same master, but since the scaling benefits of the resolution enhancing techniques of directed self-assembly in the periodic lattice of the data field are not directly applicable to certain servo and sector header patterns, it may be worthwhile to investigate alternative techniques. There are several issues in considering the impact of the alternatives, which are driven by choices in the fabrication of the servo pattern.

By including DSA features in the servo area, it is possible to scale the servo features to higher areal densities and to reduce the fabrication burden of separately integrating the servo and data areas on the same master. In this case, since the feature dimensions are similar to the data area, it is expected that the read head signal output level from the servo area in a DC erased state would be small and the SNR low. Therefore, it is advantageous to perform a servo formatting operation to magnetically polarize the servo area with bipolar magnetization. The usual method for writing the servo pattern by using all write heads simultaneously is most likely not available for BPM for two reasons. Since each disk surface in the stack is produced separately and some pattern misalignment between the surfaces is expected. Also, there is a relatively large tolerance to disk centering in the stack of about 50 μm. Therefore, the time and cost required for the servo write operation would be multiplied by the number of surfaces in the drive. In addition, the accuracy of alignment of the written servo waveform to the underlying patterned features must be a fraction of the magnetic island dimension in both the down-track and cross-track dimensions. This high degree of required positional accuracy without the benefit of fully polarized servo features may prove to be challenging.

The proposed approach that alleviates the burden of the issues described above is the non-registered chevron servo pattern. In order to optimize the fabrication process for both the servo and the data areas, the patterns are prepared on separate sub-masters. The final master is formed by imprinting the servo pattern onto the data pattern and performing an etch step to selectively remove islands in certain regions of the servo pattern areas. In addition to alleviating the burden of the fabrication integration, this approach can also completely eliminate the servo writing process. The minimum servo features can be designed with dimensions larger than the read head, so with the magnetic media in a DC erased state, the read head signal amplitude can be high with sufficient SNR. One of the drawbacks of the non-registered approach is that the misalignment between the data and the servo pattern can be equivalent to many tracks, so the relative positional information between the servo and data patterns must be determined and stored prior to writing and reading data. Also, since the servo pattern features are etched into a two-dimensional array of islands, rather than a continuous film, their edges can suffer from pixelation, causing magnetic transitions to shift away from intended locations. This introduces error into the servo signal, which is repeatable and can be calibrated. Since all imprint copies are nearly identical, the data from a single, careful calibration is valid across the entire disk population.

We have performed computer simulations of the response of a magnetic read head from a non-registered chevron servo pattern etched into a uniform bit-patterned media data island lattice. The simulations included standard analytical models for magnetic read head response such as Wallace spacing loss, read gap loss, head sensitivity function, media thickness loss and soft underlayer effect. To simulate a BPM data island lattice, a high resolution SEM image of a template was used, shown in Fig. 30(c). The down-track bit pitch of the template array was 14.5 nm and the cross-track pitch of the array was 26.5 nm. This corresponds to an AD of 1.6 Td/in². Several copies of the 1.6 μm x 1.6 μm image were seamlessly tiled to form a 10 μm x 10 μm image and superimposed a chevron servo pattern. An example of the two-dimensional read head response signal from this simulation is shown in Fig. 47. The area of the shown response is 1 μm x 0.6 μm. The down track direction is along the horizontal axis. The pitch of the chevron lines is 62 nm and the angle created with the cross track direction is 20 degrees. The media is uniformly magnetized along the normal direction.

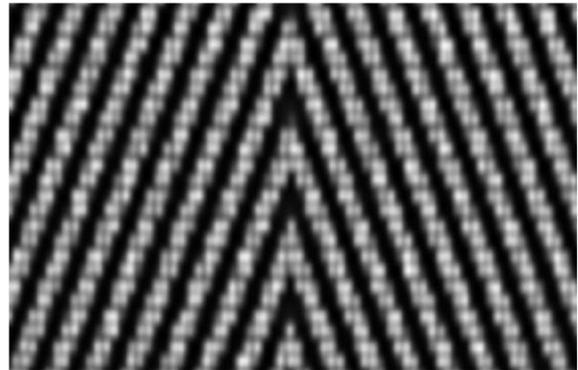

Fig. 47. Simulation of the read head response to a chevron PES pattern etched into a uniform island lattice. Area of the image is 1 μm x 0.6 μm. The horizontal axis corresponds to the data track direction. The pitch of the chevron pattern is 62 nm and the angle with respect to cross track direction is 20deg.

By using the SEM image of a real BPM template, rather than a synthesized array, all nonideal aspects of the fabrication process are included in the simulation, such as variation of line pitch, island position, size and shape. In this case, the pitch variation in the track direction was measured to be about 3.5% of the track pitch, and the pitch variation in the down track direction was about 2.7% of the bit pitch.

The cross track position estimate was obtained from the product of the chevron slope and the phase difference between the two halves of the chevron field. The position estimate error is the difference of the estimate and the ideal cross track position. An example of the error for the non-registered chevron calculation is shown in Fig. 48. The raw error, shown in the lighter line, has a large amplitude, periodic component, which is due to track pitch walking. By calculating a reference waveform, consisting of the average of the error with a period equal to twice the data pattern track pitch, and subtracting this reference waveform from each section of the position estimate error waveform of the same length, we obtain the residual error, shown in the heavier curve of Fig. 48. The standard deviation of the statistical ensemble of the position estimate error samples, which is closely matched with a Gaussian probability distribution well past 6σ, is equal to 0.3 nm.

This shows that the reference waveform subtraction method is well suited to removing most of the repeatable, periodic error components caused by systematic deviations from the ideal in the two-dimensional lattice. The residual, which is a small fraction of the raw error, is due to random position, size and shape jitter of the individual islands and can be reduced further by increasing the length of the PES field and reducing the random component of the patterned island position and size variation in the fabrication process.

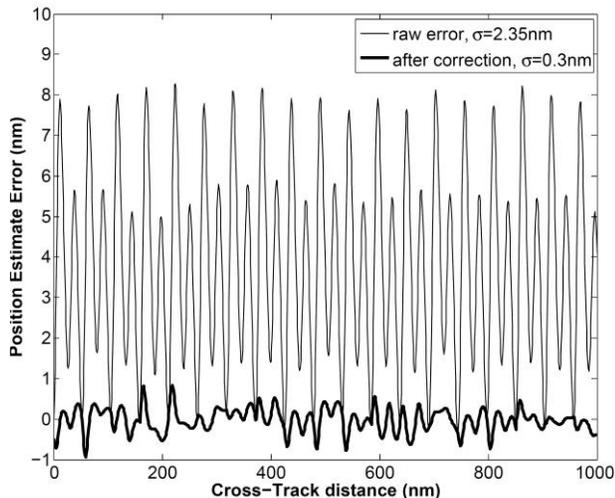

Fig. 48. Position estimate error as a function of cross track position. The lighter line shows raw error; the heavier line is the residual after reference waveform subtraction.

The residual error is fairly independent of the chevron pitch and angle, except for situations where the down-track chevron pitch is an integer multiple of the bit pitch. Under these conditions, the pixelation effect of the data lattice is not reduced since it is in phase with each chevron period.

### C. Write Synchronization

As described in Section II the recording system needs to maintain tight tolerances for the write phase in order to maintain BER below the sector failure rate threshold. Matching the write phase to the patterned islands requires an ability to accurate determine the island frequency and phase. The total budget allotted for write phase errors depends on the effective bit position jitter and the allowed sector failure rate. A good starting point for the total channel timing error requirement is a phase variation under 2% 1-sigma of the data rate across the whole sector. For a 4kB block this means a frequency accuracy of approximately $3 \times 10^{-7}$ if rephasing during the write is not employed.

A write synchronization system needs to be able to provision for several types of disturbances and distortion. The most obvious disturbance is the slow change in frequency due to spindle speed variation. More subtle are changes in the system timing due to temperature changes. In addition the frequency and phase has repeatable runout due to pattern-spindle misalignment and template distortions. This section discusses the resulting challenge brought about by template runout and the implementation and demonstration of write synchronization.

### 1) Zoning and Frequency RRO

As discussed in a previous section, the cylindrical symmetry required in an HDD disk can be imposed on a simple 2D lattice in one of two ways. In both cases one of the lattice vectors points along the radial direction aligned with the data tracks. In the first method, the magnitude of radial lattice vector increase linearly with radius, stretching the lattice along one axis. There is typically a practical limit to how much a lattice can stretch or compress, so disk made in this way are zoned to keep the range of expansion manageable, as shown in Fig. 49(a). This method is denoted as discrete zoning. In the second method, the radial lattice vector is kept constant and extra islands are added every several hundred tracks to fill in the additional space available by the increasing circumferential distance with radius. Care needs to be taken with the radial lattice vector as now it is impossible to maintain orthogonallity between the radial and circumferential lattice vectors around the whole disk. One solution is zone the disk into wedges, within which the radial lattice vector is points along a constant spatial direction as shown in Fig. 49(b). This method is denoted as continuous zoning.

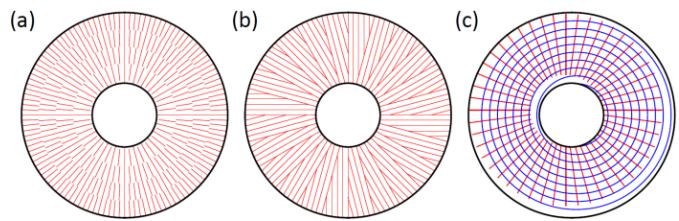

Fig. 49. (a) Discrete and (b) continuous zoning schemes. Runout between the track template and the radial template is shown in (c).

In the finished disk the radial and circumferential patterns will have runouts with respect to each other and with respect to the center of rotation of the disk as shown in Fig. 49(c). When track following the two runouts will cause a period variation of the bit frequency as seen by a write head flying of the disk. The dependence of the frequency runout on sub-template runout is different depending on whether the radial template is discrete zoned or continuous zoned. To first order in the runout the frequency varies as

$$f = f_0\left(1 + \frac{RRO_{rad}}{R_{trk}}\cos(\theta - \theta_{rad})\right), \text{discrete zone} \quad (8)$$

$$f = f_0\left(1 + \frac{RRO_{trk}}{R_{trk}}\cos(\theta - \theta_{trk})\right), \text{continuous zone} \quad (9)$$

Plots of the first order and higher order frequency runout for the two zoning schemes are shown in Fig. 50.

### 2) Write clock PLL

To reach the required timing accuracy, a PLL circuit is used to create an internal clock that is phase locked to the address marks on the disk. This circuit uses the timing and phase measured between address marks to update the write frequency and phase for the write clock. For a stable system the timing error is much smaller than a clock cycle and the timing measurement reduces to a phase difference measurement between the PLL clock and the phase of the address marks.

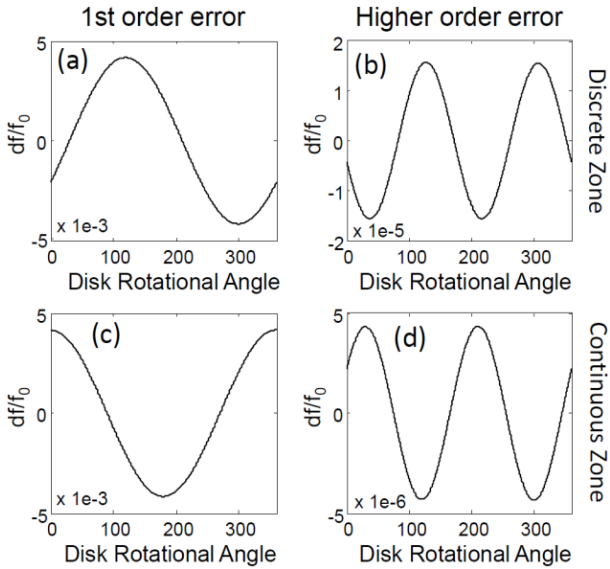

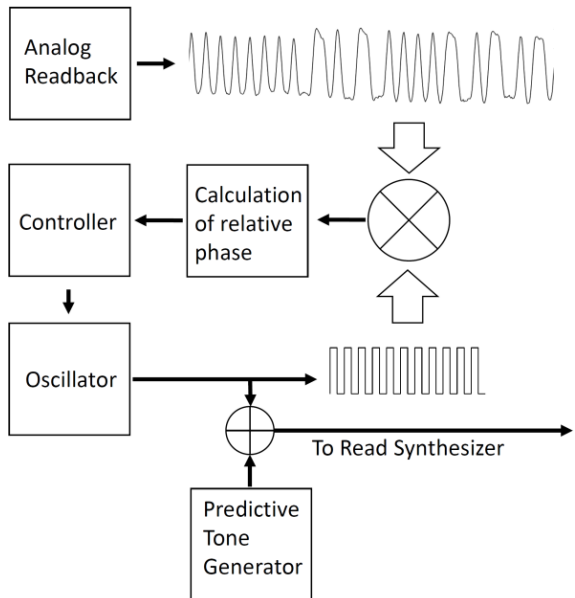

Fig. 50. Frequency RRO at radius 12 mm (near inner diameter) position due a 50 μm misalignment of the track and circumferential templates relative to the center of disk rotation. The track runout is along angle 0 and the radial runout along 120 degrees. The runout is shown as the change df of the frequency normalized by the average frequency $f_0$ around the rev. (a) & (b) show first and higher order frequency RRO for discrete zoning, while (c) & (d) show the same for continuous zoning.

Fig. 51 shows the operating principle of the PLL circuit. An oscillator with a frequency close to twice the data rate is used to drive a sampling circuit that synchronously samples the analog readback signal. The relative phase between analog readback and clock is then determined by a discrete Fourier transform. A controller then locks the frequency and the phase of the oscillator to the timing marks on the disk.

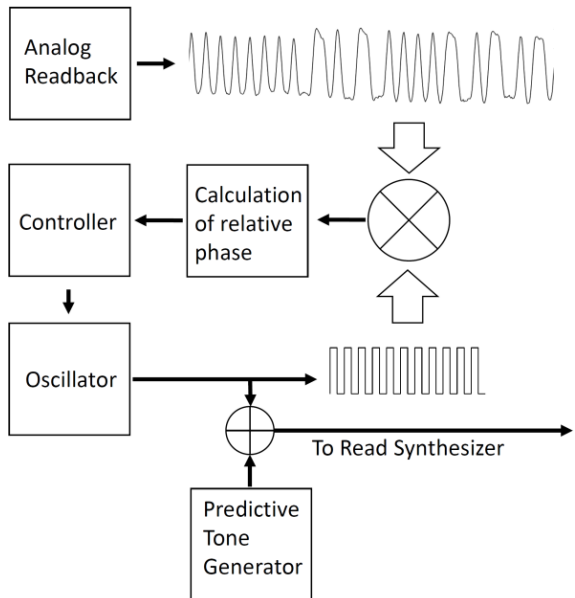

Fig. 51. Block diagram of the write synchronization system. The analog signal is synchronously demodulated to determine the relative phase with respect to the oscillator with a precision $2^{-7}$ bits.

Fig. 52 shows the timing accuracy achieved with a write synchronization system that employed a circuit similar to that shown in Fig. 52. The measurement was performed on a 200

Gd/in² BPM disk with 216 sync fields distributed around the disk. The timing jitter shown in Fig. 52 represents the sync mark to sync mark timing variation using the write clock whose frequency and phase was adjusted by the feedback circuit to keep the timing constant. Only 1.6% of the events are off by more than 5% of the bit pitch. The accuracy is 1-sigma for a deviation of 2% from the center of the bit.

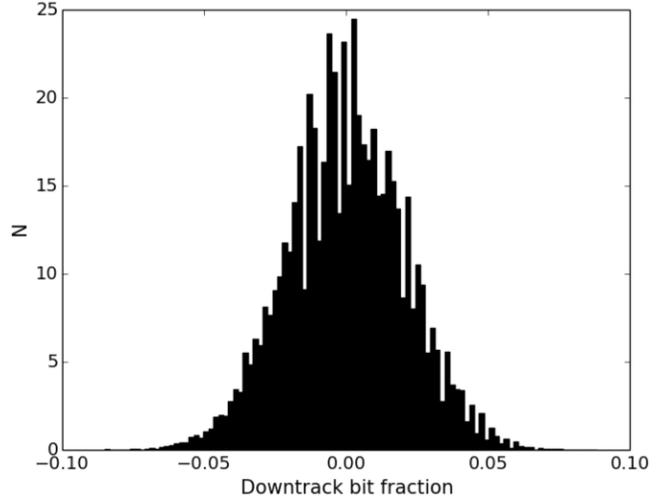

Fig. 52 Timing accuracy of the disk locked clock circuit. The accuracy is 1-σ for a deviation of 2% from the center of the bit.

To compensate for known frequency and phase variation caused by the misalignment and template distortion, a predictive tone generator can be used to shift the oscillator frequency to feed-forward a correction for the known distortion. Because of hardware constraints, we were not able to implement a predictive tone generator for the results shown here. The frequency variation around the disk measured on the same 200 Gb/in² disk is shown in Fig. 53. The first harmonic of the repeatable runout was diminished by centering the patterned tracks to the center of spindle rotation. The second harmonic of the frequency runout is clearly visible as well as a sector to sector variation (see e.g. sectors 175-200). Both of these variations stem from known pattern variations in the disk.

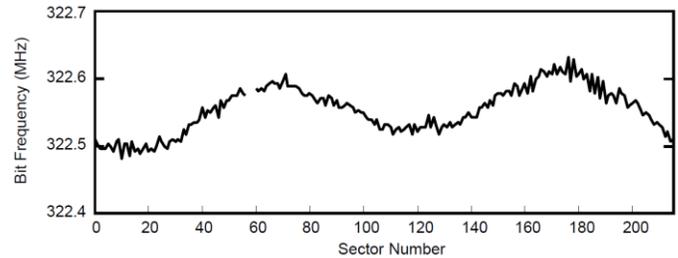

Fig. 53. Frequency variation along the track due to the residual runout. For this experiment the disk was centered within 100 nm of the spindle axis. For this reason the 2nd order term dominates and the 1st order is barely visible.

Fig. 54 shows the bit error rate as a function of the down-track write phase on a 200 Gd/in² disk, measured on a spin stand, using the synchronization circuit described in Fig. 51. The minimum of the BER is 4 x $10^{-4}$, with the residual errors roughly equally split between timing and media.

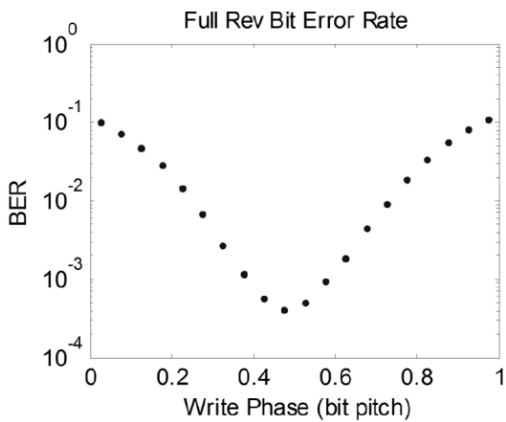

Fig. 54. Bit error rate (BER) as a function of the down-track write phase for 200 Gd/in² media.

Full surface drive operation requires the ability to match the zone frequency during seeks. This can be accomplished by adding an additional term to the predictive tone generator derived from the servo subsystem.

### D. Recording System Modeling

Due to simpler reversal dynamics in BPM as compared to PMR media, many issues in recording system modeling can be addressed without using micromagnetic modeling. In this section we introduce the concepts of switching probability contours and of the Ω-point, which are used in Section V-E in the analysis of BAR optimization. We expand on the concepts introduced in Section II-V and include in our treatment the 3D head field profile and system tolerances and uncertainties. The goal is to generate effective off-track erasure and on-track failure contours and use them to optimize BAR geometry and determine max achievable AD for the given head, media, and system parameters.

Let $BER_w$ be the maximum allowed write-addressability bit error rate [10]. The on track failure contour $C_2$ is defined as the contour outside of which the switching probability falls below $P_2 = 1 - BER_w$. The off-track erasure contour $C_1$ is defined as the contour inside of which the switching probability is above $P_1 = BER_w$. These contours, shown in Fig. 55, are computed by considering the relevant island properties (iSFD, lithographic placement tolerance, etc.), the head field profile, and the system uncertainties (TMR uncertainty, synchronization jitter, etc.). The reversal probability can be computed by comparing island Hc to the island averaged head field, or using micromagnetic simulations if higher accuracy is desired.

In determining the optimized track pitch ($TP$) the Ω-point is of particular significance, as illustrated in Fig. 55, since it measures the lateral extent of the write field in a probabilistic view that is predicated by the addressability bit error rate $BER_w$ for the adjacent track. Therefore the Ω-point can provide a critical mark from which the track pitch $TP$ can be referenced. $TP$ must be large enough to provision of TMR events. Assuming Gaussian servo fluctuations with standard deviation $\sigma_{PES}$, a certain multiple $\beta_s$ of $\sigma_{PES}$ must be included in the cross-track pitch depending on the permissible rate of sector failure rates. Aiming for a ~$10^{-12}$ chance of the write head straying into the adjacent track motivates $\beta_s \sim 7$. The TMR margin is hence approximately $\pm 7\ \sigma_{PES}$ and must be set aside relative to the nominal cross-track location of the head trajectory. Adding $\beta_s \sigma_{PES}$ in the positive y-direction yields the radial shingling set point. Another step with $\beta_s \sigma_{PES}$ in the positive y-direction gives the point $P_+$. Next we query the horizontal intercepts $\Delta_+$, $\Delta_c$, and $\Delta_-$ which are defined in Fig. 55. The down-track pitch $BP$ is given by $BP = \max(\Delta_+, \Delta_c, \Delta_-)$. Different initial choices for $TP$ generally lead to different $BP$ hence to different BAR, where BAR = $TP/BP$. For each pair of ($BP$, $TP$) the areal density $AD$ is obtained from $AD = 645.16/(BP*TP)$, where $AD$ is measured in Td/in² and $BP$ and $TP$ are in nm. This process for calculating the AD can be applied repeatedly to search for the maximum the AD as a function of BAR.

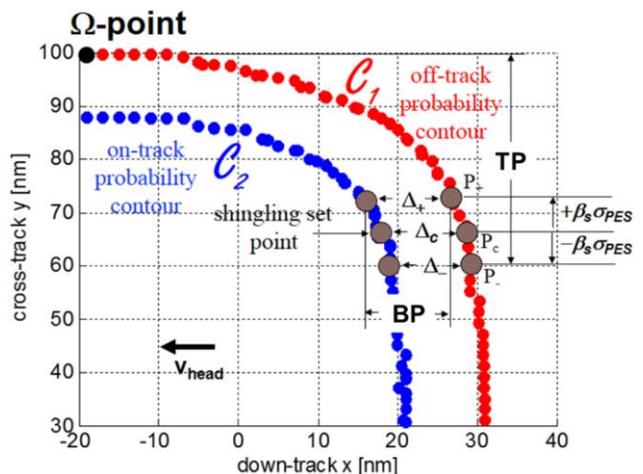

Fig. 55. Probability contours showing the shingling set point, down-track pitch BP, and cross-track pitch TP.

Note that for many write heads there is an ordering

$$\Delta_+ < \Delta_c < \Delta_-, \qquad (10)$$

which leads to the down-track bit-pitch given by $BP = \Delta_+$. The ordering of $\Delta_+$, $\Delta_c$, and $\Delta_-$ reflects the increase of the down-track gradients of $H_{eff}$ as one approaches the cross-track center of the trailing edge of the write pole. It also is a reminder of the trade-off in shingled recording relative to centered-write approaches, which allow larger down-track densities.

The above discussion shows also that the bit pitches depend on the local, high-frequency fluctuations (contained in the probability contours), on the low-frequency servo fluctuations (described by $\beta_s$) and on the global properties of the effective write-field profile (captured by the Ω-point). The general shape of $C_1$ and $C_2$ depends on the head design (shield spacing, gradients, and tails of the head field profile) and on the center and width of the switching field distribution of the BPM islands.

### E. Effect of Servo Fluctuations

In this section we use the concepts of the previous section to illustrate the impact of servo fluctuations on a particular design example at about 1-2.5 Td/in². A write head is used with wrap-around shields [18] having side-gaps and trailing gaps of 30 nm. The magnetic spacing between the air-bearing surface of the head and the top surface of the magnetic layer of the islands is 6 nm. The islands have a single 6 nm thick layer of magnetic material with and average anisotropy field $<H_k>=2<K_U/M_s>=19$ kOe, the moment density $M_s=600$ emu/cm$^3$. The intrinsic switching field distribution of the islands is 400 Oe and the dipole-induced switching field distribution of the islands is 150 Oe. The local lithographic placement errors $\sigma_x$ and $\sigma_y$ are both 1 nm, $\sigma_{sync}$ is 0.9 nm. The target bit error rate BER$_w$ is chosen to be $<10^{-3}$. The servo fluctuations are characterized by Gaussian standard deviations $\sigma_{PES}$ in the range of 0.6 to 1.5 nm. We choose $\beta_x=7$ to achieve a chance of better than about $10^{-12}$ that the mechanical excursion of the write head reaches the adjacent track and induces erroneous island writing. We note that target BER$_w$, $\sigma_{sync}$, and sector failure rate are conservative and more aggressive numbers would yiled larger areal densities.

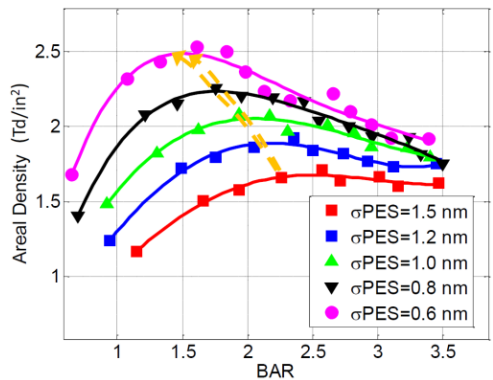

Fig. 56. AD as a function of BAR and $\sigma_{PES}$.

Under these assumptions we carried out Monte-Carlo simulations similar to [111] and computed the probability contours $C_1$ and $C_2$. Then we calculated the AD as a function of BAR as discussed in the previous section. The results are summarized in Fig. 56, where the AD is plotted as a function of BAR and of $\sigma_{PES}$. We see that the maximum areal densities are achieved for BARs in the range of about 1.4 to 2.3, depending on $\sigma_{PES}$. If $\sigma_{PES}$ gets larger the AD potential markedly decreases. Furthermore, the optimum BAR increases to accommodate the larger lateral excursions of the $\Omega$-point of the probability contours.

The concepts and considerations of this section informed the fabrication strategies for rectangular bitcells that were discussed in earlier sections.

## VI. BPM RECORDING EXPERIMENTS AT 1.6 Td/IN$^2$

The potential recording performance of the 1.6 Td/in$^2$ disk is examined using a drag tester, which scans a conventional perpendicular magnetic recording (PMR) head over the medium at 100 μm/sec while performing write and read operations [112], [113], [114], [26]. Fig. 57 shows two-dimensional (2D) readback images of the BPM disk before and after writing one track with pseudo-random data. Prior to imaging, the media was field AC-erased. Fig. 57(a) illustrates the 2D readback of the field erased BPM. The reader width is about 40 nm, which is almost twice the medium cross-track dot pitch. The reader resolution is just enough to make out areas of the medium with checker-board magnetization configuration. The configuration of the magnetization of the dots is actually evaluated using two dimensional decoding.

First, the dot positions are extracted from the absolute values of the readback image. Next, the image of a center dot and its 8 nearest neighbors is compared to the $2^9$ reference readback images of 3x3 (9-dot) rectangle patterns. For the first decoding iteration, the $2^9$ reference images are calculated from a best guess reader point-spread function. After the first decoding iteration, the 2D reader point-spread function is derived from the decoded image and is used for the following decoding iteration. The optimization of the reader point spread function and the image decoding are repeated until the decoded image results in a stable configuration of the magnetic dots. As mentioned earlier, the decoding itself is two dimensional and done using a 2D Viterbi algorithm [115]. Fig. 58 shows an example of a decoded image.

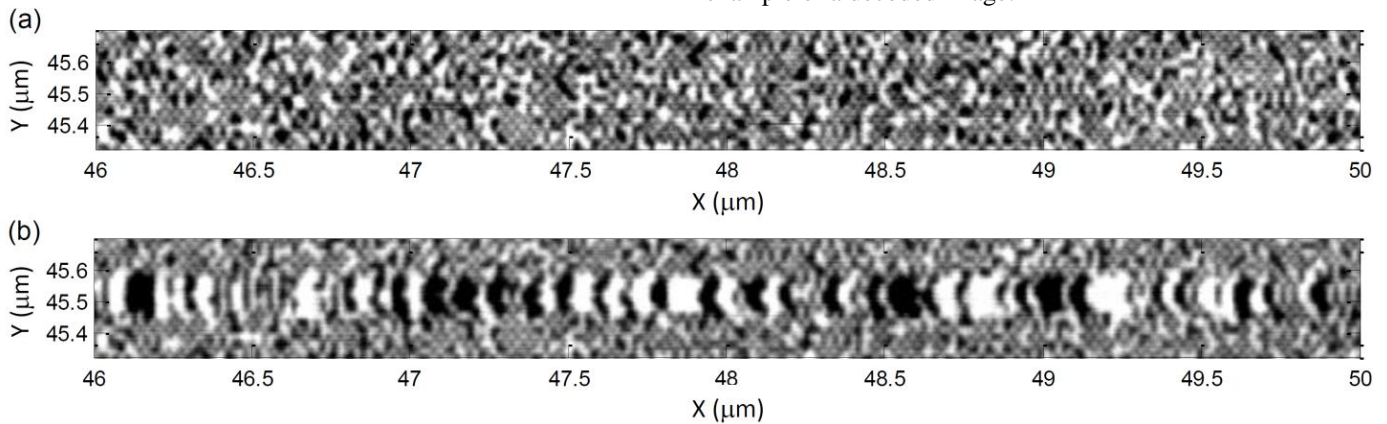

Fig. 57. (a) Two-dimensional readback image of the field AC erased 1.6 Td/in$^2$ bit patterned media disk. (b) Readback image of the same area after writing one track with pseudo-random data.

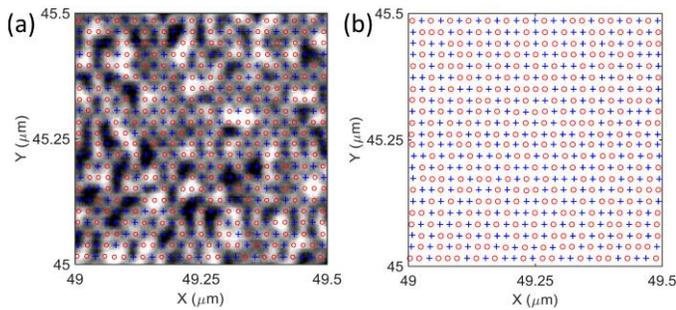

Fig. 58. (a) Readback image of the BPM medium overlaid with the bits position and their decoded magnetization state. (b) Image of the bits position and their decoded magnetization states (red dot: bit with magnetization up, blue cross: bit with magnetization down).

Fig. 57(b) shows the same medium area after writing a single track with pseudo-random data. The write process consists on pulsing the writer current every 185 µs, which is close to every dot pitch. The write pulse duration is typically 50 nsec and the write current is 25 mA. The writer width is much larger than the medium cross-track pitch: the written track is 120 nm wide corresponding to more than 5 dots cross-track. Note that the recording experiments are performed without write synchronization. The writer position relative to the dots is determined *a posteriori*.

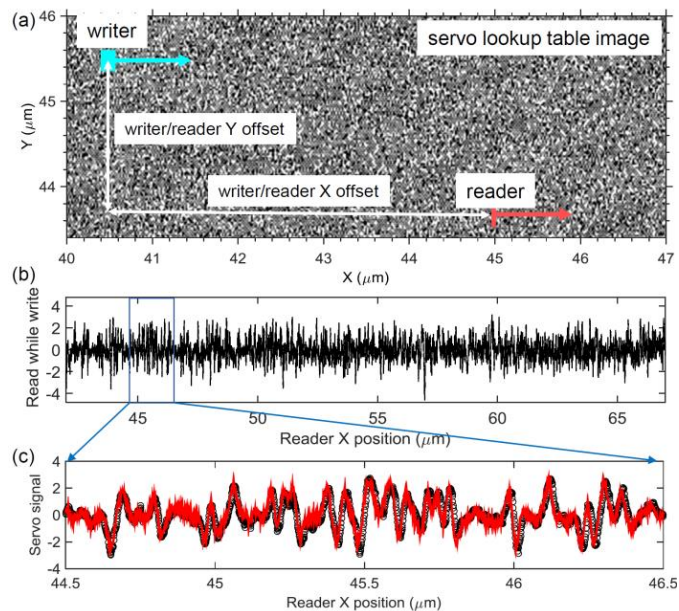

Fig. 59. (a) 2D readback image of used for finding writer position during write process (servo lookup table). (b) Readback trace recorded during a write operation. (c) Snippet of the "servo" readback trace (black circles) and best match from servo lookup table image (red line). This locates the reader during write at position Y=43.78 µm for X=45.6 µm, and allows to derive the corresponding writer position at X=39.57 µm, Y=45.52 µm.

The writer motion is not registered to the medium dots, but the signal of the reader is captured during the write process. This "servo" readback signal is compared to a lookup table readback image captured prior to the write events, as shown in Fig. 59. The lookup image is in a field AC erased state. Snippets of the "servo" readback waveform are compared to the lookup readback image. The best match with the lookup image gives the position of the reader at a specific time of the write process.

After shifting by a constant writer/reader offset, the writer position as a function of write time is obtained. In addition, the writer position relative to the dot lattice is derived for each write pulse: this defines a down-track write phase and a cross-track writer offset. From this knowledge, the configuration of the magnetic dots after a write event can be compared to the input write sequence.

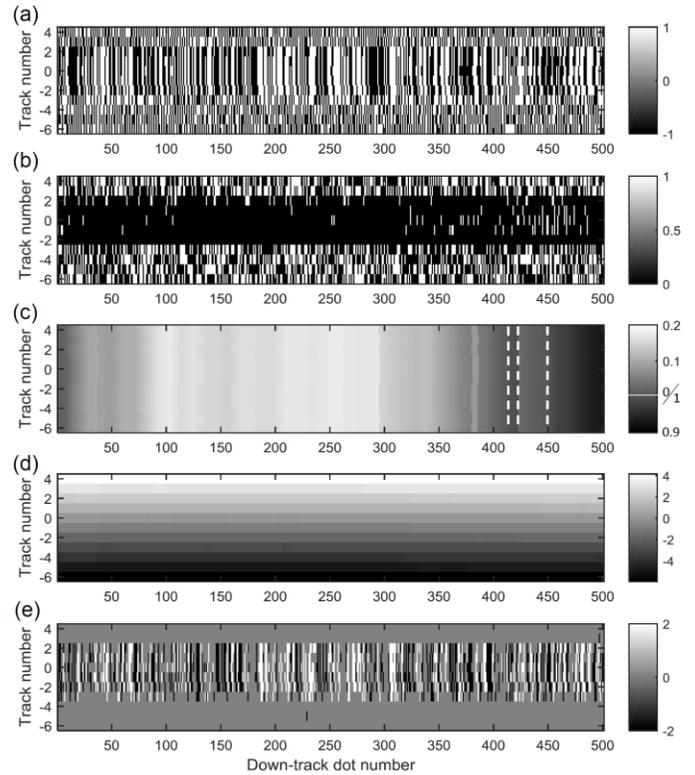

Fig. 60. (a) 2D image of the decoded dots after writing one track with pseudo random data. (b) 2D image of the dots in error after writing the track (0: no error, 1: error). (c) Writer registration in the down-track direction normalized to the dot period. For this recording, the down-track phase varies between 0 and 0.2, shifting to between 0.9 and 1 after phase shifts, which are indicated by dotted white lines. (d) Writer registration in the cross-track direction normalized to the cross-track pitch, in the range of -6 to +4 track pitches. (e) 2D image of the dots that have flipped during the write process as obtained by comparison of the decoded dots before and after writing (0: not flipped, 2: flipped to positive, -2: flipped to negative).

Fig. 60(a) shows for instance the decoded magnetization state after writing one track with pseudorandom data. From the knowledge of the writer position with time, Fig. 60(b) is derived that shows the dots that are in error, i.e. the dots that were not written properly with reference to the input pseudorandom sequence. As seen on Fig. 60(b), there are only small number of errors in the center of the track. These errors correspond to medium defects or write phase errors. Fig. 60(c) and (d) show the write phase down-track and cross-track, respectively, which measure the writer position relative to the dot lattice for each write pulse. These images, and hundreds of equivalent images, are used to generate the on-track bit error rate (OTER) map as a function of the on-track and cross-track writer/dot registration (Fig. 61(a)). The x-axis is the write phase normalized to the dot pitch. The y-axis is the cross-track offset from the center of the write pole. Outside of the written track, where the medium is not written, ½ of the dots end up in error as the initial magnetization state is random. The OTER map shows wide margins at $10^{-2}$ error rates and below in both the cross-track and down-track directions. Analysis of the error rate floor show a defect error rate of 6 x $10^{-3}$, as shown in Fig. 62(a). The OTER is consistent with the known defect rate in the sample of $10^{-3}$. In the center of the track, the OTER abruptly increases towards 0.5 at one specific write phase. This corresponds to the write location being in the center of the dot. The latter varies with cross-track offset as a result of the curvature of the effective field under the pole of the writer. Given the distributions of the dot position, switching field, dot size, and the finite field gradients, the OTER increase is not abrupt and can be fitted with an error function characterized by effective down-track jitter $\sigma_{DT}$ as shown in Fig. 62(a). $\sigma_{DT}$ follows Eq. 2 with the first term, $\sigma_{mag,DT} = \sqrt{\left(\sigma_{iSFD}^2 + \sigma_{dipole}^2\right)}/(dH/dx)$, that characterizes the media and field distributions, $\sigma_{litho,DT}$ is the lithographic placement error, and $\sigma_{sync,DT}$ is the head positioning error during write.

Fig. 60(e) shows a two-dimensional image of the dots that have flipped following a write event. This map is the difference of the decoded images before and after writing: black and white corresponds to dots that flipped, grey corresponds to dots that did not flip. Using images in Fig. 60(c-e) and hundreds of equivalent images, a write rate map can be generated as a function of the on-track and cross-track writer/dot registration. This map is shown in Fig. 61(b) and a cross-section along the cross-track direction is shown in Fig. 62(b). Statistically, half of the dots are expected to have flipped in the center of the track and almost none well outside the track. The edges of the write rate map are not abrupt because of the BPM distributions and the cross-track writer field gradients. Theses cross-track distributions are characterized by an effective cross-track jitter, which is extracted by fitting the write rate profile with error functions. $\sigma_{CT}$ follows Eq. 2, with $\sigma_{mag,CT} = \sqrt{\left(\sigma_{iSFD}^2 + \sigma_{dipole}^2\right)}/(dH/dy)$ the magnetic contribution to the jitter, $\sigma_{litho,CT}$ is the lithographic placement error, and $\sigma_{sync,CT}$ is the head positioning error during write.

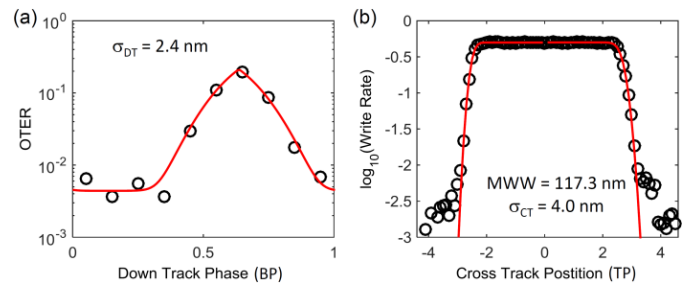

Fig. 62. (a) OTER profile vs. down-track phase in the center of the track. (b) Write rate profile vs. cross-track position averaged over all down-track phases. Circles are extracted from Fig. 61 maps. Lines are fit of the error rate and write rate with error functions.

For this medium we find that $\sigma_{DT}$ = 2.4 nm and that $\sigma_{CT}$ = 4.0 nm. These effective jitters are still dominated by the magnetic jitter. We estimate that the effective write field gradient are 300 Oe/nm down-track and 150 Oe/nm cross-track. The island switching field distribution is dominated by its intrinsic component iSFD, measured to be 540 Oe. We obtain $\sigma_{mag,DT}$ = 1.8 nm and $\sigma_{mag,CT}$ = 3.6 nm. The position distributions are estimated from the analysis of scanning electron microscopy images: $\sigma_{litho,DT}$ = 1.1 nm and $\sigma_{litho,CT}$ = 1.2 nm. This leads to $\sigma_{sync,DT}$ ~ 1 nm and $\sigma_{sync,CT}$ ~ 1.2 nm, which accounts for the piezo stage accuracy and the servo decoding scheme.

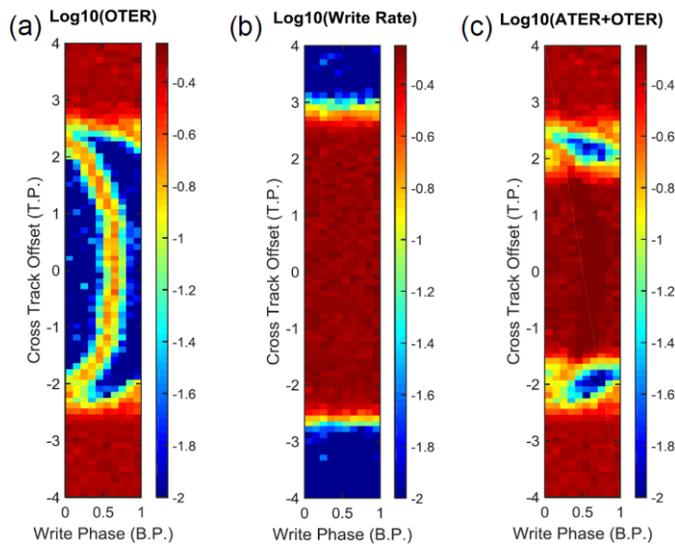

Fig. 61. (a) Map of on-track error rate as a function of the down-track and cross-track dot/writer registration. (b) Map of the write rate as a function of the dot/writer registration. (c) Map of the error rate after adjacent track writing, as a function of the down-track and cross-track dot/writer registration.

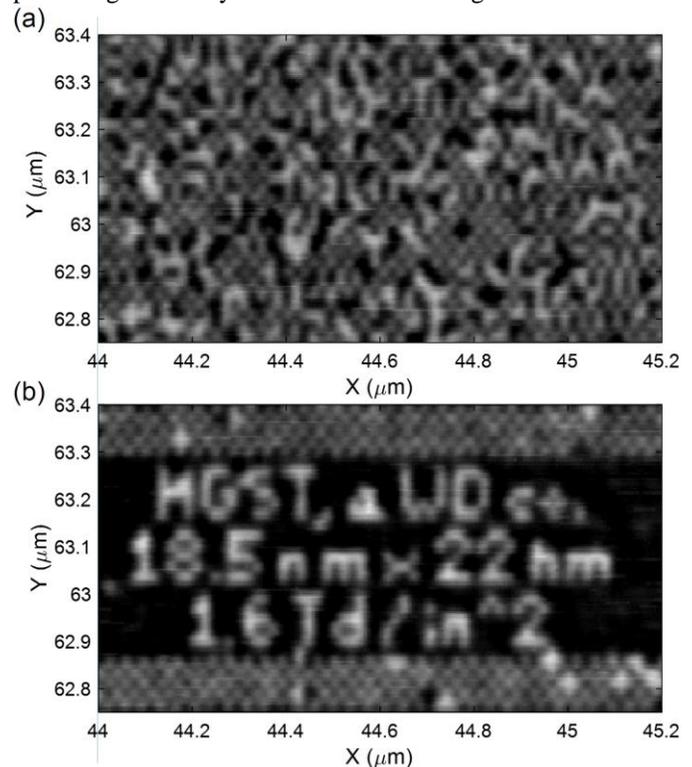

Fig. 63. Readback image (a) before and (b) after shingle writing individual dots of the 1.6 Td/in² disk. The write process was performed open loop.

The performance of the BPM system for shingle magnetic recording is evaluated by adding to the on-track error rate, errors that result from adjacent-track writing rate (ATWR). This is achieved by adding the on-track errors of Fig. 61(a) and the errors due to overwrite during adjacent track writing. The latter is the write rate shifted by one track pitch up for the lower writer edge and by one track pitch down for the upper writer edge. The combined ATWR and OTER map is shown in Fig. 61(c). It indicates that sub $10^{-2}$ BER recording is obtained at both corners of the write head with wide margins ($\sim$+/-3.5 nm down-track and $\sim$+/-4.5 nm cross-track). These experiments demonstrate that such 1.6 Td/in$^2$ medium can to be successfully recorded with shingle recording at an error rate of less than $10^{-2}$. Fig. 63 further illustrates the recording achievement by showing a 2D readback image after successfully shingle writing individual dots of the 1.6 Td/in$^2$ disk.

## VII. EXTENDIBILITY

The ultimate extendibility of BPM technology is difficult to determine as it requires a careful consideration of the ultimate extendibility of the fabrication processes, materials, components, and subsystems that will make up the BPM storage system [9], [10], [11], [12], [18], [116]. In addition, in order for the technology to be useful it must be cost competitive with alternative technologies that serve the same purpose. This section will not attempt to predict the development of supporting technologies or the emergence of competing technologies. Instead we will try to bring up the relevant issues for considering extendibility of BPM and in particular the requirements for island properties in order to meet margin requirements imposed by additional subsystems.

### A. Extendibility Assessment Framework

As discussed in a previous section, a data storage system needs to be able to write and retrieve customer data with an acceptable irretrievable sector failure rate. Data loss can occur during the write process, via thermal decay, or via failure of some other subsystem of the drive. The acceptable irretrievable sector failure rate depends on the application. A higher failure rate might be tolerable in a system with high data redundancy within the drive or among an array of independent HDDs. Likewise a system that has additional data integrity verification process can allow a larger sector failure rate.

For the write process, it is useful to think of the different system requirements as components of down-track and cross-track write registration budgets that needs to sum up to match the physical dimensions of the islands. There are four main components of these budgets: (1) effective island placement errors (2) write synchronization errors due write frequency and phase mismatch, (3) track misregistration (TMR) errors, and (4) write width and write curvature provisioning. These components were discussed in Section II and are illustrated in Fig. 64.

The effective bit placement budget reflects the extent of the probability contour for which the number of bit errors due to effective bit jitter exceeds the sector failure threshold. The effective bit placement errors are assumed to be uncorrelated from bit to bit while TMR and write synchronization errors are constant across the whole sector. The write width and write curvature provisioning stems from non-ideal matching of head write field contours to the patterned islands. A track pitch smaller that the write width of the head will be difficult to achieve if centered track writing is desired. Head field curvature reduces the gradients near the edge of the write bubble and can produce an "overwrite band" in which the write rate is 100% but the on-track bit error rate is high enough to cause sector failure. Head field curvature also necessitates compensation in the write phase for TMR. Hence TMR uncertainty factors into the write synchronization budget.

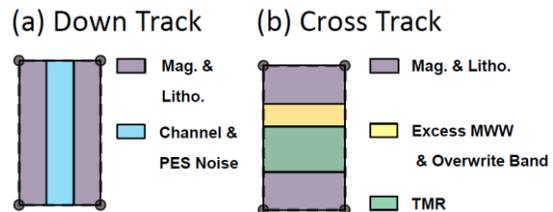

Fig. 64. Schematic of (a) down-track & (b) cross-track margin budgets. The black dots represent the locations of the islands in the unit cell.

To avoid assessing the details of different storage applications, we will assume that the TMR, synchronization, and magnetic write width (MWW) budgets capture subsystem tolerances and drive operational requirements such as the allowed sector failure rate. For example a system that can handle a $10^{-5}$ sector failure rate will have smaller TMR and write synchronization budget than one with a $10^{-12}$ sector failure rate requirement. Likewise, advances in track following technologies will reduce TMR sigmas and reduce the TMR budget, allowing other budgets to be larger or the pitches to be reduced. We assume that a sector failure will occur if the number of bits in the incorrect state exceeds 1 in 100. While advances in error correcting codes can increase the number of allowed bit errors, the impact on magnetic & lithographic budgets is roughly logarithmic in BER (see Fig. 3 in Section II). We do not consider readback errors or the feasibility of fabricating readers and creating readback schemes that have sufficient cross-track resolution to resolve the narrow track pitches.

### B. Island Tolerances and Areal Density

We examine BPM extendibility by determining the required magnetic tolerances and BAR for accommodating different imposed TMR and MWW budgets. Magnetic tolerance refers to the contribution to effective island position jitter from island iSFD, dipolar interaction broadening, and head field gradient. Lithographic placement tolerance is assumed to scale as 5% of the smallest patterned dimension. As the head field gradient can be different in the cross-track and down-track the stated magnetic tolerance can differ along the two directions. The stated tolerance is the smaller of the down-track and cross-track tolerances. The write synchronization budget is fixed at 30% of the down-track budget.

In the first scenario the TMR budget is 10 nm and the MWW is 40 nm. This scenario represents one possibility for centered track recording with a narrow conventional write pole and low sector failure rate operation. A 10 nm TMR budget indicates that +/- 5 nm TMR excursions need to be tolerated without the BER exceeding $10^{-2}$. The second scenario lifts the MWW constraint while keeping the TMR budget fixed at 10 nm. Such a scenario represents shingled magnetic recording or heat assisted recording conditions under low sector failure rates. The last scenario reduces the TMR budget to 6 nm, representing either a highly advanced servo system or a recording system that is highly tolerant of sector failure events. The choice of 6 and 10 nm TMR budgets is illustrative and we make no statements about the recording system design that would make these budgets practical.

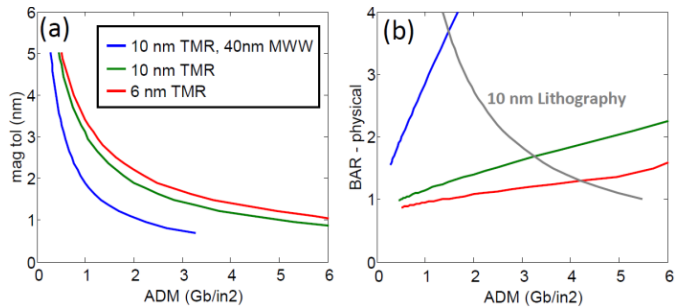

Fig. 65. (a) Required magnetic tolerance in order to meet the stated areal density with margin (ADM) for three scenarios of required TMR and MWW tolerance. (b) Corresponding BAR for the three scenarios. The gray line represents where on the BAR vs ADM curve 10 nm lithography will be required.

Fig. 65 shows the magnetic tolerance and optimized BAR as a function of customer AD in Gb/in². Higher AD not surprisingly requires tighter magnetic tolerances and optimization towards a larger BAR. For the stated magnetic tolerance at given AD, an increase or decrease in the BAR will result in a lower AD. Larger TMR and MWW budget requirements decrease the required magnetic tolerance and increase the optimized BAR. The requirement for centered track writing with finite MWW poses the biggest restriction on achievable AD as the constrained track pitch forces the bit pitch to run into lithographic limits.

The computations only show the magnetic tolerance requirements, but say nothing as to whether such tolerances or assumptions can be achieved. Current conventional recording heads can generate gradients in the range of 300-400 Oe/nm in the down-track direction. If a similar gradient is achieved in the cross-track direction, then 300-400 Oe of combined iSFD and dipole broadening will produce of magnetic tolerance of 1 nm. iSFDs in this range have been achieved at 1 Td/in² previously [26]. Dipolar broadening is greatly reduced in the recording environment due to the randomized nature of recorded data and shielding by the head shield. As a result we do not believe it will be a significant contributor to the magnetic tolerance budget.

Combining heat assisted magnetic recording with bit pattern media produces several known synergies: lower required laser power and higher thermal gradients [117]. Thermal gradients > 20 K/nm might be achievable with BP-HAMR. The required sigma of the media Curie temperature ($T_C$) of 20 K in order to achieve 1 nm magnetic tolerances is already possible in granular HAMR media [118]. As BPM allows a tighter distribution of island properties, achieving sigma $T_C$ below 20 K at high densities seems feasible.

Relaxing the 5% lithographic placement tolerance will require a tighter magnetic tolerance. The question of what lithographic tolerance and can ultimately be achieved is difficult to answer, although it should be noted that lithography innovations for scaling minimum feature size are often accompanied by solutions for making similar gains in tolerances. Current semiconductor roadmaps predict nodes out to 8 nm half pitch in the 2023 time frame. The unique requirements and lithographic approaches being pursued for BPM may provide a path for smaller features and tighter tolerances than indicated in semiconductor roadmaps. As a point of reference we draw the 10 nm full pitch lithography (5 nm half pitch) boundary for reference in Fig. 65.

### C. Thermal Stability & Writeability Limits

For recording with a conventional write pole, the magnetic recording trilemma of writeability, SNR, & thermal stability will limit the achievable AD. As discussed in a previous section, we suggest a minimum average $K_U V/K_B T$ of 67 in order to maintain 10 year thermal stability and 10% island size variations. The corresponding $K_U$ requirement for a 65 C operating temperature depends on the highest effective fields and filling factors that can be generated. Exchange coupled composite structures can improve the writeability [14]. Templated growth BPM could achieve an 80% filling factor and 8-10 nm thick magnetic layers. Independent of system requirements, thermal stability will limit AD for BPM somewhere in the 2-3 Tb/in² range without the use of energy assist [116].

For BP-HAMR thermal fluctuations during the write process, rather than long term thermal stability, can limit achievable AD. Several authors have examined these effects and concluded that the thermal fluctuation limit for BP-HAMR is anywhere between 4 and 20 Tb/in² [9], [10], [116].

## VIII. SUMMARY

BPMR has been developed into a competitive technology for future HDDs, with potential to extend AD further than any other proposed technology. The theory of BPMR is well understood and relatively simple compared to conventional recording on granular media. CoCrPt alloy provides suitable magnetic properties for BPM, and has been demonstrated to support acceptable error rate at up to 1.6 Td/in² recording density. Considerable challenges for BPM fabrication have been overcome by a combination of innovative lithographic technologies including BCP DSA, SADP, and nanoimprinting. Both etching and templated growth provide suitable pattern transfer to CoCrPt films. Overall media manufacturability, however, remains a concern. Although BPMR adds significant new complexity to the recording system in terms of write synchronization, servo, and head-disk interface, solutions are known and being demonstrated successfully.